\documentclass[10pt]{article}
\usepackage{graphicx}
\usepackage{natbib}
\usepackage{times}
\usepackage{fullpage}
\usepackage{amsmath, amssymb, amstext}

\providecommand{\norm}[1]{\lVert#1\rVert}
\newcommand{\erf}{\text{erf}}
\newcommand{\Times}[1]{\ensuremath{\times \text{10}^{#1}}}
\newcommand{\Ep}{\ensuremath{\epsilon_p}}
\newcommand{\Epi}{\ensuremath{\epsilon_{pi}}}
\newcommand{\Epdot}[1]{\ensuremath{\dot{\epsilon}_{p#1}}}
\newcommand{\Xidot}{\ensuremath{\dot{\xi}}}

\newcommand{\That}{\ensuremath{\hat{T}}}

\newcommand{\Bnabla}{\ensuremath{\boldsymbol{\nabla}}}

\newcommand{\Bsig}{\ensuremath{\boldsymbol{\sigma}}}

\newcommand{\Beta}{\ensuremath{\boldsymbol{\eta}}}

\newcommand{\Ba}{\ensuremath{\boldsymbol{a}}}

\newcommand{\Bd}{\ensuremath{\boldsymbol{d}}}

\newcommand{\Bf}{\ensuremath{\boldsymbol{f}}}

\newcommand{\Bm}{\ensuremath{\boldsymbol{m}}}

\newcommand{\Bs}{\ensuremath{\boldsymbol{s}}}

\newcommand{\Bu}{\ensuremath{\boldsymbol{u}}}
\newcommand{\Bv}{\ensuremath{\boldsymbol{v}}}

\newcommand{\Bx}{\ensuremath{\boldsymbol{x}}}

\newcommand{\BD}{\ensuremath{\boldsymbol{D}}}

\newcommand{\BG}{\ensuremath{\boldsymbol{G}}}

\newcommand{\Tr}{\ensuremath{\text{~tr}}}

\newcommand{\Tint}{\ensuremath{\text{int}}}
\newcommand{\Text}{\ensuremath{\text{ext}}}

\newcommand{\Half}{\ensuremath{\frac{1}{2}}}

\newcommand{\Grad}[1]{\ensuremath{\Bnabla #1}}

\newcommand{\Partial}[2]{\ensuremath{\frac{\partial #1}{\partial #2}}}

\begin{document}

  \title{An evaluation of plastic flow stress models for the simulation 
         of high-temperature and high-strain-rate deformation of metals}
  \author{Biswajit Banerjee
          \footnote{Phone: 1-801-585-5239, Fax: 1-801-585-0039, 
               Email: banerjee@eng.utah.edu}\\
           Department of Mechanical Engineering, University of Utah, \\
           50 S Central Campus Dr., MEB 2110, 
           Salt Lake City, UT 84112, USA}
  \maketitle
  \raggedright
  \setlength{\parskip}{8pt}

  \begin{abstract}
  Phenomenological plastic flow stress models are used extensively in the
  simulation of large deformations of metals at high strain-rates and
  high temperatures.  Several such models exist and it is difficult to 
  determine the applicability of any single model to the particular 
  problem at hand.  Ideally, the models are based on the underlying (subgrid)
  physics and therefore do not need to be recalibrated for every regime
  of application.  In this work we compare the Johnson-Cook, 
  Steinberg-Cochran-Guinan-Lund, Zerilli-Armstrong, 
  Mechanical Threshold Stress, and Preston-Tonks-Wallace plasticity models.  
  We use OFHC copper as the comparison material because it is well
  characterized.  First, we determine parameters for the specific heat model, 
  the equation of state, shear modulus models, and melt temperature models.
  These models are evaluated and their range of applicability is identified.
  We then compare the flow stresses predicted by the five flow stress models
  with experimental data for annealed OFHC copper and quantify modeling errors.
  Next, Taylor impact tests are simulated, comparison metrics
  are identified, and the flow stress models are evaluated on the basis of 
  these metrics.  The material point method is used for these computations.
  We observe that the all the models are quite accurate at low temperatures and
  any of these models could be used in simulations.  However, at high 
  temperatures and under high-strain-rate conditions, their accuracy can vary
  significantly.
  \end{abstract}

\section{Introduction}
  The Uintah computational framework (\citet{Dav2000}) was developed
  to provide tools for the simulation of multi-physics problems such as
  the interaction of fires with containers, the explosive deformation
  and fragmentation of metal containers, impact and penetration of materials,
  dynamics deformation of air-filled metallic foams, and other such 
  situations.  Most of these situations involve high strain-rates.  In some
  cases there is the additional complication of high temperatures.  This work 
  arose out of the need to validate the Uintah code and to quantify 
  modeling errors in the subgrid scale physics models.  

  Plastic flow stress models and the associated specific heat, shear modulus,
  melting temperature, and equation of state models are subgrid scale models
  of complex deformation phenomena.  It is unreasonable to expect that 
  any one model will be able to capture all the subgrid scale physics under
  all possible conditions.  We therefore evaluate a number of models which
  are best suited to the regime of interest to us.  This regime consists of 
  strain-rates between 10$^3$ /s and 10$^6$ /s and temperatures between
  230 K and 800 K.  We have observed that the combined effect of high 
  temperature and high strain-rates has been glossed over in most other 
  similar works (for example \citet{Zerilli87,Johnson88a,Zocher00}).  Hence 
  we examine the temperature dependence of plastic deformation at high 
  strain-rates in some detail in this paper. 

  In this paper, we attempt to quantify the modeling errors that we get when we 
  model large-deformation plasticity (at high strain-rates and high
  temperatures) with five recently developed models.  These models are
  the Johnson-Cook model (\citet{Johnson83}), the Steinberg-Cochran-Guinan-Lund 
  model (\citet{Steinberg80,Steinberg89}), the Zerilli-Armstrong model
  (\citet{Zerilli87}), the Mechanical Threshold Stress model 
  (\citet{Follans88}), and the Preston-Tonks-Wallace model (\citet{Preston03}).
  We also evaluate the associated shear modulus models of \citet{Varshni70},
  \citet{Steinberg80}, and \citet{Nadal03}.  The melting temperature models
  of \citet{Steinberg80} and \citet{Burakovsky00} are also examined.
  A temperature-dependent specific heat relation is used to compute specific
  heats and a form of the Mie-Gr{\"u}neisen equation of state that assumes
  a linear slope for the Hugoniot curve are also evaluated.  We suggest that
  the model that is most appropriate for a given set of conditions can be 
  chosen with greater confidence once the modeling errors are quantified, 

  The most common approach for determining modeling error is the 
  comparison of predicted uniaxial stress-strain curves with experimental
  data.  For high strain-rate conditions, flyer plate impact tests provide
  further one-dimensional data that can be used to evaluate plasticity models.  
  Taylor impact tests (\citet{Taylor48}) can be use to obtain two-dimensional
  estimates of modeling errors.  We restrict ourselves to comparing uniaxial
  tests and Taylor impact tests in this paper; primarily because 
  high-temperature flyer plate impact experimental data are not readily 
  available in the literature.  We simulate uniaxial tests and Taylor 
  impact tests with the Material Point Method (\citet{Sulsky94,Sulsky95}).
  The model parameters that we use in these simulations are, for the most
  part, the values that are available in the literature.  We do not 
  recalibrate the models to fit the experimental data that we use for our
  comparisons.  For simplicity, we use annealed OFHC copper as the material for
  which we evaluate all the models because this material is well-characterized. 
  A similar exercise for various tempers of 4340 steel can be found elsewhere 
  (\citet{Banerjee05b}).

  Most comparisons between experimental data and simulations involve 
  the visual estimation of errors.  For example, two stress-strain curves or
  two Taylor specimen profiles are overlaid on a graph and the viewer 
  estimates the difference between the two.  We extend this approach 
  by providing quantitative estimates of the error and providing metrics
  with which such estimates can be made.  The metrics are discussed and
  the models are evaluated on the basis of these metrics.

  The organization of this paper is as follows.  Section~\ref{sec:models}
  discusses the specific heat model, the equation of state, the melting 
  temperature models, and the shear modulus models.  Flow stress models
  are discussed in Section~\ref{sec:flow} and evaluated on the basis of
  one-dimensional tension and shear tests.  Section~\ref{sec:taylor} discusses
  experimental data, metrics, and simulations of Taylor impact tests.
  Conclusions are presented in Section~\ref{sec:conclude}.

\section{Models}\label{sec:models}
  In most computations involving plastic deformation, the specific heat,
  the shear modulus, and the melting temperature are assumed to be constant.
  However, the shear modulus is known to vary with temperature and pressure.
  The melting temperature can increase dramatically at the large pressures 
  experienced during high strain-rate deformation.  In some materials, the
  specific heat can also change significantly with change in temperature.
  If the range of temperatures and strain-rates is small then these variations
  can be ignored.  However, if a simulation involves a change in strain-rate
  from quasistatic to explosive, and a change in temperature from ambient
  values to values that are close to the melt temperature, the temperature-
  and pressure-dependence of these physical properties has to be taken into
  consideration.  

  The models used in our simulations are discussed in this section.  The
  material response is assumed to be isotropic.  The stress is decomposed
  into a volumetric and a deviatoric part.  The volumetric part of the 
  stress is computed using the equation of state.  The deviatoric part of the
  stress is computed using an additive decomposition of the rate of 
  deformation into elastic and plastic parts, the von Mises yield condition, 
  and a flow stress model.  The variable shear modulus is used to update both 
  the elastic and plastic parts of the stress and is also used by some of the 
  flow stress models.  The melting temperature model is used to determine if 
  the material has melted locally and also feeds into one of the shear modulus 
  models.  The increase in temperature due to the dissipation of plastic work
  is computed using the variable specific heat model.  We stress 
  physically-based models in this work because these can usually be used in 
  a larger range of conditions than empirical models and need less
  recalibration.
  
  Copper shows significant strain hardening, strain-rate sensitivity, and 
  temperature dependence of plastic flow behavior.  The material is quite
  well characterized and a significant amount of experimental data are 
  available for copper in the open literature.  Hence it is invaluable for 
  testing the accuracy of plasticity models and validating codes that simulate 
  plasticity.  In this work, we have only considered fully annealed 
  oxygen-free high conductivity (OFHC) copper and electrolytic tough pitch 
  (ETP) copper.

  \subsection{Adiabatic Heating, Specific Heat, Thermal Conductivity}
  A part of the plastic work done is converted into heat and used to update the 
  temperature of a particle.  The increase in temperature ($\Delta T$) due to 
  an increment in plastic strain ($\Delta\epsilon_p$) is given by the equation
  \begin{equation}
    \Delta T = \cfrac{\chi\sigma_y}{\rho C_p} \Delta \epsilon_p
  \end{equation}
  where $\chi$ is the Taylor-Quinney coefficient (\citet{Taylor34}), 
  and $C_p$ is the specific heat.  The value of the Taylor-Quinney coefficient 
  is assumed to be 0.9 in all our simulations (see \citet{Ravi01} for more 
  details on the variation of $\chi$ with strain and strain-rate).
  The specific heat is also used in the estimation of the change in 
  internal energy required by the Mie-Gr{\"u}neisen equation of state.  

  The specific heat ($C_p$) versus temperature ($T$) model used in our
  simulations of copper has the form shown below.  The units of $C_p$ 
  are J/kg-K and the units of $T$ are degrees K. 
  \begin{align}
    C_p & = \begin{cases}
              0.0000416~T^3 - 0.027~T^2 + 6.21~T - 142.6 & 
                 \text{for}~~ T < 270 \text{K} \\
              0.1009~T + 358.4 & 
                 \text{for}~~ T \ge 270 \text{K} 
          \end{cases} \label{eq:CuSpHeat}
  \end{align}
  
\begin{figure}[p]
    \centering
    \scalebox{0.50}{\includegraphics{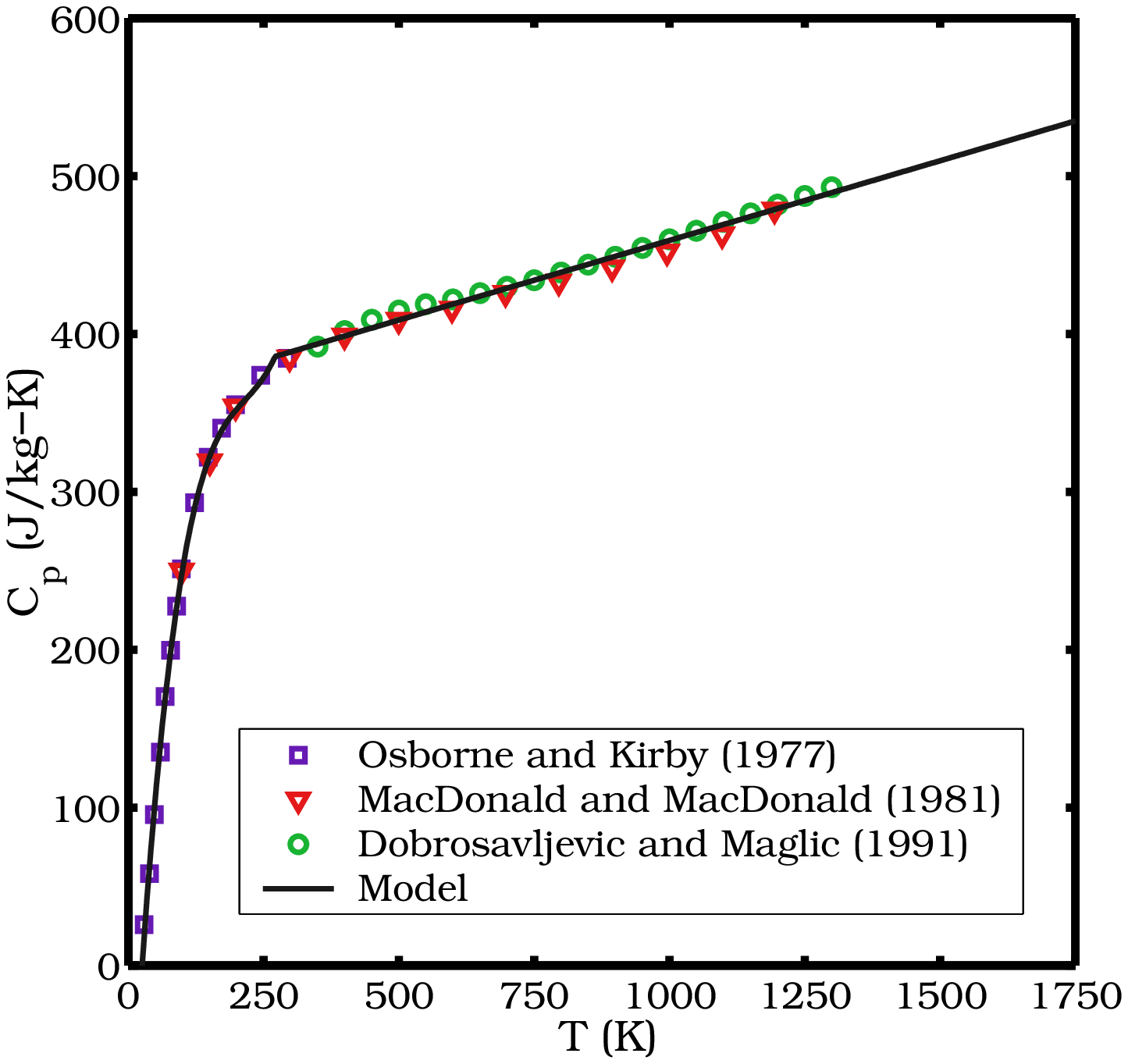}}
    \caption{Variation of the specific heat of copper with temperature.
             The solid line shows the values predicted by the model.
             Symbols show experimental data from \citet{Osborne77},
             \citet{MacDonald81}, and \citet{Dobro91}.}
    \label{fig:CuSpHeat}
  \end{figure}

  A constant specific heat (usually assumed to be 414 J/kg-K) is not 
  appropriate at temperatures below 250 K and temperatures above 700 K, as 
  can be seen from Figure~\ref{fig:CuSpHeat}.
  The specific heat predicted by our model (equation (\ref{eq:CuSpHeat})) 
  is shown as a solid line in the figure.  This model is used to compute
  the specific heat in all the simulations described in this paper.
  
  The heat generated at a material point is conducted away at the end of a 
  time step using the transient heat equation.  The thermal conductivity of
  the material is assumed to be constant in our calculations.  The effect of 
  conduction on material point temperature is negligible for the high 
  strain-rate problems simulated in this work.  We have assumed a constant 
  thermal conductivity of 386 W/(m-K) for copper which is the
  value at 500 K and atmospheric pressure.  

  \subsection{Equation of State}
  The hydrostatic pressure ($p$) is calculated using a temperature-corrected 
  Mie-Gr{\"u}neisen equation of state of the form used by \citet{Zocher00} 
  (see also \citet{Wilkins99}, p.61)
  \begin{equation} \label{eq:EOSMG}
   p = \frac{\rho_0 C_0^2 (\eta -1)
              \left[\eta - \frac{\Gamma_0}{2}(\eta-1)\right]}
             {\left[\eta - S_{\alpha}(\eta-1)\right]^2} + \Gamma_0 E;\quad
   \eta = \cfrac{\rho}{\rho_0}
  \end{equation}
  where $C_0$ is the bulk speed of sound, 
  $\rho_0$ is the initial density, $\rho$ is the current density, 
  $\Gamma_0$ is the Gr{\"u}neisen's gamma at reference state, 
  $S_{\alpha} = dU_s/dU_p$ is a linear Hugoniot slope coefficient, 
  $U_s$ is the shock wave velocity, $U_p$ is the particle velocity, and
  $E$ is the internal energy per unit reference specific volume.  
  
  The change in internal energy is computed using
  \begin{equation}
    E = \frac{1}{V_0} \int C_v dT \approx \frac{C_v (T-T_0)}{V_0}
  \end{equation}
  where $V_0 = 1/\rho_0$ is the reference specific volume at temperature 
  $T = T_0$, and $C_v$ is the specific heat at constant volume.  In our
  simulations, we assume that $C_p$ and $C_v$ are equal. 

  The hydrostatic pressure is used to compute the volumetric part of the 
  Cauchy stress tensor in our simulations.  The parameters that we use
  in the Mie-Gr{\"u}neisen equation of state are shown in 
  Table~\ref{tab:CuEOS}.
  \begin{table}[p]
    \caption{Parameters used in the Mie-Gr{\"u}neisen EOS for copper.
             The bulk speed of sound and the slope of the linear fit to 
             the Hugoniot for copper are from \citet{Mitchell81}.  
             The value of the Gr{\"u}neisen gamma is from \citet{MacDonald81}.
             }
    \vspace{10pt}
    \begin{minipage}[t]{6in}
      \centering
      \begin{tabular}{ccccccccccc}
         \hline
         $C_0$ (m/s) & $S_{\alpha}$ & $\Gamma_0$ ($T <$ 700 K) & 
              $\Gamma_0$ ($T \ge$ 700 K) \\
         \hline
         3933 & 1.5          & 1.99   & 2.12 \\ 
         \hline
      \end{tabular}
    \end{minipage}
    \label{tab:CuEOS}
  \end{table}

\begin{figure}[p]
    \centering
    \scalebox{0.50}{\includegraphics{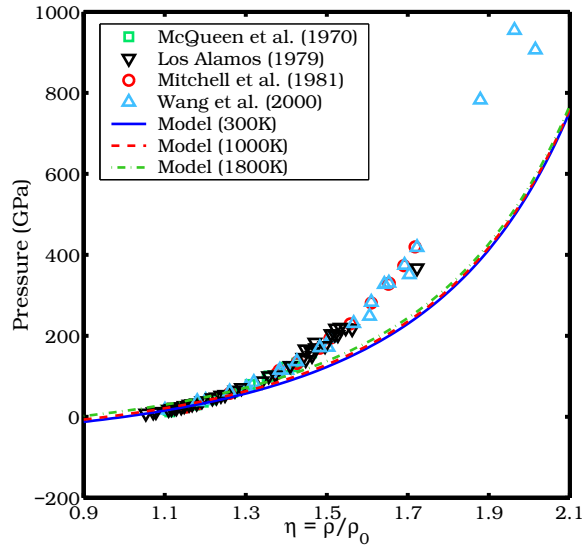}}
    \caption{The pressure predicted by the Mie-Gr{\"u}neisen equation
             of state for copper as a function of compression.  The continuous
             lines show the values predicted by the model for three
             temperatures.  The symbols show experimental data obtained from 
             \citet{McQueen70}, \citet{LAHugoniot79}, \citet{Mitchell81},
             and \citet{Wang00}.  The original sources of the experimental
             data can be found in the above citations.}
    \label{fig:CuEOS}
  \end{figure}

  Figure~\ref{fig:CuEOS} shows plots of the pressure predicted by the
  Mie-Gr{\"u}neisen equation of state at three different
  temperatures.  The reference temperature for these calculations is
  300 K.  An initial density $\rho_0$ of 8930 kg/m$^3$ has been used 
  in the model calculations.  The predicted pressures can be compared
  with pressures obtained from experimental shock Hugoniot data (shown
  by symbols in Figure~\ref{fig:CuEOS}).  The model equation of state 
  performs well for compressions less than 1.3.  The pressures are 
  underestimated at higher compression.  We rarely reach compressions greater
  than 1.2 in our simulations.  Therefore, the model that we have used is 
  acceptable for our purposes.
  
  \subsection{Melting Temperature}
  The melting temperature model is used to determine the pressure-dependent
  melt temperature of copper.  This melt temperature is used to 
  compute the shear modulus and to flag the state (solid or liquid) 
  of a particle.  Two melting temperature models are evaluated in this paper.  
  These are the Steinberg-Cochran-Guinan (SCG) melt model and the 
  Burakovsky-Preston-Silbar (BPS) melt model.
  
  \subsubsection{The Steinberg-Cochran-Guinan (SCG) melt model} 
  The Steinberg-Cochran-Guinan (SCG) melt model (\citet{Steinberg80}) 
  is a relation between the melting temperature ($T_m$) and the applied
  pressure.  This model is based on a modified Lindemann law and has the form
  \begin{equation} \label{eq:TmSCG}
    T_m(\rho) = T_{m0} \exp\left[2a\left(1-\frac{1}{\eta}\right)\right]
              \eta^{2(\Gamma_0-a-1/3)}; \quad
    \eta = \frac{\rho}{\rho_0}
  \end{equation}
  where $T_{m0}$ is the melt temperature at $\eta = 1$, 
  $a$ is the coefficient of the first order volume correction to 
  Gr{\"u}neisen's gamma ($\Gamma_0$).  

  \subsubsection{The Burakovsky-Preston-Silbar (BPS) melt model} 
  An alternative melting relation that is based on dislocation-mediated
  phase transitions is the Burakovsky-Preston-Silbar (BPS) model
  (\citet{Burakovsky00}).  The BPS model has the form
  \begin{align}
    T_m(p) & = T_m(0)
      \left[\cfrac{1}{\eta} + 
            \cfrac{1}{\eta^{4/3}}~\cfrac{\mu_0^{'}}{\mu_0}~p\right]~; 
    \quad
    \eta = \left(1 + \cfrac{K_0^{'}}{K_0}~p\right)^{1/K_0^{'}} 
    \label{eq:TmBPS}\\
    T_m(0) & = \cfrac{\kappa\lambda\mu_0~v_{WS}}{8\pi\ln(z-1)~k_b}
               \ln\left(\cfrac{\alpha^2}{4~b^2\rho_c(T_m)}\right)
  \end{align}
  where $p$ is the pressure, $\eta$ is the compression (determined using
  the Murnaghan equation of state), 
  $\mu_0$ is the shear modulus at room temperature and zero pressure, 
  $\mu_0^{'} := \partial\mu/\partial p$ is the derivative of the shear modulus 
  at zero pressure, $K_0$ is the bulk modulus at room temperature and
  zero pressure, $K_0^{'} := \partial K/\partial p$ is the derivative of the 
  bulk modulus at zero pressure, $\kappa$ is a constant, $\lambda := b^3/v_{WS}$
  where $b$ is the magnitude of the Burgers vector, $v_{WS}$ is the 
  Wigner-Seitz volume, $z$ is the coordination number, $\alpha$ is a 
  constant, $\rho_c(T_m)$ is the critical density of dislocations, and
  $k_b$ is the Boltzmann constant.

  \subsubsection{Evaluation of melting temperature models} 
\begin{figure}[p]
    \centering
    \scalebox{0.60}{\includegraphics{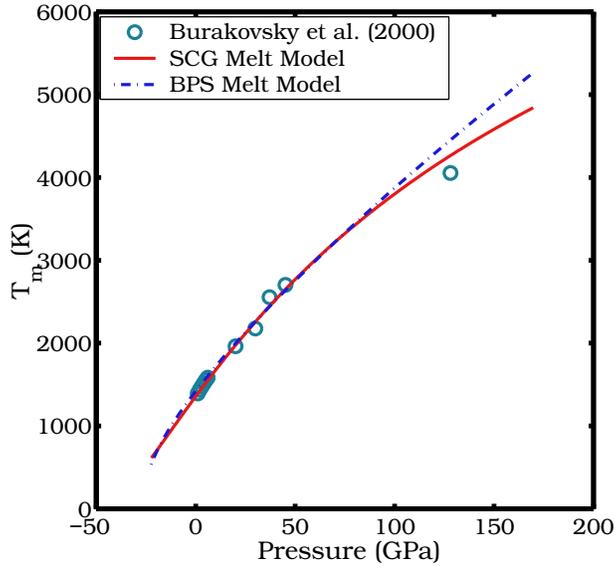}}
    \caption{The melting temperature of copper as a function of pressure.
             The lines show values predicted by the SCG and BPS models. 
             The open circles show experimental data obtained from 
             \citet{Burakovsky00}. The original sources of the experimental
             data can be found in the above citation.}
    \label{fig:CuTm}
  \end{figure}

  Table~\ref{tab:CuTm} shows the parameters used in the melting temperature
  models of copper.  
  \begin{table}[p]
    \centering
    \caption{Parameters used in melting temperature models for copper.
             The parameter $T_{m0}$ used in the SCG model is from 
             \citet{Guinan74}.  The value of $\Gamma_0$ is from 
             \citet{MacDonald81}.  The value of $a$ has been chosen to fit
             the experimental data.  The values of the initial bulk and shear 
             moduli and their derivatives in the BPS model are from
             \citet{Guinan74}.  The remaining parameters for the BPS model 
             are from ~\citet{Burakovsky00a} and \citet{Burakovsky00b}.}
    \vspace{10pt}
    \begin{tabular}{ccccccccccc}
       \hline
       \multicolumn{11}{l}{Steinberg-Cochran-Guinan (SCG) model} \\
       \hline
       $T_{m0}$ (K) & $\Gamma_0$ & $a$ \\
       \hline
       1356.5   & 1.99       & 1.5 \\
       \hline
       \hline
       \multicolumn{11}{l}{Burakovsky-Preston-Silbar (BPS) model} \\
       \hline
       $K_0$ (GPa) & $K_0^{'}$ & $\mu_0$ (GPa) & $\mu_0^{'}$ &
       $\kappa$ & $z$ & $b^2\rho_c(T_m)$ & $\alpha$ & $\lambda$ & 
       $v_{WS}$ & $a$ (nm) \\
       \hline
       137 & 5.48 & 47.7 & 1.4 & 1.25 & 12 & 0.64 & 2.9 & 1.41 & 
       $a^3/4$  & 3.6147 \\ 
       \hline
    \end{tabular}
    \label{tab:CuTm}
  \end{table}
  Figure~\ref{fig:CuTm} shows a comparison of the two melting temperature
  models along with experimental data from \citet{Burakovsky00} (shown as 
  open circles).  An initial density $\rho_0$ of 8930 kg/m$^3$ has been used 
  in the model calculations.

  Both models predict the melting temperature quite accurately for pressures
  below 50 GPa.  The SCG model predicts melting temperatures that
  are closer to experimental values at higher pressures.  However, the data
  at those pressures are sparse and should probably be augmented before
  conclusions regarding the models can be made.  In any case, the pressures
  observed in our computations are usually less than 100 GPa and hence either
  model would suffice.  We have chosen to use the SCG model for
  our copper simulations because the model is more computationally efficient
  than the BPS model. 
  
  \subsection{Shear Modulus} \label{sec:ModelShear}
  The shear modulus of copper decreases with temperature and is also 
  pressure-dependent.  The value of the shear modulus at room temperature
  is around 150\% of the value close to melting.  Hence, if we use the
  room temperature value of shear modulus for high temperature simulations
  we will overestimate the shear stiffness.  This leads to the inaccurate 
  estimation of the plastic strain-rate in radial return algorithms for 
  elastic-plastic simulations.  On the other hand, if the pressure-dependence
  of the shear modulus is neglected, modeling errors can accumulate for
  simulations involving shocks.

  Three models for the shear modulus ($\mu$) have been used in our 
  simulations.  The MTS shear modulus model was developed by \citet{Varshni70}
  and has been used in conjunction with the Mechanical Threshold Stress
  (MTS) flow stress model~(\citet{Chen96,Goto00}).  
  The Steinberg-Cochran-Guinan (SCG)
  shear modulus model was developed by \citet{Guinan74} and has been used
  in conjunction with the Steinberg-Cochran-Guinan-Lund (SCGL) flow stress 
  model.  The Nadal and LePoac (NP) shear modulus model (\citet{Nadal03}) 
  is a recently developed model that uses Lindemann theory to determine the 
  temperature dependence of shear modulus and the SCG model for pressure
  dependence.

  \subsubsection{MTS Shear Modulus Model}
  The MTS shear modulus model has the form (\citet{Varshni70,Chen96})
  \begin{equation} \label{eq:MTSShear}
    \mu(T) = \mu_0 - \frac{D}{exp(T_0/T) - 1}
  \end{equation}
  where $\mu_0$ is the shear modulus at 0K, and $D, T_0$ are material
  constants.  The shortcoming of this model is that it does not
  include any pressure-dependence of the shear modulus and is probably not
  applicable for high pressure applications.  However, the MTS shear modulus 
  model does capture the flattening of the shear modulus-temperature curve 
  at low temperatures that is observed in experiments.

  \subsubsection{SCG Shear Modulus Model}
  The Steinberg-Cochran-Guinan (SCG) shear modulus 
  model (\citet{Steinberg80,Zocher00}) is pressure dependent and
  has the form
  \begin{equation} \label{eq:SCGShear}
    \mu(p,T) = \mu_0 + \Partial{\mu}{p} \frac{p}{\eta^{1/3}} +
         \Partial{\mu}{T}(T - 300) ; \quad
    \eta = \rho/\rho_0
  \end{equation}
  where, $\mu_0$ is the shear modulus at the reference state($T$ = 300 K, 
  $p$ = 0, $\eta$ = 1), $p$ is the pressure, and $T$ is the temperature.
  When the temperature is above $T_m$, the shear modulus is instantaneously
  set to zero in this model.

  \subsubsection{NP Shear Modulus Model}
  The Nadal-Le Poac (NP) shear modulus model (\citet{Nadal03}) is a modified 
  version of the SCG model.  The empirical temperature dependence of the 
  shear modulus in the SCG model is replaced with an equation based on 
  Lindemann melting theory.  In addition, the instantaneous drop
  in the shear modulus at melt is avoided in this model.
  The NP shear modulus model has the form
  \begin{equation} \label{eq:NPShear}
    \mu(p,T) = \frac{1}{\mathcal{J}(\That)}
      \left[
        \left(\mu_0 + \Partial{\mu}{p} \cfrac{p}{\eta^{1/3}} \right)
        (1 - \That) + \frac{\rho}{Cm}~k_b~T\right]; \quad
    C := \cfrac{(6\pi^2)^{2/3}}{3} f^2
  \end{equation}
  where
  \begin{equation}
    \mathcal{J}(\That) := 1 + \exp\left[-\cfrac{1+1/\zeta}
        {1+\zeta/(1-\That)}\right] \quad
       \text{for} \quad \That:=\frac{T}{T_m}\in[0,1+\zeta],
  \end{equation}
  $\mu_0$ is the shear modulus at 0 K and ambient pressure, $\zeta$ is
  a material parameter, $k_b$ is the Boltzmann constant, $m$ is the atomic
  mass, and $f$ is the Lindemann constant.

  \subsubsection{Evaluation of shear modulus models}
\begin{figure}[p]
    \centering
    \scalebox{0.45}{\includegraphics{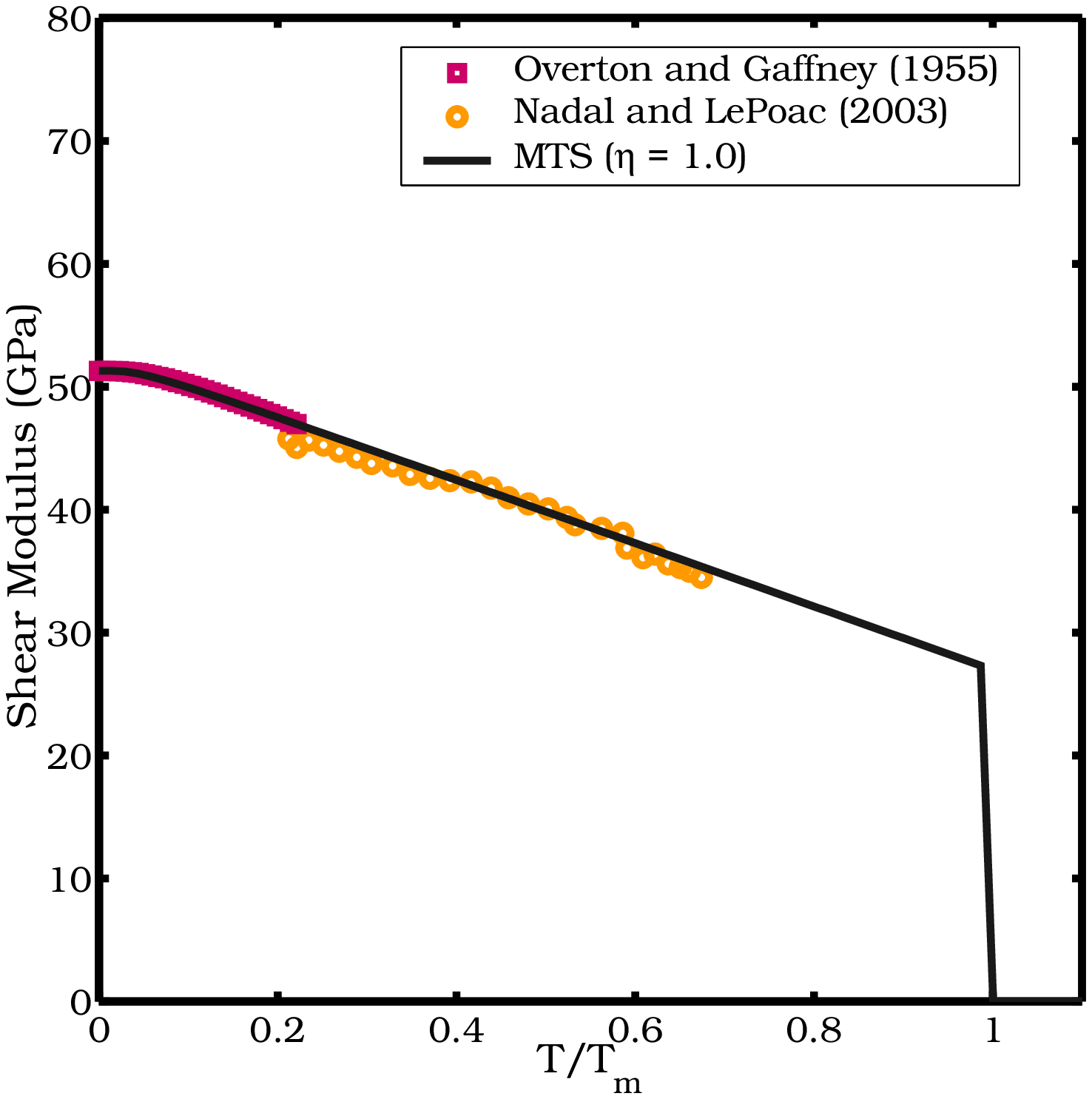}
                    \includegraphics{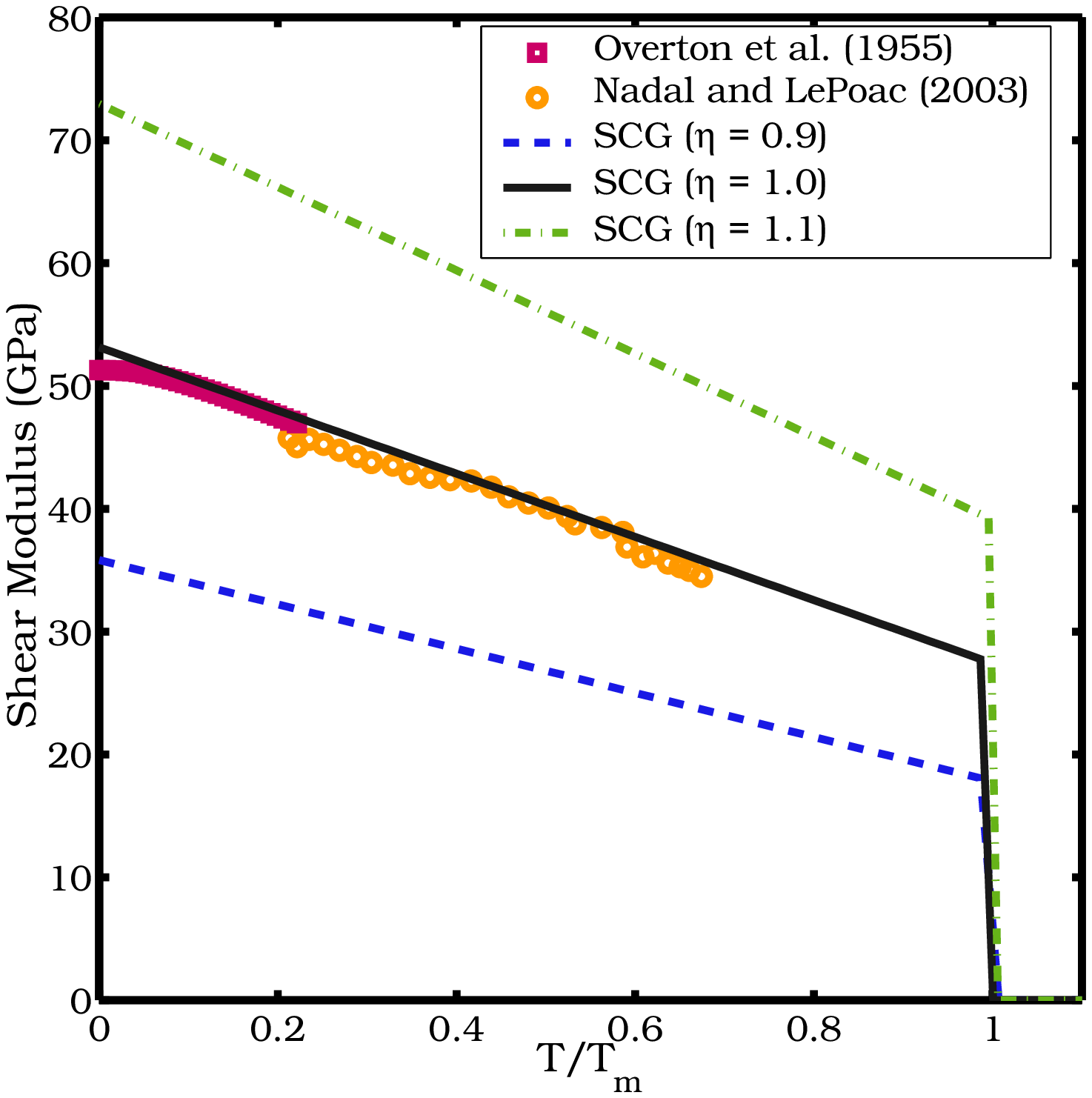}}\\
    (a) MTS Shear Model \hspace{1.5in} 
    (b) SCG Shear Model \\
    \vspace{12pt}
    \scalebox{0.45}{\includegraphics{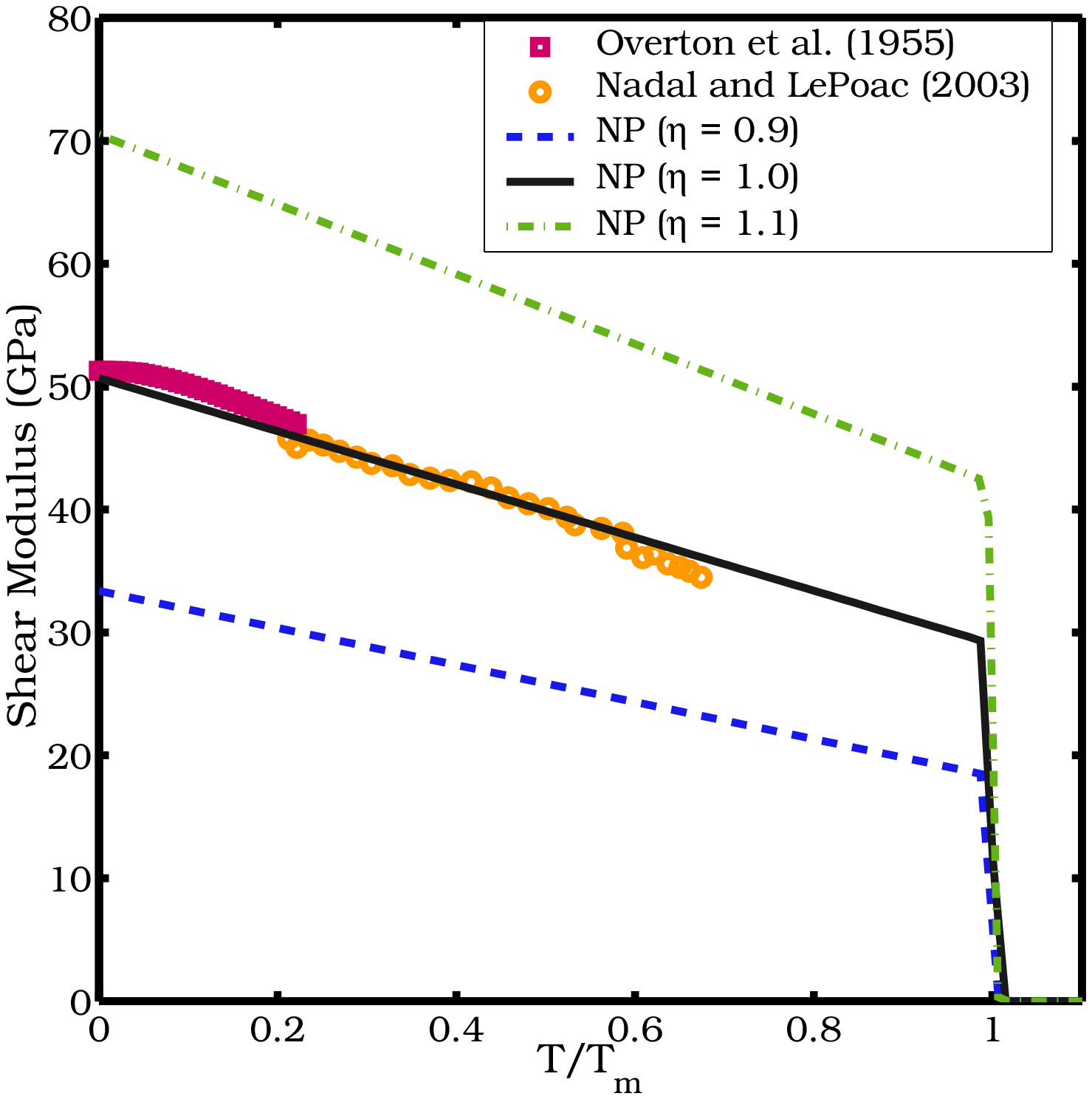}} \\
    (c) NP Shear Model \\
    \caption{Shear modulus of copper as a function of temperature and pressure.
             The symbols represent experimental data from \citet{Overton55} and 
             \citet{Nadal03}.  The lines show values of the shear
             modulus at different compressions ($\eta = \rho/\rho_0$).}
    \label{fig:CuShear}
  \end{figure}

  The parameters used in the three shear modulus models are given in 
  Table~\ref{tab:CuShear}.  
  \begin{table}[p]
    \centering
    \caption{Parameters used in shear modulus models for copper.  
             The parameters for the MTS model have been chosen to fit the 
             experimental data. The parameters for the SCG model are from
             \citet{Guinan74}.  The NP model parameters are from 
             \citet{Nadal03}.}
    \vspace{10pt}
    \begin{tabular}{ccccccc}
       \hline
       \hline
       \multicolumn{4}{l}{MTS shear modulus model} & 
       \multicolumn{3}{l}{SCG shear modulus model} \\
       \hline
       $\mu_0$ (GPa) & $D$ (GPa) & $T_0$ (K) &  &
       $\mu_0$ (GPa) & $\partial\mu/\partial p$ 
                     & $\partial\mu/\partial T$ (GPa/K) \\
       \hline
       51.3          & 3.0      & 165        &  & 
       47.7          & 1.3356   & 0.018126 \\
       \hline
       \hline
       \multicolumn{5}{l}{NP shear modulus model} \\
       \hline
       $\mu_0$ (GPa) & $\partial\mu/\partial p$ & $\zeta$ & $C$ & $m$ (amu) \\
       \hline
       50.7          & 1.3356             & 0.04    & 0.057 & 63.55 \\
       \hline
    \end{tabular}
    \label{tab:CuShear}
  \end{table}
  Figure~\ref{fig:CuShear}(a) shows the shear modulus predicted by the MTS
  shear modulus model at zero hydrostatic pressure.  It can be seen that 
  the model fits the low temperature data quite well.
  The shear moduli predicted by the SCG and NP shear models are shown 
  in Figure~\ref{fig:CuShear}(b) and Figure~\ref{fig:CuShear}(c), respectively.
  The SCG shear model predicts slightly different moduli than the NP model
  at different values of compression.  Both models fit the experimental
  data quite well except at very low temperatures (at which the MTS model
  performs best).  We have not be able to validate the pressure dependence
  of the shear modulus at high temperatures due to lack of experimental data.
  An initial density of 8930 kg/m$^3$ has been used in the model calculations.
  
  \section{Flow Stress Models}\label{sec:flow}
  We have explored five temperature and strain-rate dependent 
  models that can be used to compute the flow stress:
  \begin{enumerate}
    \item the Johnson-Cook model 
    \item the Steinberg-Cochran-Guinan-Lund model. 
    \item the Zerilli-Armstrong model.
    \item the Mechanical Threshold Stress model. 
    \item the Preston-Tonks-Wallace model.
  \end{enumerate}
  The Johnson-Cook (JC) model (\citet{Johnson83}) is purely empirical and
  is the most widely used of the five.  However, this model exhibits an
  unrealistically small strain-rate dependence at high temperatures.  The
  Steinberg-Cochran-Guinan-Lund (SCGL) model (\citet{Steinberg80,Steinberg89})
  is semi-empirical.  The model is purely empirical and strain-rate
  independent at high strain-rates.  A dislocation-based extension based
  on \citet{Hoge77} is used at low strain-rates.  The SCGL model is used
  extensively by the shock physics community.  The Zerilli-Armstrong (ZA)
  model (\citet{Zerilli87}) is a simple physically-based model that has 
  been used extensively.  A more complex model that is based on ideas
  from dislocation dynamics is the Mechanical Threshold Stress (MTS) model
  (\citet{Follans88}).  This model has been used to model the plastic 
  deformation of copper, tantalum (\citet{Chen96}), alloys of steel 
  (\citet{Goto00,Banerjee05b}), and aluminum alloys (\citet{Puchi01}).
  However, the MTS model is limited to strain-rates less than around 10$^7$ /s.
  The Preston-Tonks-Wallace (PTW) model (\citet{Preston03}) is also 
  physically based and has a form similar to the MTS model.  However, the
  PTW model has components that can model plastic deformation in the 
  overdriven shock regime (strain-rates greater that 10$^7$ /s).  Hence 
  this model is valid for the largest range of strain-rates among the 
  five flow stress models.

  \subsection{JC Flow Stress Model}
  The Johnson-Cook (JC) model (\citet{Johnson83}) is purely empirical and gives 
  the following relation for the flow stress ($\sigma_y$) 
  \begin{equation}\label{eq:JCmodel}
    \sigma_y(\Ep,\Epdot{},T) = 
    \left[A + B (\Ep)^n\right]\left[1 + C \ln(\Epdot{}^{*})\right]
    \left[1 - (T^*)^m\right]
  \end{equation}
  where $\Ep$ is the equivalent plastic strain, $\Epdot{}$ is the 
  plastic strain-rate, and $A, B, C, n, m$ are material constants.

  The normalized strain-rate and temperature in equation (\ref{eq:JCmodel})
  are defined as 
  \begin{equation}
    \Epdot{}^{*} := \cfrac{\Epdot{}}{\Epdot{0}} \qquad\text{and}\qquad
    T^* := \cfrac{(T-T_0)}{(T_m-T_0)}
  \end{equation}
  where $\Epdot{0}$ is a user defined plastic strain-rate, 
  $T_0$ is a reference temperature, and $T_m$ is a reference melt temperature.  
  For conditions where $T^* < 0$, we assume that $m = 1$.

  \subsection{SCGL Flow Stress Model}
  The Steinberg-Cochran-Guinan-Lund (SCGL) model is a semi-empirical model
  that was developed by \citet{Steinberg80} for high strain-rate 
  situations and extended to low strain-rates and bcc materials by
  \citet{Steinberg89}.  The flow stress in this model is given by
  \begin{equation}\label{eq:SCGLmodel}
    \sigma_y(\Ep,\Epdot{},T) = 
     \left[\sigma_a f(\Ep) + \sigma_t (\Epdot{}, T)\right]
     \frac{\mu(p,T)}{\mu_0}; \quad 
      \sigma_a f \le \sigma_{\text{max}} ~~\text{and}~~
      \sigma_t \le \sigma_p
  \end{equation}
  where $\sigma_a$ is the athermal component of the flow stress,
  $f(\Ep)$ is a function that represents strain hardening,
  $\sigma_t$ is the thermally activated component of the flow stress,
  $\mu(p,T)$ is the pressure- and temperature-dependent shear modulus, 
  and $\mu_0$ is the shear modulus at standard temperature and pressure.  
  The saturation value of the athermal stress is $\sigma_{\text{max}}$.
  The saturation of the thermally activated stress is the Peierls stress
  ($\sigma_p$).  The shear modulus for this model is usually computed
  with the SCG shear modulus model.

  The strain hardening function ($f$) has the form
  \begin{equation}
    f(\Ep) = [1 + \beta(\Ep + \Epi)]^n
  \end{equation}
  where $\beta, n$ are work hardening parameters, and $\Epi$ is the 
  initial equivalent plastic strain.  

  The thermal component ($\sigma_t$) is computed using a bisection 
  algorithm from the following equation (citet{Hoge77,Steinberg89}).
  \begin{equation}
    \Epdot{} = \left[\frac{1}{C_1}\exp\left[\frac{2U_k}{k_b~T}
      \left(1 - \frac{\sigma_t}{\sigma_p}\right)^2\right] + 
      \frac{C_2}{\sigma_t}\right]^{-1}; \quad
    \sigma_t \le \sigma_p
  \end{equation}
  where $2 U_k$ is the energy to form a kink-pair in a dislocation segment
  of length $L_d$, $k_b$ is the Boltzmann constant, $\sigma_p$ is the Peierls
  stress. The constants $C_1, C_2$ are given by the relations
  \begin{equation}
    C_1 := \frac{\rho_d L_d a b^2 \nu}{2 w^2}; \quad
    C_2 := \frac{D}{\rho_d b^2}
  \end{equation}
  where $\rho_d$ is the dislocation density, $L_d$ is the length of a 
  dislocation segment, $a$ is the distance between Peierls valleys, 
  $b$ is the magnitude of the Burgers' vector, $\nu$ is the Debye frequency,
  $w$ is the width of a kink loop, and $D$ is the drag coefficient.

  \subsection{ZA Flow Stress Model}
  The Zerilli-Armstrong (ZA) model (\citet{Zerilli87,Zerilli93,Zerilli04}) 
  is based on simplified dislocation mechanics.  The general form of the
  equation for the flow stress is
  \begin{equation}\label{eq:ZAmodel}
    \sigma_y(\Ep,\Epdot{},T) = 
      \sigma_a + B\exp(-\beta(\Epdot{}) T) + 
                           B_0\sqrt{\Ep}\exp(-\alpha(\Epdot{}) T) ~.
  \end{equation}
  In this model, $\sigma_a$ is the athermal component of the flow stress 
  given by
  \begin{equation}
    \sigma_a := \sigma_g + \frac{k_h}{\sqrt{l}} + K\Ep^n,
  \end{equation}
  where $\sigma_g$ is the contribution due to solutes and initial dislocation
  density, $k_h$ is the microstructural stress intensity, $l$ is the 
  average grain diameter, $K$ is zero for fcc materials, 
  $B, B_0$ are material constants.  

  In the thermally activated terms, the functional forms of the exponents 
  $\alpha$ and $\beta$ are 
  \begin{equation}
    \alpha = \alpha_0 - \alpha_1 \ln(\Epdot{}); \quad
    \beta = \beta_0 - \beta_1 \ln(\Epdot{}); 
  \end{equation}
  where $\alpha_0, \alpha_1, \beta_0, \beta_1$ are material parameters that
  depend on the type of material (fcc, bcc, hcp, alloys).  The Zerilli-Armstrong
  model has been modified by \citet{Abed05} for better performance at high 
  temperatures.  However, we have not used the modified equations in our
  computations.

  \subsection{MTS Flow Stress Model}
  The Mechanical Threshold Stress (MTS) model 
  (\citet{Follans88,Goto00a,Kocks01}) has the form
  \begin{equation}\label{eq:MTSmodel}
    \sigma_y(\Ep,\Epdot{},T) = 
      \sigma_a + (S_i \sigma_i + S_e \sigma_e)\frac{\mu(p,T)}{\mu_0} 
  \end{equation}
  where $\sigma_a$ is the athermal component of mechanical threshold stress,
  $\sigma_i$ is the component of the flow stress due to intrinsic barriers 
  to thermally activated dislocation motion and dislocation-dislocation 
  interactions, $\sigma_e$ is the component of the flow stress due to 
  microstructural evolution with increasing deformation (strain hardening), 
  ($S_i, S_e$) are temperature and strain-rate dependent scaling factors, and  
  $\mu_0$ is the shear modulus at 0 K and ambient pressure, 

  The scaling factors take the Arrhenius form
  \begin{align}
    S_i & = \left[1 - \left(\frac{k_b~T}{g_{0i}b^3\mu(p,T)}
    \ln\frac{\Epdot{0i}}{\Epdot{}}\right)^{1/q_i}
    \right]^{1/p_i} \\
    S_e & = \left[1 - \left(\frac{k_b~T}{g_{0e}b^3\mu(p,T)}
    \ln\frac{\Epdot{0e}}{\Epdot{}}\right)^{1/q_e}
    \right]^{1/p_e}
  \end{align}
  where $k_b$ is the Boltzmann constant, $b$ is the magnitude of the Burgers' 
  vector, ($g_{0i}, g_{0e}$) are normalized activation energies, 
  ($\Epdot{0i}, \Epdot{0e}$) are constant reference strain-rates, and
  ($q_i, p_i, q_e, p_e$) are constants.  

  The strain hardening component of the mechanical threshold stress 
  ($\sigma_e$) is given by an empirical modified Voce law
  \begin{equation}\label{eq:MTSsige}
    \frac{d\sigma_e}{d\Ep} = \theta(\sigma_e)
  \end{equation}
  where
  \begin{align}
    \theta(\sigma_e) & = 
       \theta_0 [ 1 - F(\sigma_e)] + \theta_{IV} F(\sigma_e) \\
    \theta_0 & = a_0 + a_1 \ln \Epdot{} + a_2 \sqrt{\Epdot{}} - a_3 T \\
    F(\sigma_e) & = 
      \cfrac{\tanh\left(\alpha \cfrac{\sigma_e}{\sigma_{es}}\right)}
      {\tanh(\alpha)}\\
    \ln(\cfrac{\sigma_{es}}{\sigma_{0es}}) & =
    \left(\frac{kT}{g_{0es} b^3 \mu(p,T)}\right)
    \ln\left(\cfrac{\Epdot{}}{\Epdot{0es}}\right)
  \end{align}
  and $\theta_0$ is the hardening due to dislocation accumulation, 
  $\theta_{IV}$ is the contribution due to stage-IV hardening,
  ($a_0, a_1, a_2, a_3, \alpha$) are constants,
  $\sigma_{es}$ is the stress at zero strain hardening rate, 
  $\sigma_{0es}$ is the saturation threshold stress for deformation at 0 K,
  $g_{0es}$ is a constant, and $\Epdot{0es}$ is the maximum strain-rate.  Note
  that the maximum strain-rate is usually limited to about $10^7$/s.

  \subsection{PTW Flow Stress Model}
  The Preston-Tonks-Wallace (PTW) model (\citet{Preston03}) attempts to 
  provide a model for the flow stress for extreme strain-rates 
  (up to $10^{11}$/s) and temperatures up to melt.  A linear Voce hardening
  law is used in the model.  The PTW flow stress is given by
  \begin{equation}\label{eq:PTWmodel}
    \sigma_y(\Ep,\Epdot{},T) = 
       \begin{cases}
         2\left[\tau_s + \alpha\ln\left[1 - \varphi
          \exp\left(-\beta-\cfrac{\theta\Ep}{\alpha\varphi}\right)\right]\right]
         \mu(p,T) & \text{thermal regime} \\
         2\tau_s\mu(p,T) & \text{shock regime}
       \end{cases}
  \end{equation}
  with 
  \begin{equation}
    \alpha := \frac{s_0 - \tau_y}{d}; \quad
    \beta := \frac{\tau_s - \tau_y}{\alpha}; \quad
    \varphi := \exp(\beta) - 1
  \end{equation}
  where $\tau_s$ is a normalized work-hardening saturation stress,
  $s_0$ is the value of $\tau_s$ at 0K,
  $\tau_y$ is a normalized yield stress, $\theta$ is the hardening constant
  in the Voce hardening law, and $d$ is a dimensionless material
  parameter that modifies the Voce hardening law.  

  The saturation stress and the yield stress are given by
  \begin{align}
    \tau_s & = \max\left\{s_0 - (s_0 - s_{\infty})
       \erf\left[\kappa
         \That\ln\left(\cfrac{\gamma\Xidot}{\Epdot{}}\right)\right],
       s_0\left(\cfrac{\Epdot{}}{\gamma\Xidot}\right)^{s_1}\right\} \\
    \tau_y & = \max\left\{y_0 - (y_0 - y_{\infty})
       \erf\left[\kappa
         \That\ln\left(\cfrac{\gamma\Xidot}{\Epdot{}}\right)\right],
       \min\left\{
         y_1\left(\cfrac{\Epdot{}}{\gamma\Xidot}\right)^{y_2}, 
         s_0\left(\cfrac{\Epdot{}}{\gamma\Xidot}\right)^{s_1}\right\}\right\} 
  \end{align}
  where $s_{\infty}$ is the value of $\tau_s$ close to the melt temperature,
  ($y_0, y_{\infty}$) are the values of $\tau_y$ at 0K and close to melt,
  respectively, $(\kappa, \gamma)$ are material constants, $\That = T/T_m$,
  ($s_1, y_1, y_2$) are material parameters for the high strain-rate
  regime, and
  \begin{equation}
    \Xidot = \frac{1}{2}\left(\cfrac{4\pi\rho}{3M}\right)^{1/3}
             \left(\cfrac{\mu(p,T)}{\rho}\right)^{1/2}
  \end{equation}
  where $\rho$ is the density, and $M$ is the atomic mass.

  \subsection{Evaluation of flow stress models}
  In this section, we evaluate the flow stress models on the basis of
  one-dimensional tension and compression tests.  The high rate tests have 
  been simulated using the explicit Material Point 
  Method~\cite{Sulsky94,Sulsky95} (see Appenedix~\ref{appA}) in conjunction 
  with the stress update algorithm given in Appendix~\ref{appB}.  The 
  quasistatic tests have been simulated with a fully implicit version of the 
  Material Point Method (\citet{Guilkey03}) with an implicit stress update 
  (\citet{Simo98}).  Heat conduction is performed at all strain-rates.  As 
  expected, we obtain nearly isothermal conditions for the quasistatic tests 
  and nearly adiabatic conditions for the high strain-rate tests.  We have
  used a constant thermal conductivity of 386 W/(m-K) for copper which is the
  value at 500 K and atmospheric pressure.  To damp out large oscillations 
  in high strain-rate tests, we use a three-dimensional form of the von 
  Neumann artificial viscosity (\citet{Wilkins99}, p.29).  The viscosity
  factor takes the form
  \begin{equation}\label{eq:artificialVisc}
    q = C_0~\rho~l~\sqrt{\cfrac{K}{\rho}}\left|\Tr{\BD}\right| +
        C_1~\rho~l^2~\left(\Tr{\BD}\right)^2
  \end{equation}
  where $C_0$ and $C_1$ are constants, $\rho$ is the mass density,
  $K$ is the bulk modulus, $\BD$ is the rate of deformation tensor, and
  $l$ is a characteristic length (usually the grid cell size).  We have
  used $C_0 = 0.2$ and $C_1 = 2.0$ in all our simulations.  The 
  temperature-dependent specific heat model, the Mie-Gr{\"u}neisen equation 
  of state, and the SCG melting temperature model have been used in all the 
  following simulations.
  
  The predicted stress-strain curves are compared with experimental data for 
  annealed OFHC copper from tension tests (\cite{Nemat04} (p. 241-242)) and 
  compression tests (\cite{Samanta71}).
  The data are presented in form of true stress versus true strain.  Note that
  detailed verification has been performed to confirm the correct implementation
  of the models withing the Uintah code.  Also note that the high strain-rate
  experimental data are suspect for strains less than 0.1.  This is because
  the initial strain-rate fluctuates substantially in Kolsky-Hopkinson bar
  experiments.  
  
  \subsubsection{Johnson-Cook Model.}
  The parameters that we have used in the Johnson-Cook (JC) flow stress model 
  of annealed copper are given in Table~\ref{tab:CuJC}.  We have used the
  NP shear modulus model in simulations involving the JC model.
  \begin{table}[p]
    \centering
    \caption{Parameters used in the Johnson-Cook model for copper
             (\citet{Johnson85}).}
    \vspace{10pt}
    \begin{tabular}{cccccccc}
       \hline
       $A$ (MPa) & $B$ (MPa) & $C$ & $n$ & $m$ & $\Epdot{0}$ (/s) & $T_0$ (K) & 
       $T_m$ (K) \\
       \hline
       90 & 292 & 0.025 & 0.31 & 1.09 & 1.0 & 294 & 1356 \\
       \hline
    \end{tabular}
    \label{tab:CuJC}
  \end{table}

  The Johnson-Cook model is independent of pressure.  Hence, the predicted 
  yield stress is the same in compression and tension.  The use of a variable 
  specific heat model leads to a reduced yield stress at 77 K for high strain 
  rates.  However, the effect is relatively small.  At high temperatures, the 
  effect of the higher specific heat is to reduce the rate of increase of 
  temperature with increase in plastic strain.  This effect is also small.  The
  temperature dependence of the shear modulus does not affect the yield stress.
  However, it has a small effect on the value of the plastic strain-rate.

\begin{figure}[p]
    \centering
    \scalebox{0.45}{\includegraphics{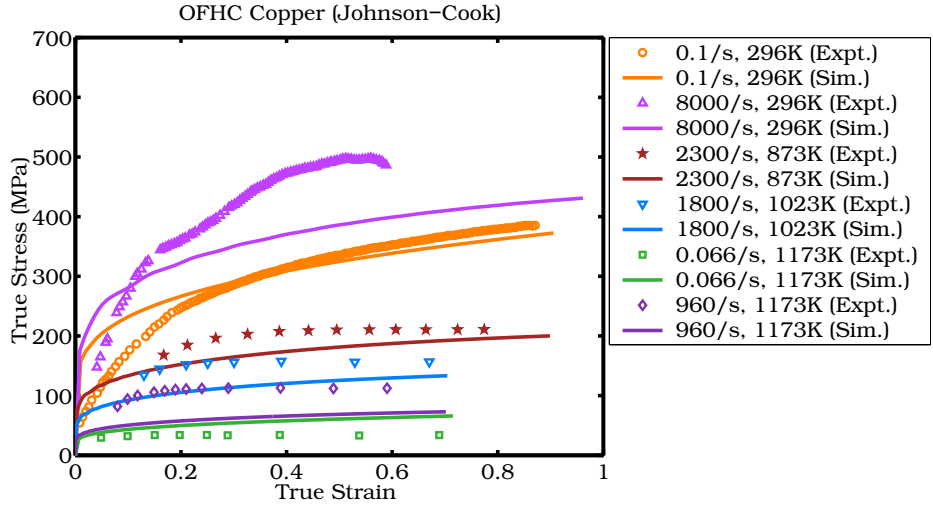}} \\
    (a) Various strain-rates and temperatures. \\
    \vspace{12pt}
    \scalebox{0.45}{\includegraphics{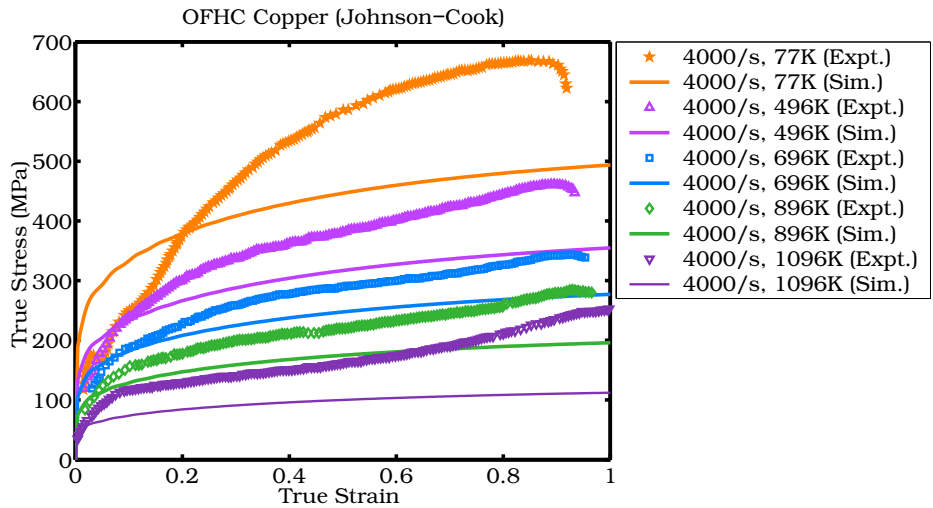}}\\
    (b) Various temperatures at 4000/s strain-rate.
    \caption{Predicted values of yield stress from the Johnson-Cook model.
             The experimental data at 873 K, 1023 K, and 1173 K are from 
             \citet{Samanta71} and represent compression tests.  
             The remaining experimental data are from tension tests
             in \citet{Nemat04}. 
             The solid lines are the predicted values.}
    \label{fig:Cu1DJC}
  \end{figure}

  The solid lines in Figures~\ref{fig:Cu1DJC}(a) and (b) show predicted values
  of the yield stress for various strain-rates and temperatures.  The symbols
  show the experimental data.  The Johnson-Cook model overestimates the
  initial yield stress for the quasistatic (0.1/s strain-rate), room temperature
  (296 K), test.  The rate of hardening is underestimated by the model for
  the room temperature test at 8000/s.  The strain-rate dependence of the 
  yield stress is underestimated at high temperature (see the data at 1173 K
  in Figure~\ref{fig:Cu1DJC}(a)).  For the tests at a strain-rate of 4000/s 
  (Figure~\ref{fig:Cu1DJC}(b)), the yield stress is consistently underestimated
  by the Johnson-Cook model.  
  
  \subsubsection{Steinberg-Cochran-Guinan-Lund Model.}
  The parameters used in the Steinberg-Cochran-Guinan-Lund (SCGL) model of 
  annealed OFHC copper are listed in Table~\ref{tab:CuSCGL}.  We have used
  the SCG shear modulus model in simulations involving the SCGL model.  We 
  could alternatively have used the NP shear modulus model.  However, we use 
  the SCG model to highlight a problem with the equivalence of 
  $\partial \mu/\partial T$ and $\partial \sigma_y/\partial T$ that is
  assumed by the SCGL model.  A bisection algorithm is used to determine the 
  thermally activated part of the flow stress for low strain-rates (less than 
  1000/s).
  \begin{table}[p]
    \centering
    \caption{Parameters used in the Steinberg-Cochran-Guinan-Lund model 
             for copper. The parameters for the athermal part of the
             SCGL model are from ~\citet{Steinberg80}.  The parameters
             for the thermally activated part of the model are from a
             number of sources.  The estimate for the Peierls stress 
             is based on \citet{Hobart65}.}
    \vspace{10pt}
    \begin{tabular}{ccccccccc}
       \hline
       $\sigma_a$ (MPa) & $\sigma_{\text{max}}$ (MPa) & $\beta$ & $\Epi$ (/s) & 
       $n$ & $C_1$ (/s) & $U_k$ (eV) & $\sigma_p$ (MPa) & $C_2$ (MPa-s) \\
       \hline
       125  & 640 & 36 & 0.0 & 0.45 & 0.71$\times$10$^6$ & 0.31 & 20 & 
       0.012 \\
       \hline
    \end{tabular}
    \label{tab:CuSCGL}
  \end{table}

\begin{figure}[p]
    \centering
    \scalebox{0.45}{\includegraphics{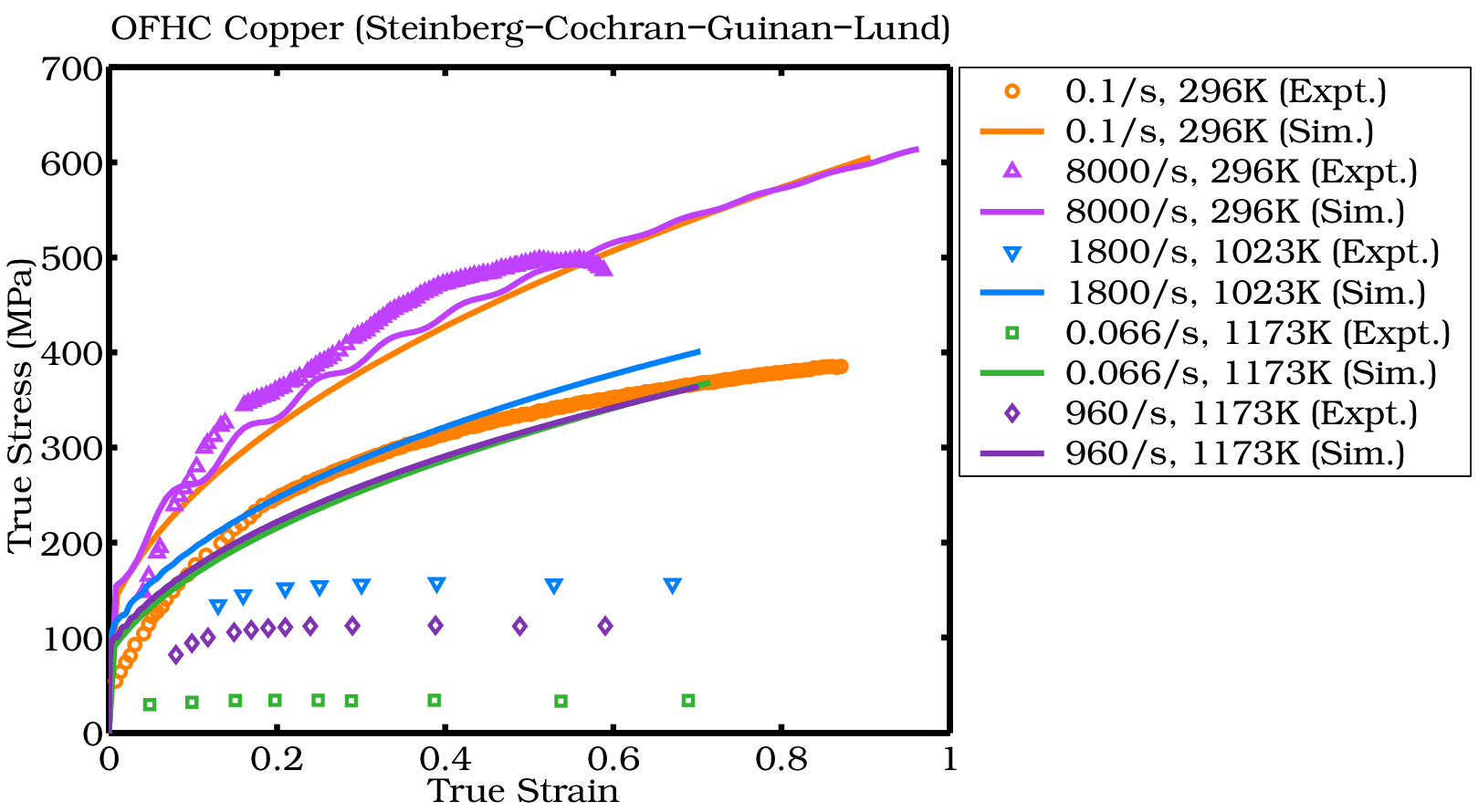}} \\
    (a) Various strain-rates and temperatures. \\
    \vspace{12pt}
    \scalebox{0.45}{\includegraphics{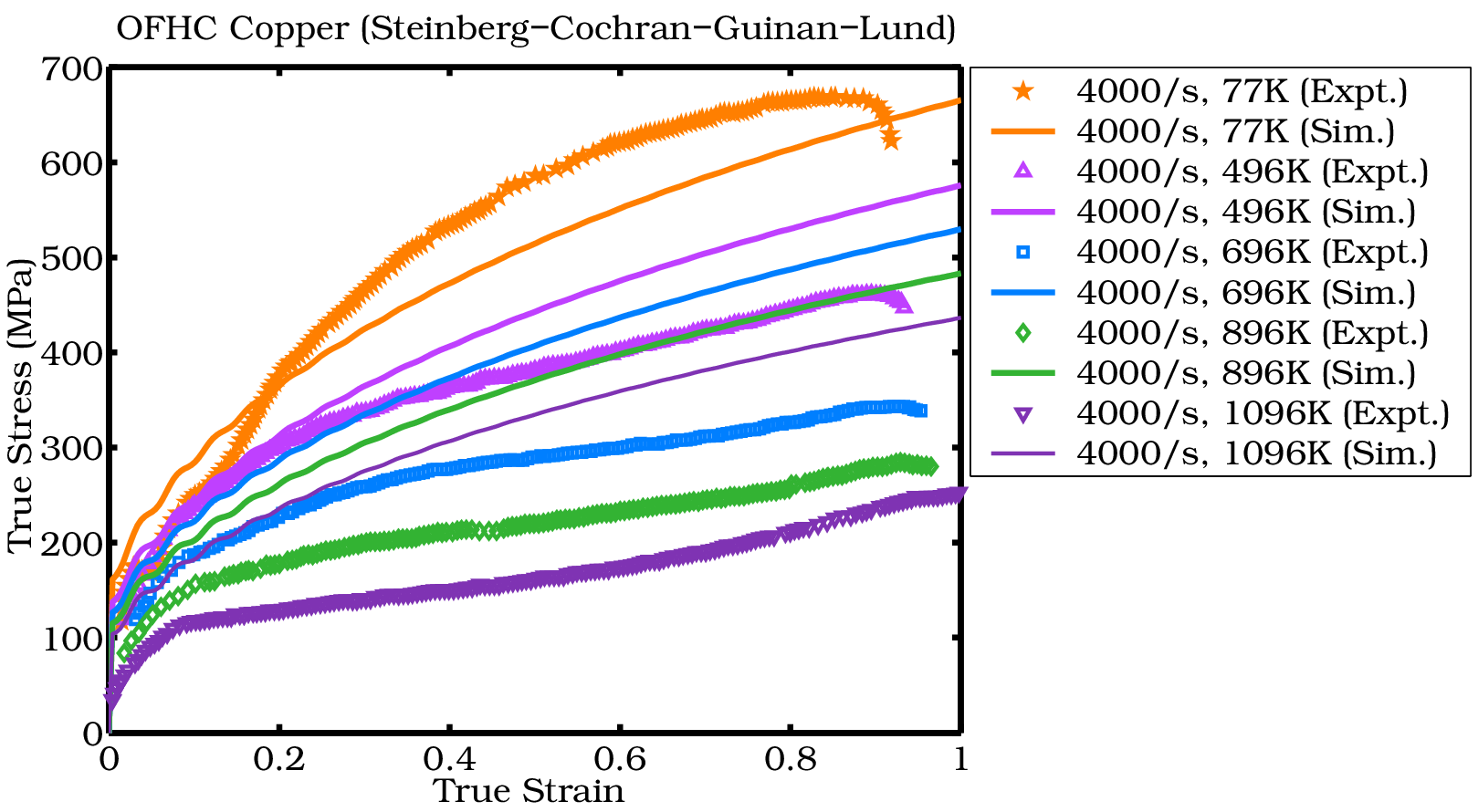}}\\
    (b) Various temperatures at 4000/s strain-rate.
    \caption{Predicted values of yield stress from the 
             Steinberg-Cochran-Guinan-Lund model.  
             Please see the caption of Figure~\ref{fig:Cu1DJC} for the
             sources of the experimental data.}
    \label{fig:Cu1DSCG}
  \end{figure}

  The solid lines in Figures~\ref{fig:Cu1DSCG}(a) and (b) show the flow 
  stresses predicted by the SCGL model.  Clearly, the softening associated
  with increasing temperature is underestimated by the SCGL model though
  the yield stress at 8000/s is predicted reasonably accurately. 
  For the tests at 4000/s shown in Figure~\ref{fig:Cu1DSCG}(b), the SCGL
  modes performs progressively worse with increasing temperature.  

  Overall, at low temperatures, the high strain-rate predictions from the SCGL
  model match the experimental data best.  This is not surprising since the
  original model by ~\citet{Steinberg80} (SCG) was rate-independent and 
  designed for high strain-rate applications.  However, the low strain 
  rate extension by ~\citet{Steinberg89} does not lead to good predictions
  of the yield stress of OFHC copper at low temperatures.

  The high temperature response of the SCGL model is dominated by the shear
  modulus model; in particular, the derivative of the shear modulus with 
  respect to temperature.  From Figure~\ref{fig:CuShear}(b) we can see 
  that a value of -0.018126 GPa/K for $\partial\mu/\partial T$ matches the
  experimental data quite well.  \citet{Steinberg80} assume that the values
  of $(\partial\sigma_y/\partial T)/\sigma_{y0}$ and 
  $(\partial\mu/\partial T)/\mu_0$ (-3.8$\times$10$-4$ /K) are comparable.  
  That does not appear to be the case for OFHC copper.

  If we extract the yield stresses at a strain of 0.2 from the experimental
  data shown in Figure~\ref{fig:Cu1DSCG}(b), we get the following values of
  temperature and yield stress for a strain-rate of 4000/s: (77 K, 380 MPa); 
  (496 K, 300 MPa); (696 K, 230 MPa); (896 K, 180 MPa); (1096 K, 130 MPa).  
  A straight line fit to the data shows that the value of 
  $\partial\sigma_y/\partial T$ is -0.25 MPa/K.  The yield stress at 300 K
  can be calculated from the fit to be approximately 330 MPa.  This gives
  a value of -7.6$\times$10$^-4$ /K for
  $(\partial\sigma_y/\partial T)/\sigma_{y0}$; approximately double the 
  slope of the shear modulus versus temperature curve.  Hence, a shear 
  modulus derived from a shear modulus model cannot be used as a multiplier
  to the yield stress in equation (\ref{eq:SCGLmodel}).  Instead, the original
  form of the SCG model (\citet{Steinberg80}) must be used, with the term
  $(\partial\mu/\partial T)/\mu_0$ replaced by 
  $(\partial\sigma_y/\partial T)/\sigma_{y0}$ 
  in the expression for yield stress.

\begin{figure}[p]
    \centering
    \scalebox{0.45}{\includegraphics{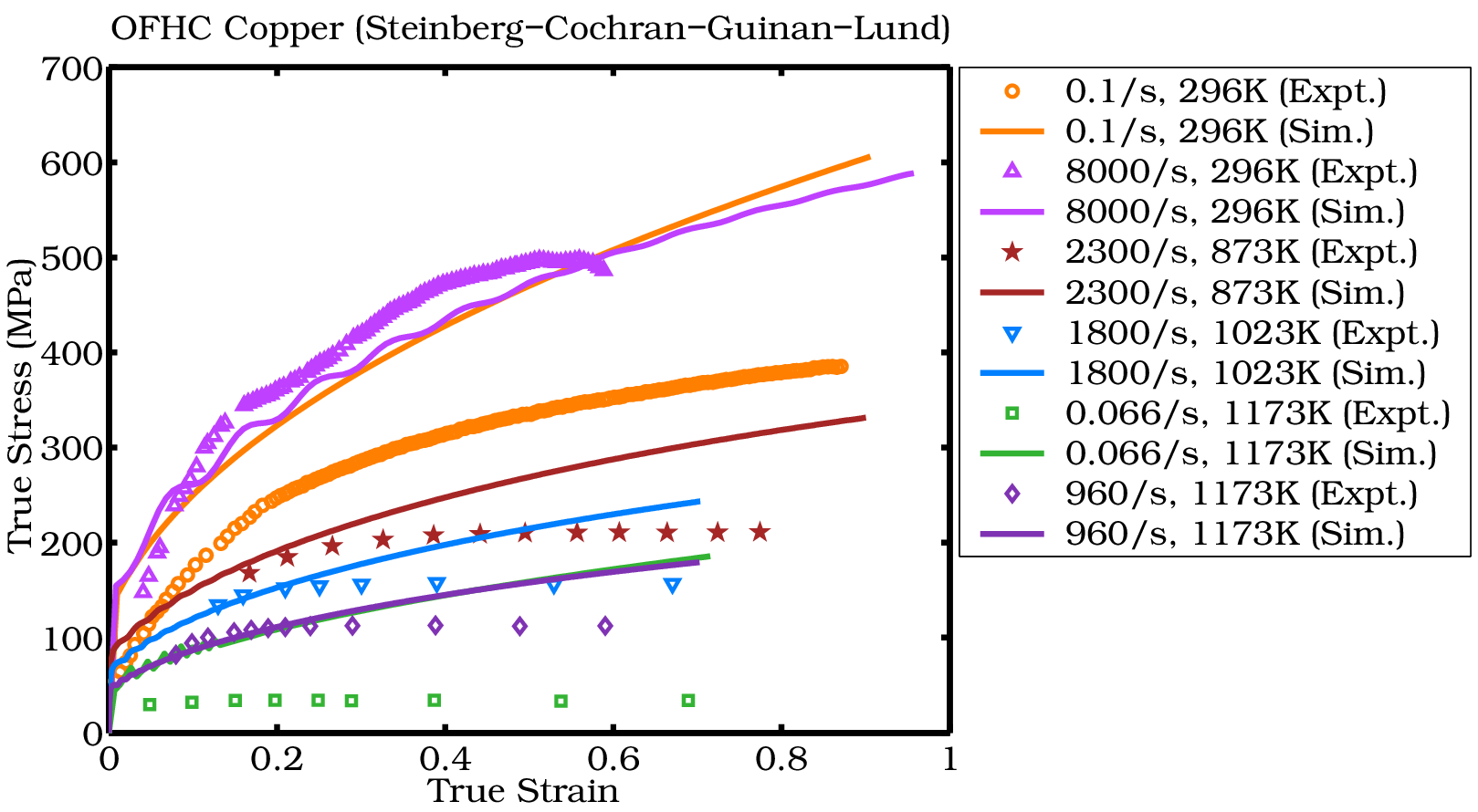}} \\
    (a) Various strain-rates and temperatures. \\
    \vspace{12pt}
    \scalebox{0.45}{\includegraphics{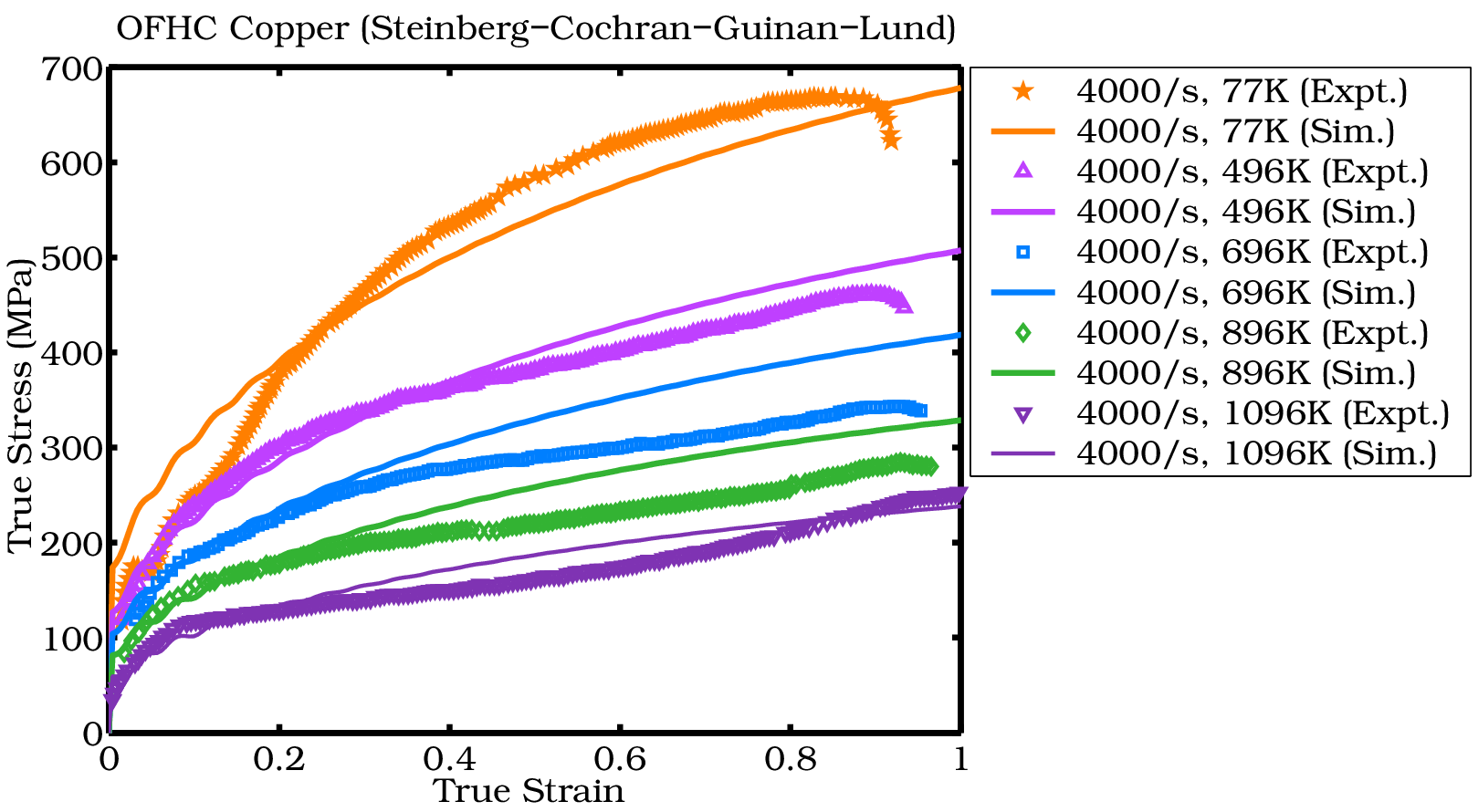}}\\
    (b) Various temperatures at 4000/s strain-rate.
    \caption{Predicted values of yield stress from the modified
             Steinberg-Cochran-Guinan-Lund model.
             Please see the caption of Figure~\ref{fig:Cu1DJC} for the
             sources of the experimental data.}
    \label{fig:Cu1DSCGMod}
  \end{figure}

  Figures~\ref{fig:Cu1DSCGMod}(a) and (b) show the predicted yield stresses from
  the modified SCGL model.
  These plots show that there is a considerable improvement in the 
  prediction of the temperature dependence of yield stress if the value of
  $(\partial\sigma_y/\partial T)/\sigma_{y0}$ is used instead of
  $(\partial\mu/\partial T)/\mu_0$.  However, the strain-rate dependence
  of OFHC copper continues to be poorly modeled by the SCGL model.

  \subsubsection{Zerilli-Armstrong Model.}
  In contrast to the Johnson-Cook and the Steinberg-Cochran-Guinan models, the
  Zerilli-Armstrong (ZA) model for yield stress is based on dislocation 
  mechanics and hence has some physical basis.  The parameters used for the
  ZA model are given in Table~\ref{tab:CuZA}.  We have used the NP
  shear modulus model in our simulations that involve the ZA model. 
  \begin{table}[p]
    \centering
    \caption{Parameters used in the Zerilli-Armstrong model 
             for copper (\citet{Zerilli87}).}
    \vspace{10pt}
    \begin{tabular}{cccccc}
       \hline
       $\sigma_g$ (MPa)& $k_h$ (MPa-mm$^{1/2}$)& $l$ (mm)& $K$ (MPa) & $n$ \\
       \hline
       46.5 & 5.0 & 0.073 & 0.0 & 0.5 \\
       \hline
       \hline
       $B$ (MPa) & $\beta_0$ (/K) & $\beta_1$ (s/K) &
       $B_0$ (MPa) & $\alpha_0$ (/K) & $\alpha_1$ (s/K) \\
       \hline
       0.0 & 0.0 & 0.0 & 890 & 0.0028 & 0.000115 \\
       \hline
    \end{tabular}
    \label{tab:CuZA}
  \end{table}

\begin{figure}[p]
    \centering
    \scalebox{0.45}{\includegraphics{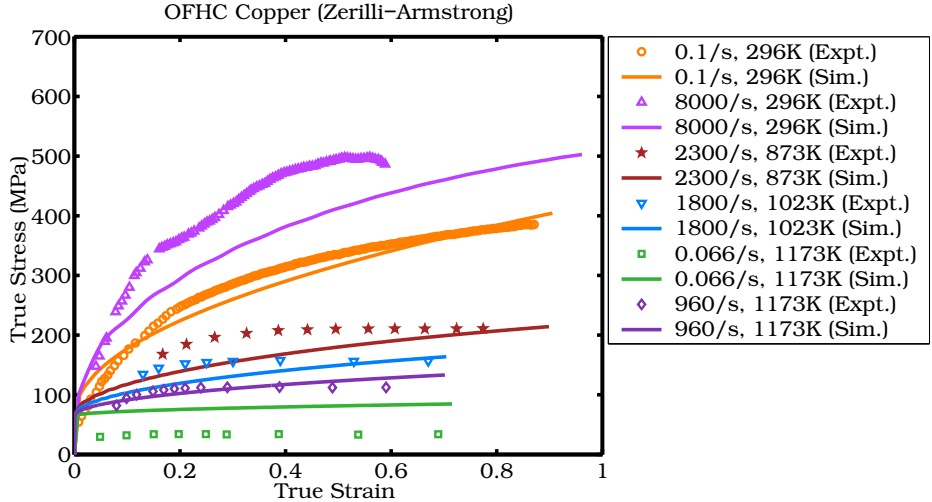}} \\
    (a) Various strain-rates and temperatures. \\
    \vspace{12pt}
    \scalebox{0.45}{\includegraphics{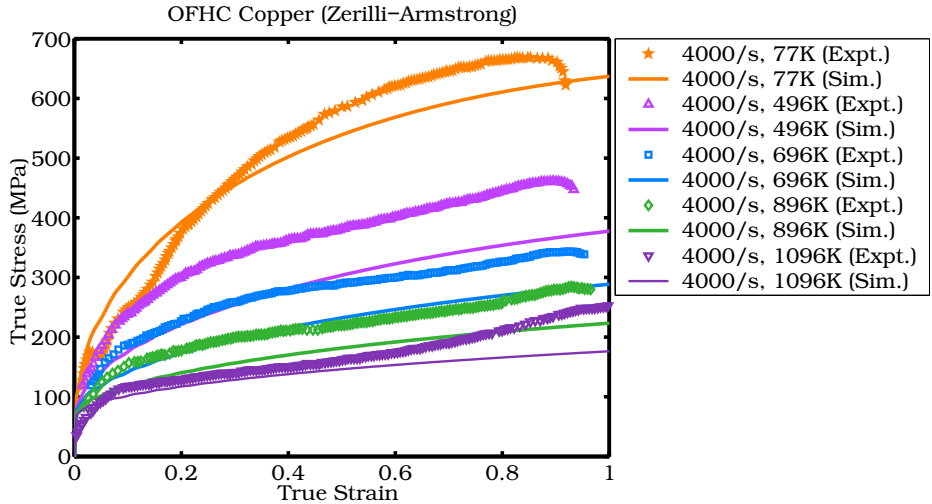}}\\
    (b) Various temperatures and 4000/s strain-rate.
    \caption{Predicted values of yield stress from the Zerilli-Armstrong
             model.  The symbols represent experimental data.  The solid lines 
             represented the computed stress-strain curves.  
             Please see the caption of Figure~\ref{fig:Cu1DJC} for the
             sources of the experimental data.}
    \label{fig:Cu1DZA}
  \end{figure}

  Figures~\ref{fig:Cu1DZA}(a) and (b) show the yield stresses predicted by
  the ZA model.  
  From Figure~\ref{fig:Cu1DZA}(a), we can see that the ZA model predicts the
  quasistatic, room temperature yield stress quite accurately.  However, the
  room temperature yield stress at 8000/s is underestimated.  The initial
  yield stress is overestimated at high temperatures; as are the saturation 
  stresses.

  Stress-strain curves at 4000/s are shown in Figure~\ref{fig:Cu1DZA}(b).  In
  this case, the ZA model predicts reasonable initial yield stresses.  However,
  the decrease in yield stress with increasing temperature is overestimated.
  We notice that the predicted yield stress at 496 K overlaps the experimental
  data for 696 K, while the predicted stress at 696 K overlaps the experimental
  data at 896 K.

  \subsubsection{Mechanical Threshold Stress Model.}
  The Mechanical Threshold Stress (MTS) model is different from the three
  previous models in that the internal variable that evolves in time is
  a stress ($\sigma_e$).  The value of the internal variable is 
  calculated for each value of plastic strain by integrating equation 
  (\ref{eq:MTSsige}) along a constant temperature and strain-rate path.  An
  unconditionally stable and second-order accurate midpoint integration scheme 
  has been used to determine the value of $\sigma_e$.  Alternatively, an 
  incremental update of the internal variable could be done using quantities
  from the previous timestep.  The integration of the evolution equation
  is no longer along a constant temperature and strain-rate path in that case.
  We have found that two alternatives give us similar values of $\sigma_e$ 
  in the simulations that we have performed.  The incremental update of
  the value of $\sigma_e$ is considerably faster than the full update along
  a constant temperature and strain-rate path.

  The parameters for the MTS model are shown in Table~\ref{tab:CuMTS}.
  The pressure-independent MTS shear modulus model has been used in 
  simulations that use the MTS flow stress model.  The reason for this choice is
  that the parameters of the model have been fit with such a shear modulus
  model.  If the shear modulus model is changed, certain parameters of the
  model will have to be changed to reflect the difference.  
  \begin{table}[p]
    \centering
    \caption{Parameters used in the Mechanical Threshold Stress model 
             for copper (\citet{Follans88}).}
    \begin{tabular}{ccccccc}
       \hline
       $\sigma_a$ (MPa) & $b$ (nm) & $\sigma_i$ (MPa) & $g_{0i}$ &
       $\Epdot{0i}$ (/s) & $p_i$ & $q_i$ \\
       \hline
       40 & 0.256 & 0 & 1 & 1 & 1 & 1 \\
       \hline
       \hline
       $g_{0e}$ & $\Epdot{0e}$ (/s) & $p_e$ & $q_e$ & $\sigma_{0es}$ (MPa) &
       $g_{0es}$ & $\Epdot{0es}$ (/s) \\
       \hline
       1.6 & 1.0$\Times{7}$ & 2/3 & 1 & 770 & 0.2625 & 1.0$\Times{7}$ \\
       \hline
       \hline
       $\alpha$ & $a_0$ (MPa) & $a_1$ (MPa-log(s)) & $a_2$ (MPa-s$^{1/2}$) & 
       $a_3$ (MPa/K) & $\theta_{IV}$ (MPa) \\
       \hline
       2 & 2390 & 12 & 1.696 & 0 & 0 \\
       \hline
    \end{tabular}
    \label{tab:CuMTS}
  \end{table}

\begin{figure}[p]
    \centering
    \scalebox{0.45}{\includegraphics{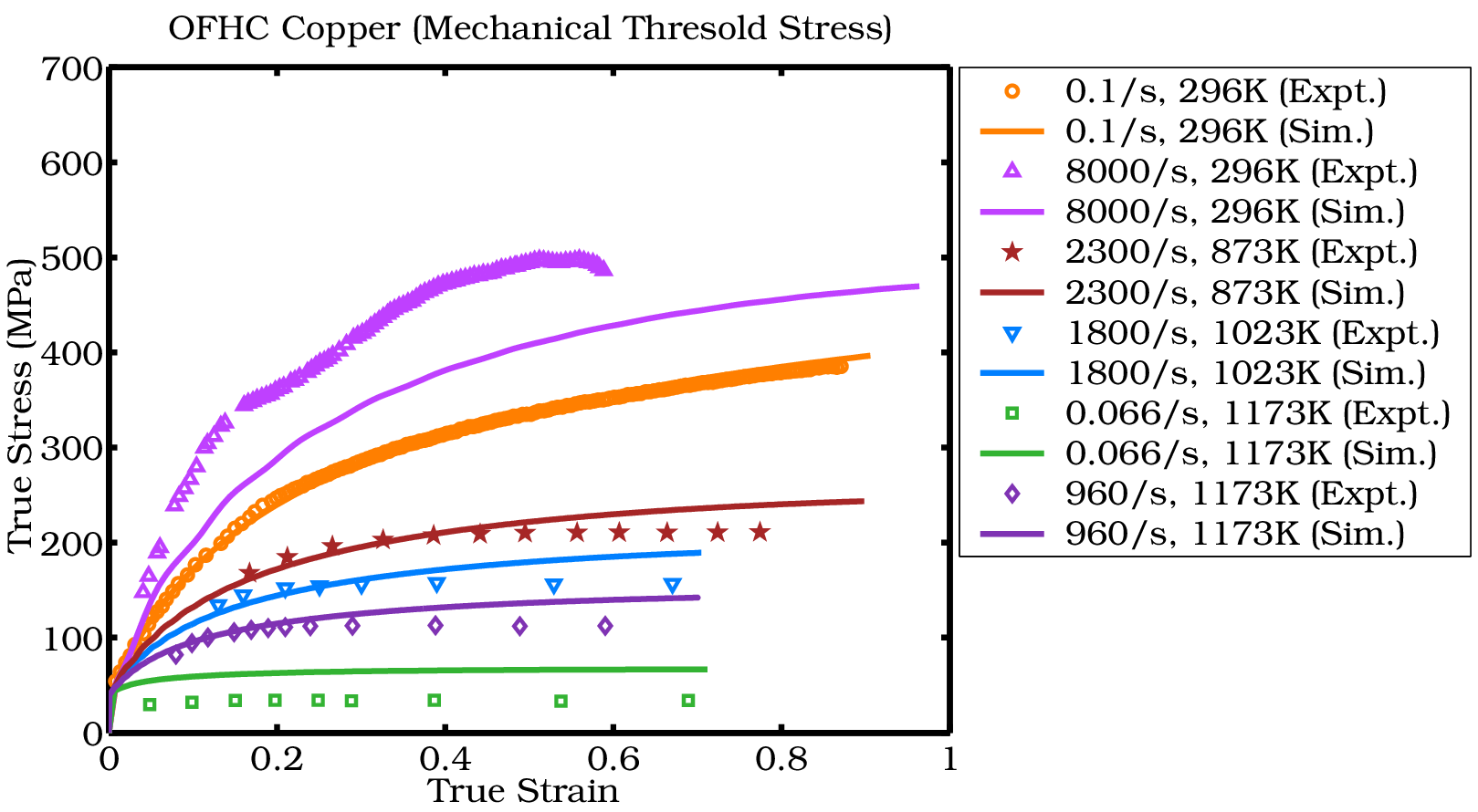}} \\
    (a) Various strain-rates and temperatures. \\
    \vspace{12pt}
    \scalebox{0.45}{\includegraphics{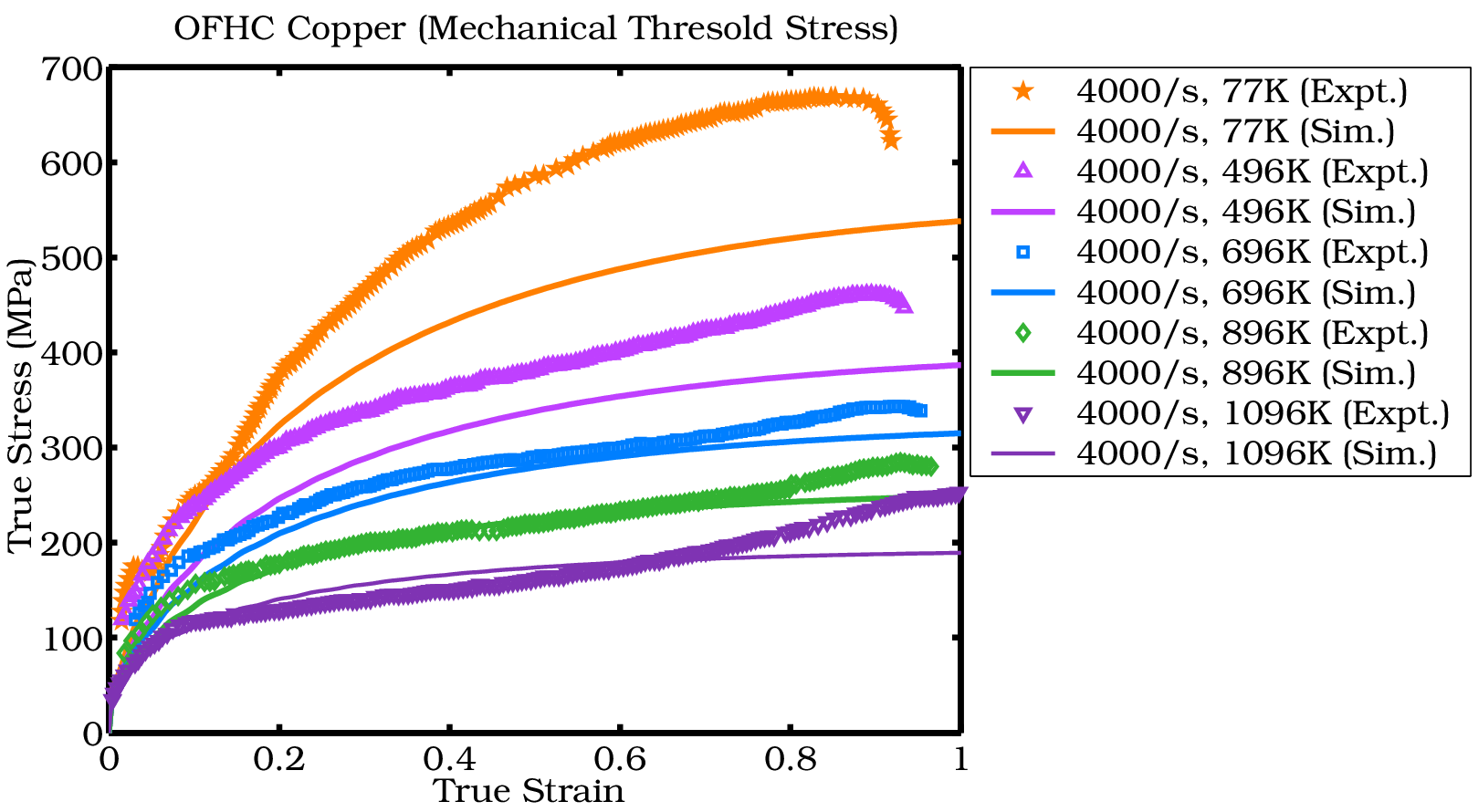}}\\
    (b) Various temperatures at 4000/s strain-rate.
    \caption{Predicted values of yield stress from the Mechanical Threshold
             Stress model.
             Please see the caption of Figure~\ref{fig:Cu1DJC} for the
             sources of the experimental data.}
    \label{fig:Cu1DMTS}
  \end{figure}

  Figures~\ref{fig:Cu1DMTS}(a) and (b) show the experimental values of yield 
  stress for OFHC copper versus those computed with the MTS model.  

  From Figure~\ref{fig:Cu1DMTS}(a), we can see that the yield stress predicted 
  by the MTS model almost exactly matches the experimental data at 296 K for 
  a strain-rate of 0.1/s.  The yield stress for the test conducted at 296 K
  and at 8000/s is underestimated.  Though reasonably accurate yield stresses
  are predicted at 1023 K and 1800/s, the experimental curves exhibit 
  earlier saturation than the model predicts.  The same is true at 873 K and
  2300 /s.  The predicted yield stress is higher for the quasistatic test at 
  1173 K than that observed experimentally.  However, the higher rate test at 
  the same temperature matches the experiments quite well except for a higher 
  amount of strain hardening at large strains.

  The variation of yield stress with temperature at a strain-rate of 
  4000/s is shown in Figure~\ref{fig:Cu1DMTS}(b).  The figure shows
  that the yield stress is underestimated by the MTS model at all
  temperatures except 1096 K.  The experimental data shows stage III or  
  stage IV hardening which is not predicted by the MTS model 
  that we have used.
  
  \subsubsection{Preston-Tonks-Wallace Model.}
  The Preston-Tonks-Wallace (PTW) model attempts to provide a single approach
  to model both thermally activated glide and overdriven shock regimes.
  The overdriven shock regime includes strain-rates greater than 10$^7$.
  The PTW model, therefore, extends the possibility of modeling plasticity
  beyond the range of validity of the MTS model.  We have not conducted 
  a simulations of overdriven shocks in this paper.  However, the PTW model 
  explicitly accounts for the rapid increase in yield stress at strain
  rates above 1000 /s.  Hence the model is a good candidate for the range of 
  strain-rates and temperatures of interest to us.  The PTW model parameters 
  used in our simulations are shown in Table~\ref{tab:CuPTW}.  In addition,
  we use the NP shear modulus model in all simulations involving the PTW yield 
  stress model. 
  \begin{table}[p]
    \centering
    \caption{Parameters used in the Preston-Tonks-Wallace yield stress model
             for copper (\citet{Preston03}).}
    \vspace{10pt}
    \begin{tabular}{cccccccc}
       \hline
       $s_0$ & $s_{\infty}$ & $y_0$ & $y_{\infty}$ & $d$ & $\kappa$ & $\gamma$ &
       $\theta$ \\
       \hline
       0.0085 & 0.00055 & 0.0001 & 0.0001 & 2 & 0.11 & 0.00001 & 0.025 \\
       \hline
       \hline
       $M$ (amu) & $s_1$ & $y_1$ & $y_2$ \\
       \hline
       63.546 & 0.25 & 0.094 & 0.575 \\
       \hline
    \end{tabular}
    \label{tab:CuPTW}
  \end{table}
  
\begin{figure}[p]
    \centering
    \scalebox{0.45}{\includegraphics{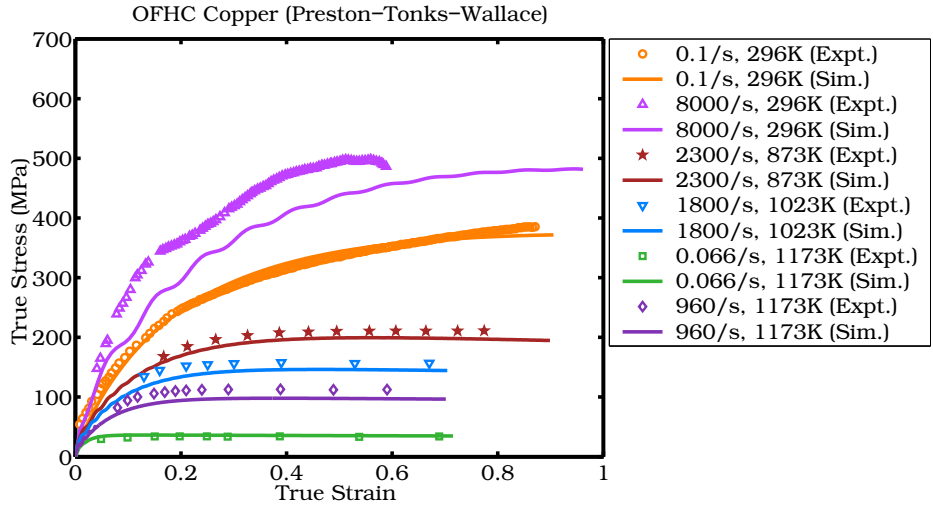}} \\
    (a) Various strain-rates and temperatures. \\
    \vspace{12pt}
    \scalebox{0.45}{\includegraphics{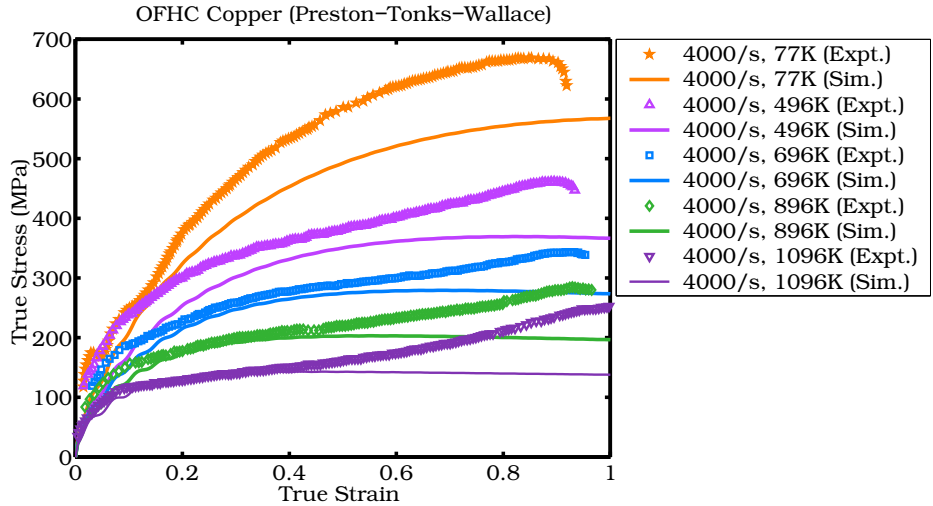}}\\
    (b) Various temperatures at 4000/s strain-rate.
    \caption{Predicted values of yield stress from the Preston-Tonks-Wallace
             model.
             Please see the caption of Figure~\ref{fig:Cu1DJC} for the
             sources of the experimental data.}
    \label{fig:Cu1DPTW}
  \end{figure}

  Experimental yield stresses are compared with those predicted by the PTW 
  model in Figures~\ref{fig:Cu1DPTW}(a) and (b).  The solid lines in the 
  figures are the predicted values while the symbols represent experimental 
  data.  From Figure~\ref{fig:Cu1DPTW}(a) we can see that the predicted yield 
  stress at 0.1/s and 296 K matches the experimental data quite well.  The error
  in the predicted yield stress at 296 K and 8000/s is also smaller than 
  that for the MTS flow stress model.  The experimental data at 873 K, 1023 K,
  and 1173 K were used by ~\citet{Preston03} to fit the model parameters.
  Hence it is not surprising that the predicted yield stresses match the
  experimental data better than any other model.

  The temperature-dependent yield stresses at 4000/s are shown in 
  Figure~\ref{fig:Cu1DPTW}(b).  In this case, the predicted values at 77 K
  are lower than the experimental values.  However, for higher temperatures,
  the predicted values match the experimental data quite well for strains
  less than 0.4.  At higher strains, the predicted yield stress saturates while
  the experimental data continues to show a significant amount of hardening.
  The PTW model predicts better values of yield stress for the compression
  tests while the MTS model performs better for the tension tests.
  
  \subsubsection{Errors in the flow stress models.}
  In this section, we use the difference between the predicted and the 
  experimental values of the flow stress as a metric to compare the various 
  flow stress models.  The error in the true stress is calculated using 
  \begin{equation}\label{eq:error1D}
    \text{Error}_{\sigma} = 
      \left(\cfrac{\sigma_{\text{predicted}}}{\sigma_{\text{expt.}}} - 1\right)
      \times 100 ~.
  \end{equation}

  A detailed discussion of the differences between the predicted and 
  experimental true stress for one-dimensional tests can be found 
  elsewhere (\citet{Banerjee05c}).  In this paper we summarize these
  differences in the form of error statistics as shown in 
  Tables~\ref{tab:Cu1D1err} and \ref{tab:Cu1D2err}.  Only true strains 
  greater than 0.1 have been considered in the generation of these statistics.
  The statistics in Tables~\ref{tab:Cu1D1err} and \ref{tab:Cu1D2err} 
  clearly show that no single model is consistently better than the 
  other models under all conditions.  
  \begin{table}[p]
    \centering
    \caption{Comparison of the error in the yield stress predicted
             by the five flow stress models at various strain-rates
             and temperatures.} 
    \begin{tabular}{cclccccc}
       \hline
       Temp. (K) & Strain Rate (/s) & Error 
                 & JC (\%) & SCGL (\%) & ZA (\%) & MTS (\%) & PTW (\%) \\
       \hline
       296 & 0.1 & Max.       &  32  & 55 &   3 &  2   &  3   \\
           &     & Min.       &  -4  & 31 & -10 & -4   & -6   \\
           &     & Mean       &  0.2 & 41 &  -4 &  0.2 &  0.5 \\
           &     & Median     &  -3  & 41 &  -5 &  0.6 &  1.1 \\
           &     & Std. Dev.  &  6   &  7 &   4 &  1.3 &  2.3 \\
       \hline
       296 & 8000 & Max.       &  1.1 &   3 & -10 & -12 & -6 \\
           &      & Min.       &  -22 & -12 & -21 & -29 & -29\\
           &      & Mean       &  -17 &  -6 & -17 & -19 & -14\\
           &      & Median     &  -20 &  -7 & -18 & -18 & -13\\
           &      & Std. Dev.  &  6   &   3 &   2 &   3 &  4\\
       \hline
       873 & 2300 & Max.       &  -7  & 49 &  -3 &  13 & -5 \\
           &      & Min.       &  -18 &  6 & -24 &  -5 & -7 \\
           &      & Mean       &  -13 & 26 & -15 &   4 & -6 \\
           &      & Median     &  -13 & 25 & -16 &   4 & -6 \\
           &      & Std. Dev.  &  4   & 16 &   7 &   7 &  0.5\\
       \hline
       1023 & 1800 & Max.       &  -16 & 53 &   3 &  20   & -7 \\
            &      & Min.       &  -30 & -3 & -22 &  -7   & -13 \\
            &      & Mean       &  -25 & 17 & -13 &   4   & -10 \\
            &      & Median     &  -27 & 11 & -17 &   1.5 & -9 \\
            &      & Std. Dev.  &  5   & 21 &   9 &  10   &  2\\
       \hline
       1173 & 0.066 & Max.       &  93 & 440 & 149 & 99 & 7 \\
            &       & Min.       &  39 & 186 & 119 & 81 & 3 \\
            &       & Mean       &  64 & 297 & 132 & 90 & 5 \\
            &       & Median     &  61 & 275 & 131 & 92 & 6 \\
            &       & Std. Dev.  &  20 &  93 &  12 &  6 & 1.4 \\
       \hline
       1173 & 960 & Max.       &  -37 & 50 &  14 & 24   & -13 \\
            &     & Min.       &  -49 & -8 &  -8 & -0.1 & -17 \\
            &     & Mean       &  -45 & 12 &  -2 & 9    & -15 \\
            &     & Median     &  -47 &  4 &  -6 & 6    & -14 \\
            &     & Std. Dev.  &  4   & 20 &   8 & 9    &  1\\
       \hline
    \end{tabular}
    \label{tab:Cu1D1err}
  \end{table}
  \begin{table}[p]
    \centering
    \caption{Comparison of the error in the yield stress predicted
             by the five flow stress models for a strain-rate of 
             4000/s.} 
    \begin{tabular}{cclccccc}
       \hline
       Temp. (K) & Strain Rate (/s) & Error 
                  & JC (\%) & SCGL (\%) & ZA (\%) & MTS (\%) & PTW (\%) \\
       \hline
       77 & 4000 & Max.       & 34 & 26 & 24 & -5  & -8 \\
          &      & Min.       & -28 & -8 & -9 & -22 & -17 \\
          &      & Mean       & -14 & -8 & -2 & -18 & -15 \\
          &      & Median     & -21 & -4 & -6 & -19 & -15 \\
          &      & Std. Dev.  & 16 &  9 &  9 &  5  &  2\\
       \hline
       496 & 4000 & Max.       & -2 & 11 & -17 & -11 & -8 \\
           &      & Min.       & -24 & -7 & -27 & -26 & -29 \\
           &      & Mean       & -17 &  3 & -22 & -15 & -14 \\
           &      & Median     & -17 &  5 & -21 & -14 & -13 \\
           &      & Std. Dev.  & 5 &  5 &   3 &   3 & 5\\
       \hline
       696 & 4000 & Max.       & -2 & 22 & -16 & -3 & -4 \\
           &      & Min.       & -20 & -2 & -25 & -16 & -20 \\
           &      & Mean       & -14 & 13 & -20 & -6 & -9 \\
           &      & Median     & -15 & 15 & -19 & -6 & -7 \\
           &      & Std. Dev.  & 4 &  7 &   3 &  3 & 5\\
       \hline
       896 & 4000 & Max.       & -16 & 20 & -17 & 3 & -2\\
           &      & Min.       & -32 & -9 & -24 & -15 & -30\\
           &      & Mean       & -23 & 13 & -20 & -3 & -13 \\
           &      & Median     & -21 & 16 & -20 & -2 & -11\\
           &      & Std. Dev.  & 4 &  7 &   2 & 5 & 9\\
       \hline
       1096 & 4000 & Max.       & -35 &  17 & -8  & 12 & 4\\
            &      & Min.       & -56 & -13 & -30 & -25 & -45 \\
            &      & Mean       & -42 &   7 & -15 & -1.4 & -18\\
            &      & Median     & -39 &   9 & -12 & 3 & -15 \\
            &      & Std. Dev.  & 7 &   8 &   7 & 12 & 16 \\
       \hline
    \end{tabular}
    \label{tab:Cu1D2err}
  \end{table}

  We can further simplify our evaluation
  by considering a single metric that encapsulates much of the information
  in these tables.  Table~\ref{tab:Cu1DAbsErr} shows comparisons based on
  one such simplified error metric.  We call this metric the average maximum 
  absolute (MA) error.  The maximum absolute (MA) error is defined as the sum 
  of the absolute mean error and the standard deviation of the error.  The 
  extreme values of the error are therefore ignored by the metric and only 
  values that are within one standard deviation of the mean are considered.
  \begin{table}[p]
    \centering
    \caption{Comparison of average "maximum" absolute (MA) errors in yield 
             stresses predicted by the five flow stress models for 
             various conditions.}
    \begin{tabular}{lccccc}
       \hline
       Condition & \multicolumn{5}{c}{Average MA Error (\%)} \\ 
                 & JC & SCGL & ZA & MTS & PTW \\
       \hline
       All Tests                       & 36 &  64 & 33 & 23 & 17 \\
       Tension Tests                   & 25 &  20 & 19 & 14 & 18 \\
       Compression Tests               & 45 & 126 & 50 & 35 & 10 \\
       High Strain Rate ($\ge$ 100 /s) & 29 &  22 & 20 & 15 & 18 \\
       Low  Strain Rate ($<$ 100 /s)   & 45 & 219 & 76 & 49 &  5 \\
       High Temperature ($\ge$ 800 K)  & 43 &  90 & 40 & 27 & 16 \\
       Low  Temperature ($<$ 800 K)    & 20 &  20 & 17 & 15 & 14 \\
       \hline
    \end{tabular}
    \label{tab:Cu1DAbsErr}
  \end{table}

  From Table~\ref{tab:Cu1DAbsErr} we observe that the least average MA
  error for all the tests is 17\% while the greatest average MA error is
  64\%.  The PTW model performs best while the SCGL model performs worst.
  In order of increasing error, the models may be arranged as PTW, MTS,
  ZA, JC, and SCGL.  

  If we consider only the tension tests, we see that the MTS model performs 
  best with an average MA error of 14\%.  The Johnson-Cook model does the
  worst at 25\% error.  For the compression tests, the PTW model does best
  with an error of 10\% compared to the next best, the MTS model with a 
  35\% error.  The SCGL error shows an average MA error of 126\% for these
  tests.  

  For the high strain-rate tests, the MTS model performs better
  than the PTW model with an average MA error of 15\% (compared to 18\%
  for PTW).  The low strain-rate tests are predicted best by the PTW
  model (5 \%) and worst by the SCGL model (219 \%).  Note that this average
  error is based on two tests at 296 K and 1173 K and may not be representative 
  for intermediate temperatures.  

  The PTW model shows an average MA error
  of 16\% for the high temperature tests compared to 27\% for the MTS model.
  The SCGL model again performs the worst.  Finally, the low temperature
  tests ($<$ 800 K) are predicted best by the PTW model.   The other models
  also perform reasonably well under these conditions.  

  From the above comparisons, the Preston-Tonks-Wallace and the 
  Mechanical Threshold Stress models clearly stand out as 
  reasonably accurate over the largest range of strain-rates and temperatures.
  To further improve our confidence in the above conclusions, we perform a 
  similar set of comparisons with Taylor impact test data in the next 
  section.
  
  Note that we could potentially recalibrate all the models to get a better 
  fit to the experimental data and render the above comparisons void.  However, 
  it is likely that the average user of such models in computational codes 
  will use parameters that are readily available in the literature with the 
  implicit assumption is that published parameters provide the best possible 
  fit to experimental data.  Hence, exercises such as ours provide 
  useful benchmarks for the comparative evaluation of various flow stress 
  models.

\section{Taylor impact simulations} \label{sec:taylor}
  The Taylor impact test (\citet{Taylor48}) was originally devised as a 
  means of determining the dynamic yield strength of solids.  The test
  involves the impact of a flat-nosed cylindrical projectile on a hard
  target at normal incidence.  The test was originally devised to determine 
  the yield strengths of materials at high strain-rates.  However, that use
  of the test is limited to peak strains of around 0.6 at the center of the 
  specimen (\citet{Johnson88a}).  For higher strains and strain-rates, the 
  Taylor test is more useful as a means of validating high strain-rate
  plasticity models in numerical codes (\citet{Zerilli87}).

  The attractiveness of the Taylor impact test arises because of the 
  simplicity and inexpensiveness of the test.  A flat-ended cylinder is 
  fired on a target at a relatively high velocity and the final deformed shape 
  is measured.  The drawback of this test is that intermediate states of 
  the cylinder are relatively difficult to measure.  

  In this section, we compare the deformed profiles of Taylor cylinders
  from experiments with profiles that we obtain from our simulations.
  The experimental profiles are from the open literature and have been 
  digitized at a high resolution.  The errors in digitization are of the order 
  of 2\% to 5\% depending on the clarity of the image.  Our simulations
  use the Uintah code and the Material Point Method (see \ref{appA} and
  \ref{appB}).

  All our simulations are three-dimensional and model a quarter of the 
  cylinder.  We have used 8 material points per cell (64 material points
  per cell for simulations at 1235 K), a 8 point interpolation
  from material points to grid, and a cell spacing of 0.3 mm.  A cell spacing of
  0.15 mm gives essentially the same final deformed profile 
  (\citet{Banerjee05c}).  The anvil is modeled as a rigid material.  Contact 
  between the cylinder and the anvil is assumed to be frictionless.  The 
  effect of frictional contact has been discussed elsewhere 
  (\citet{Banerjee05c}).  We have not included the effect of damage accumulation
  due to void nucleation and growth in these simulations.  Details of such
  effects can be found in \citet{Banerjee05c}.

  Our simulations were run for 150 $\mu$s - 200 $\mu$s depending on the 
  problem.  These times were sufficient for the cylinders to rebound from the
  anvil and to stop undergoing further plastic deformation.  However, small 
  elastic deformations continue to persist as the stress waves reflect from 
  the surfaces of the cylinder.

  We have performed a systematic and extensive set of verification and 
  validation tests to determine the accuracy of the Material Point Method
  and its implementation within Uintah 
  (~\citet{Banerjee04b,Banerjee04c,Banerjee04d,Banerjee05a,Banerjee05b,Banerjee05c}).
  A number of materials and conditions have been explored in the process.
  We are, therefore, reasonably confident in the results of our simulations.

  \subsection{Metrics} \label{sec:metric}
  The systematic verification and validation of computational codes and 
  the associated material models requires the development and utilization
  of appropriate comparison metrics (see \citet{Oberkampf02,Babuska04}).  
  In this section we discuss a few geometrical metrics that can be used in 
  the context of Taylor impact tests.  Other metrics such as the surface 
  temperature and the time of impact may also be used if measured values 
  are available.

\begin{figure}[p]
    \centering
    \scalebox{0.50}{\includegraphics{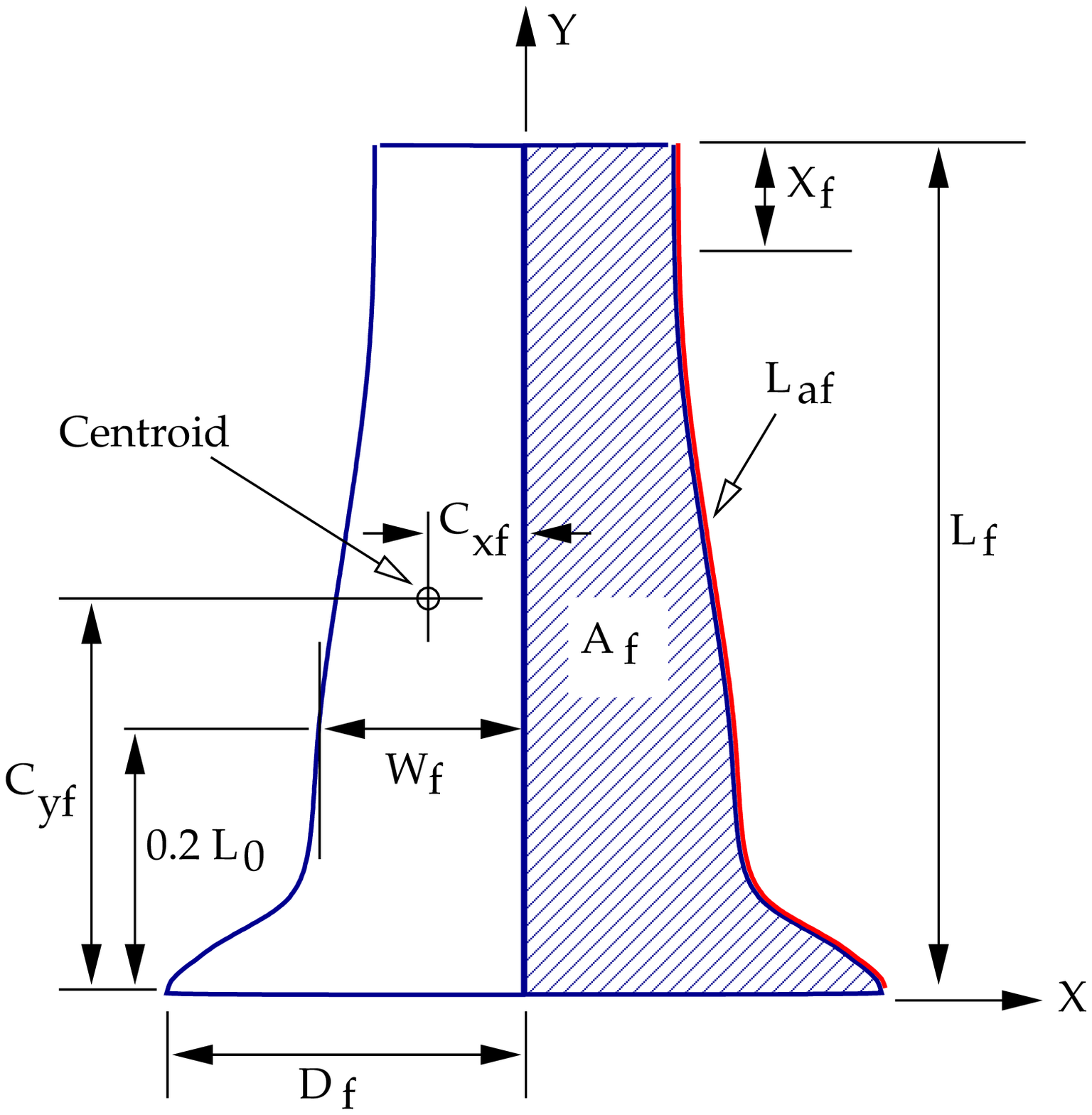}}\\
    \caption{Geometrical metrics used to compare profiles of Taylor 
             impact specimens.}
    \label{fig:TaylorMetrics}
  \end{figure}

  In most papers on the simulation of Taylor impact tests, a plot of the 
  deformed configuration is superimposed on the experimental data and a 
  visual judgement of accuracy is made.  However, when the number of Taylor 
  tests is large, it is not possible to present sectional/plan views for 
  all the tests and numerical metrics are preferable.  Some such metrics that 
  have been used to compare Taylor impact tests are 
  (see Figure~\ref{fig:TaylorMetrics}) :
  \begin{enumerate}
    \item The final length of the deformed cylinder ($L_f$)
          (\citet{Wilkins73,Gust82,Jones87,Johnson88a,House95}).
    \item The diameter of the mushroomed end of the cylinder ($D_f$)
          (\citet{Johnson88a,House95}).
    \item The length of the elastic zone in the cylinder ($X_f$)
          (\citet{Jones87,House95}).
    \item The bulge at a given distance from the deformed end ($W_f$)
          (\citet{Johnson88a}).
  \end{enumerate}
  Contours of plastic strain have also been presented in a number of
  works on Taylor impact.  However, such contours are not of much use
  when comparing simulations with experiments (though they are useful
  when comparing two stress update algorithms).

  The above metrics are inadequate when comparing the secondary bulges in
  two Taylor cylinders.  We consider some additional geometrical metrics 
  that act as a substitute for detailed pointwise geometrical comparisons 
  between two Taylor test profiles.  These are
  (see Figure~\ref{fig:TaylorMetrics}) :
  \begin{enumerate}
    \item The final length of a axial line on the surface of the 
          cylinder ($L_{af}$).
    \item The area of the cross-sectional profile of the deformed 
          cylinder ($A_f$).
    \item The volume of the deformed cylinder ($V_f$).
    \item The location of the centroid of the deformed cylinder in
          terms of a orthonormal basis with origin at deformed end
          ($C_{xf}$, $C_{yf}$).
    \item The moments of inertia of the cross section of the deformed
          cylinder about the basal plane ($I_{xf}$) and an axial plane
          ($I_{yf}$).
  \end{enumerate}
  Higher order moments should also be computed so that we can dispense
  with arbitrary measures such as $W_f$.
  The numerical formulas used to compute the area, volume, centroid, and
  moments of inertia are given in Appendix~\ref{app:Metric}.
  
  \subsection{Experimental data}
  In this section, we show plots of the experimentally determined values of 
  some of the metrics discussed in the previous section.  Quantities with
  subscript '0' represent initial values.  The abscissa in each plot 
  is a measure of the total energy density in the cylinder.  The internal
  energy density has been added to the kinetic energy to separate the high
  temperature and low temperature data.

\begin{figure}[p]
    \centering
    \scalebox{0.50}{\includegraphics{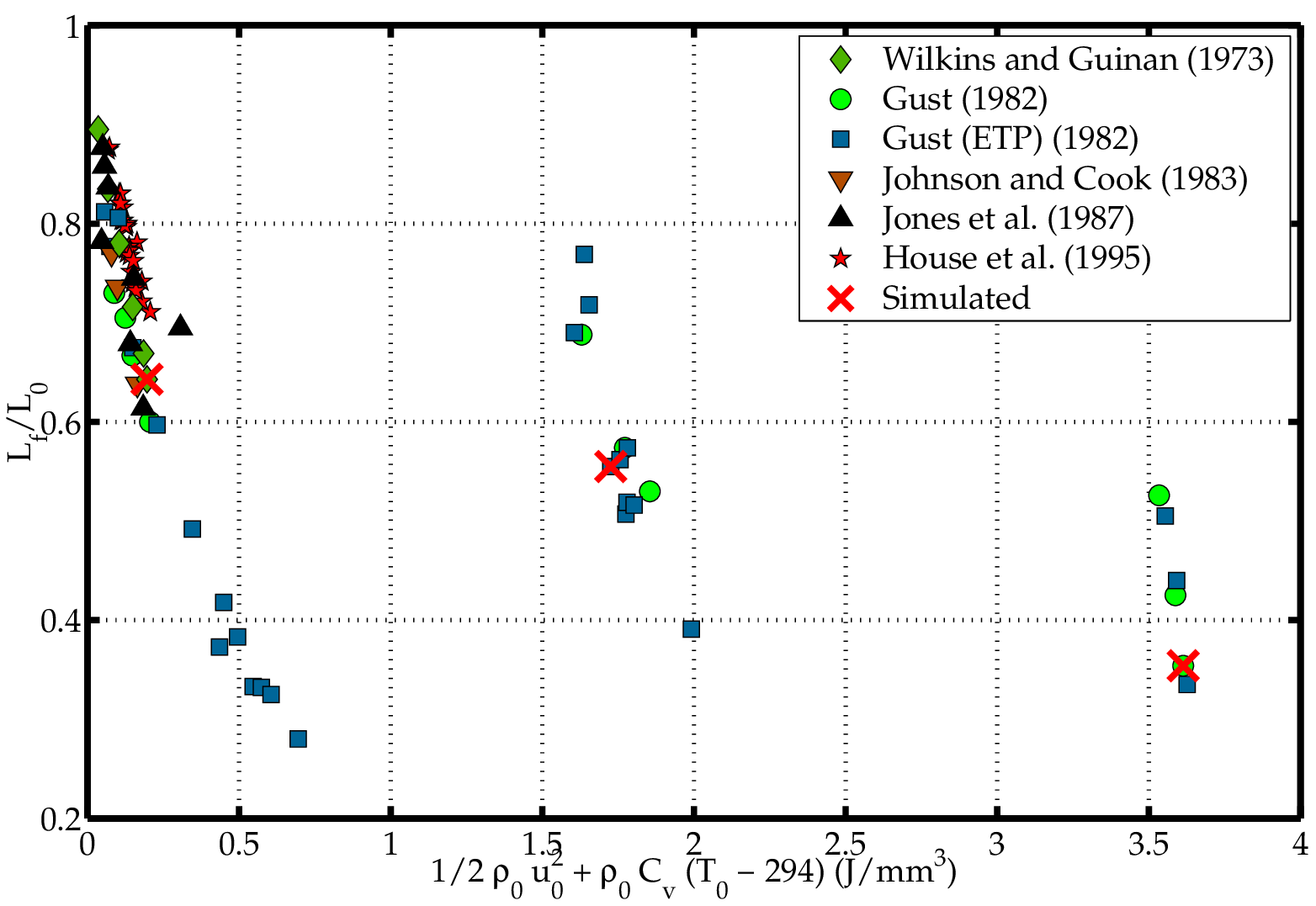}}\\
    \caption{Ratio of final length to initial length of copper
             Taylor cylinders for various conditions.  The data are from
             \citet{Wilkins73,Gust82,Johnson83,Jones87} and 
             \citet{House95}.}
    \label{fig:CuExptTaylorLf}
  \end{figure}

  Figure~\ref{fig:CuExptTaylorLf} shows the length ratios ($L_f/L_0$) for 
  a number of Taylor impact tests.  The figure indicates the following:
  \begin{enumerate}
    \item The ratio ($L_f/L_0$) is essentially independent of the 
          initial length and diameter of the cylinder.
    \item There is a linear relationship between the ratio ($L_f/L_0$)
          and the initial kinetic energy density.
    \item As temperature increases, the absolute value of the slope of this 
          line increases.
    \item The deformation of OFHC (Oxygen Free High Conductivity)
          cannot be distinguished from that of ETP (Electrolytic Tough Pitch)
          copper from this plot.
  \end{enumerate}
  We have chosen to do detailed comparisons between experiment and simulation
  for the three tests marked with crosses on the figure.  These tests 
  represent situations in which fracture has not been observed in the 
  cylinders and cover the range of temperatures of interest to us.

\begin{figure}[p]
    \centering
    \scalebox{0.50}{\includegraphics{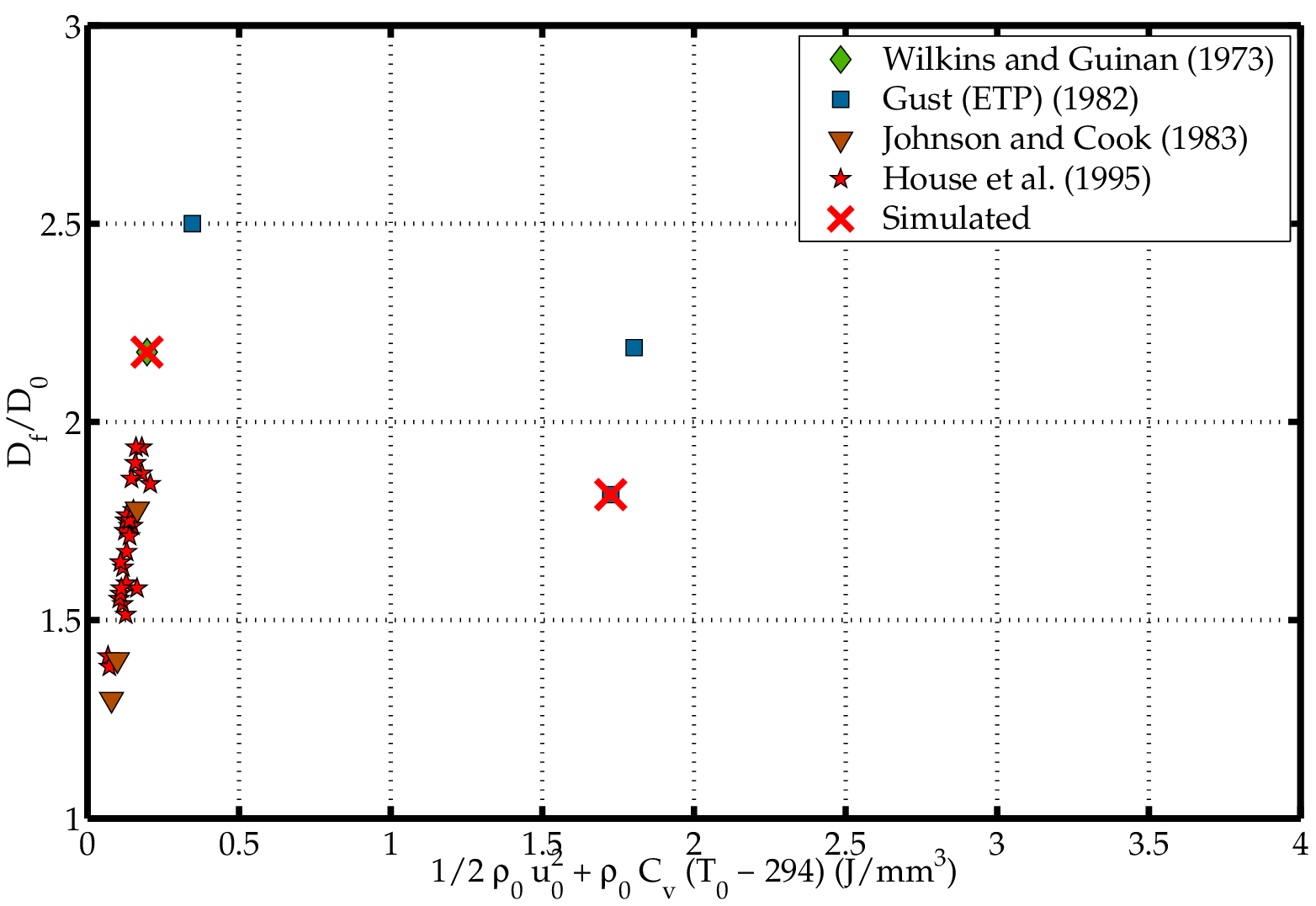}}\\
    \caption{Ratio of final length to initial length of copper
             Taylor cylinders for various conditions.  The data are from
             \citet{Wilkins73,Gust82,Johnson83} and 
             \citet{House95}.}
    \label{fig:CuExptTaylorDf}
  \end{figure}

  The ratio of the diameter of the deformed end to the original diameter
  ($D_f/D_0$) for some of these tests is plotted as a function of the energy 
  density in Figure~\ref{fig:CuExptTaylorDf}.  A linear relation similar 
  to that for the length is observed.

\begin{figure}[p]
    \centering
    \scalebox{0.50}{\includegraphics{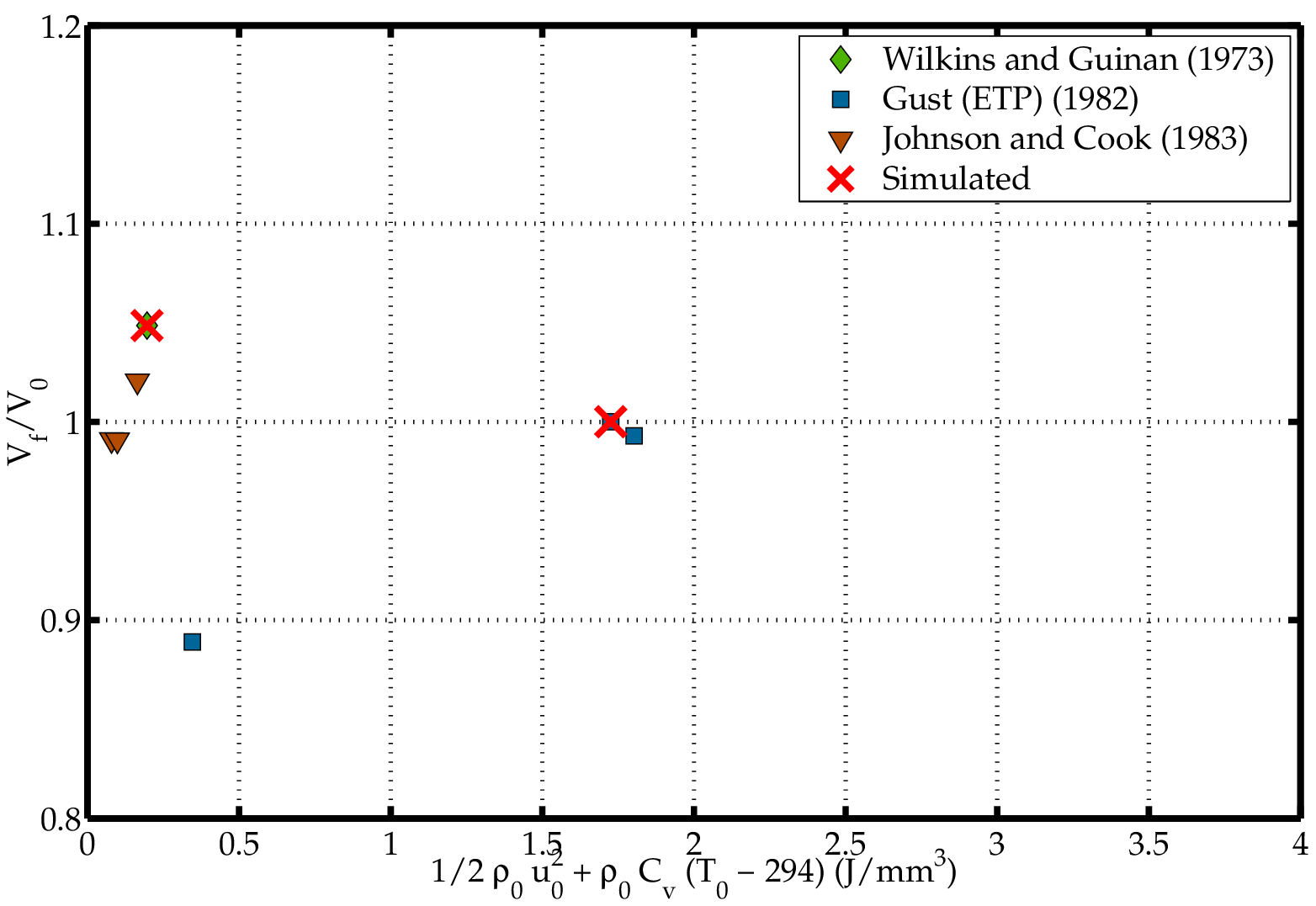}}\\
    \caption{Ratio of the final volume to initial volume of copper
             Taylor cylinders for various conditions.  The data are from
             \citet{Wilkins73,Gust82} and \citet{Johnson83}.}
    \label{fig:CuExptTaylorVf}
  \end{figure}

  The volume of the cylinder should be preserved during the Taylor
  test if isochoric plasticity holds.  Figure~\ref{fig:CuExptTaylorVf} shows 
  the ratio of the final volume to the initial volume ($V_f/V_0$) as a 
  function of the energy density.  We can see that the volume is preserved for
  three of the tests but not for the rest.  This discrepancy may be due
  to errors in digitization of the profile.

  \subsection{Evaluation of flow stress models}
  In this section we present results from simulations of three Taylor tests on 
  copper, compute validation metrics, and compare these metrics with 
  experimental data.  Table~\ref{tab:CuInitData} shows the initial dimensions, 
  velocity, and temperature of the three specimens that we have simulated.  All
  three specimens had been annealed before experimental testing.
  \begin{table}[t]
    \caption{Initial data for copper simulations.}
    \begin{tabular}{lllllll}
       \hline
       Test & Material
            & Initial & Initial
            & Initial & Initial 
            & Source \\
            & 
            & Length & Diameter 
            & Velocity & Temp. \\
            & 
            & ($L_0$ mm) & ($D_0$ mm)
            & ($V_0$ m/s) & ($T_0$ K)\\
       \hline
        Cu-1 & OFHC Cu & 23.47 & 7.62 & 210 & 298   & \citet{Wilkins73} \\
        Cu-2 & ETP  Cu & 30    & 6.00 & 188 & 718   & \citet{Gust82} \\
        Cu-3 & ETP  Cu & 30    & 6.00 & 178 & 1235  & \citet{Gust82} \\
       \hline
    \end{tabular}
    \label{tab:CuInitData}
  \end{table}

  \subsubsection{Test Cu-1}
\begin{figure}[p]
    \begin{minipage}[t]{3in}
      \centering
      \scalebox{0.4}{\includegraphics{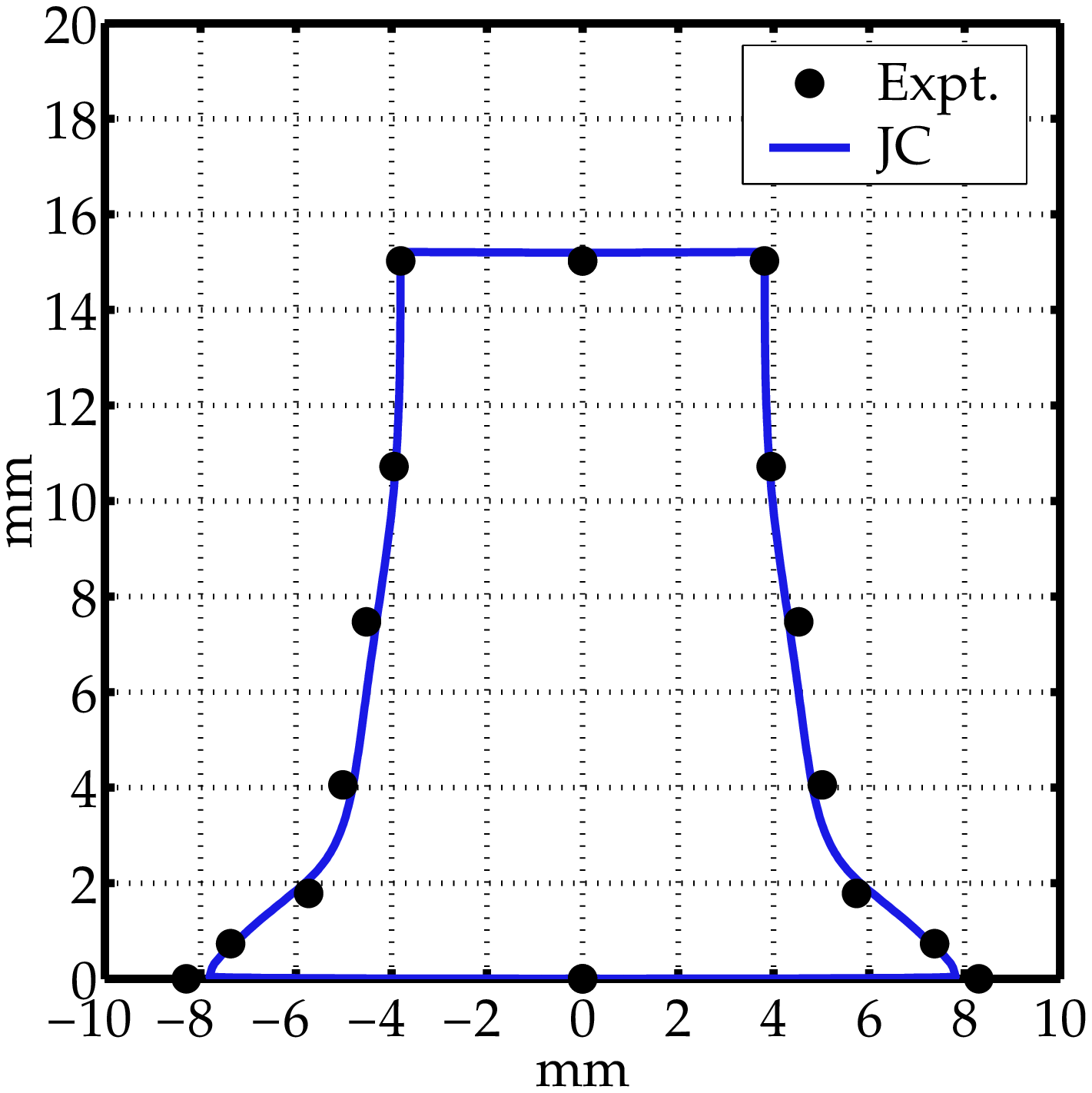}}\\
      (a) Johnson-Cook.
    \end{minipage}
    \begin{minipage}[t]{3in}
      \centering
      \scalebox{0.4}{\includegraphics{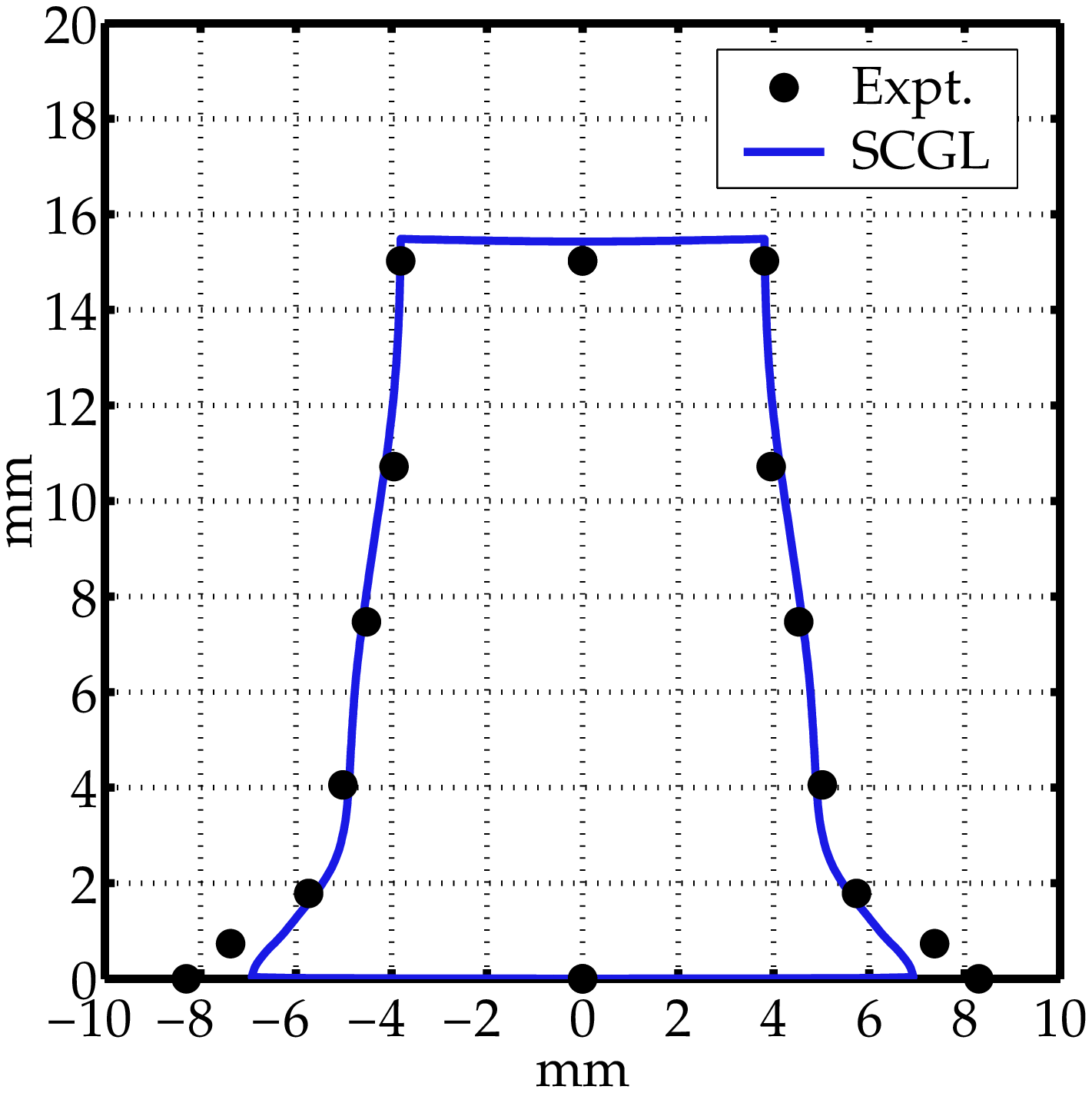}}\\
      (b) Steinberg-Cochran-Guinan-Lund.
    \end{minipage}
    \vspace{15pt}
    \begin{minipage}[t]{3in}
      \centering
      \scalebox{0.4}{\includegraphics{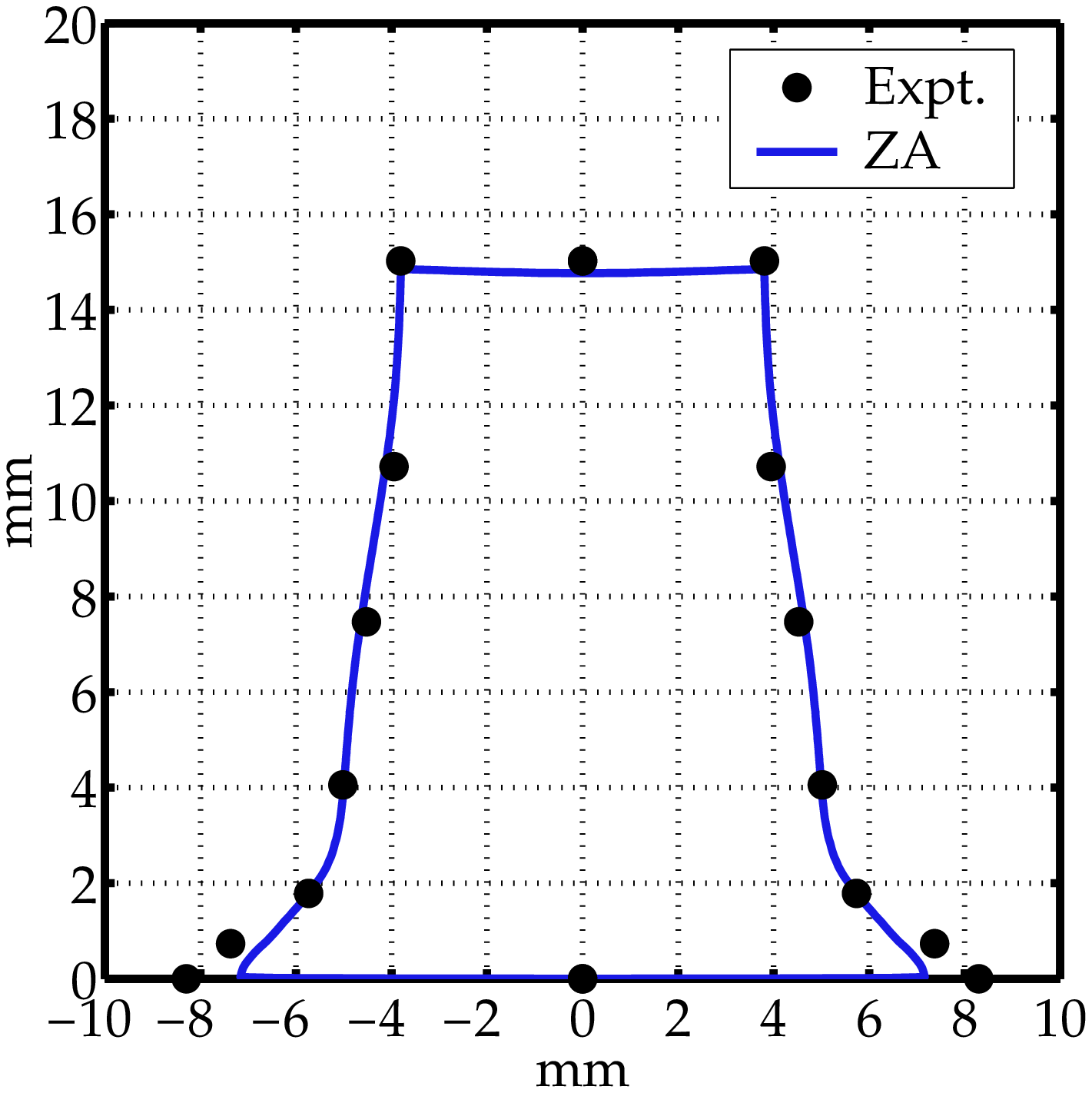}}\\
      (c) Zerilli-Armstrong
    \end{minipage}
    \begin{minipage}[t]{3in}
      \centering
      \scalebox{0.4}{\includegraphics{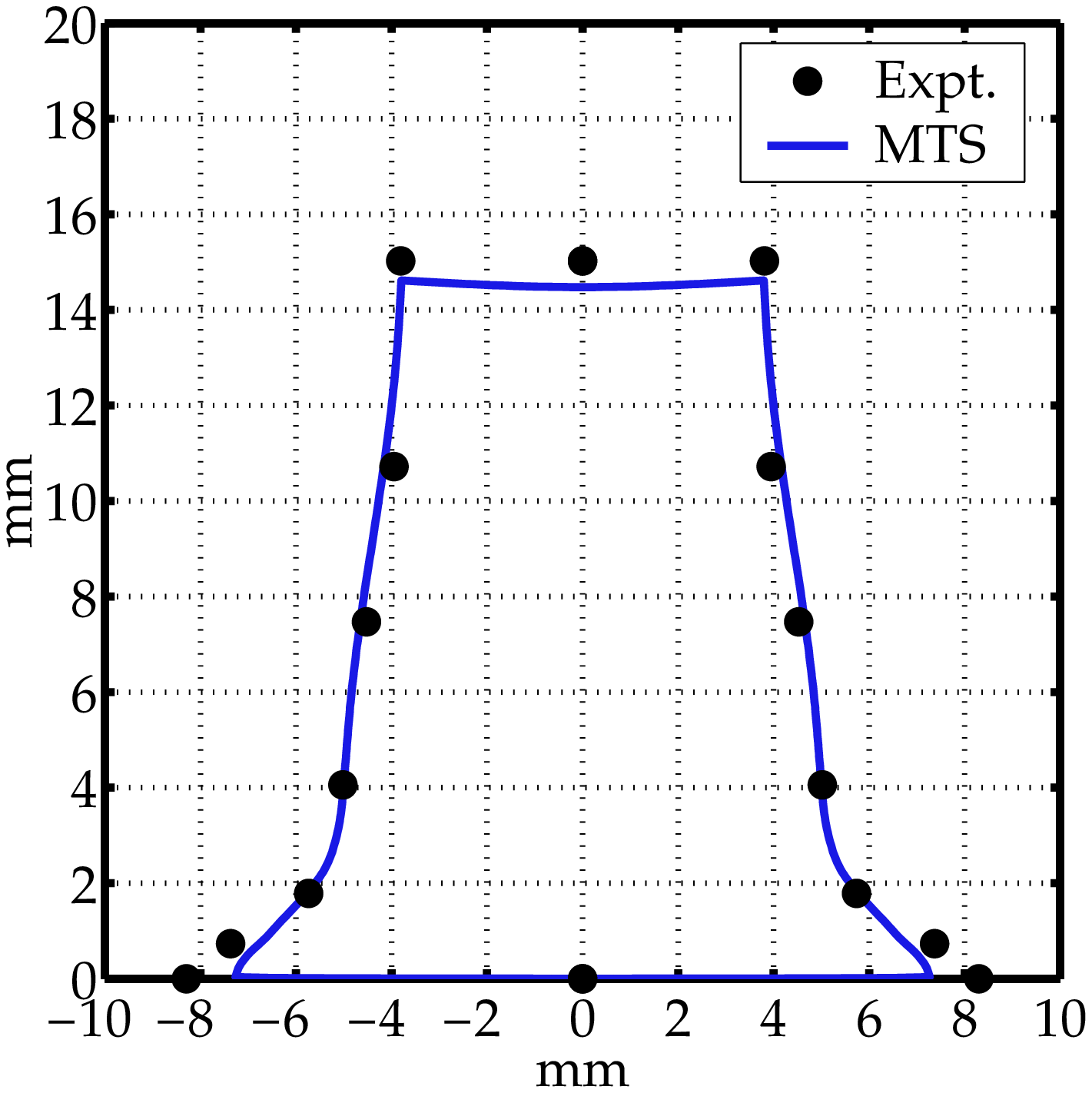}}\\
      (d) Mechanical Threshold Stress.
    \end{minipage}
    \vspace{15pt}
    \begin{minipage}[t]{3in}
      \centering
      \scalebox{0.4}{\includegraphics{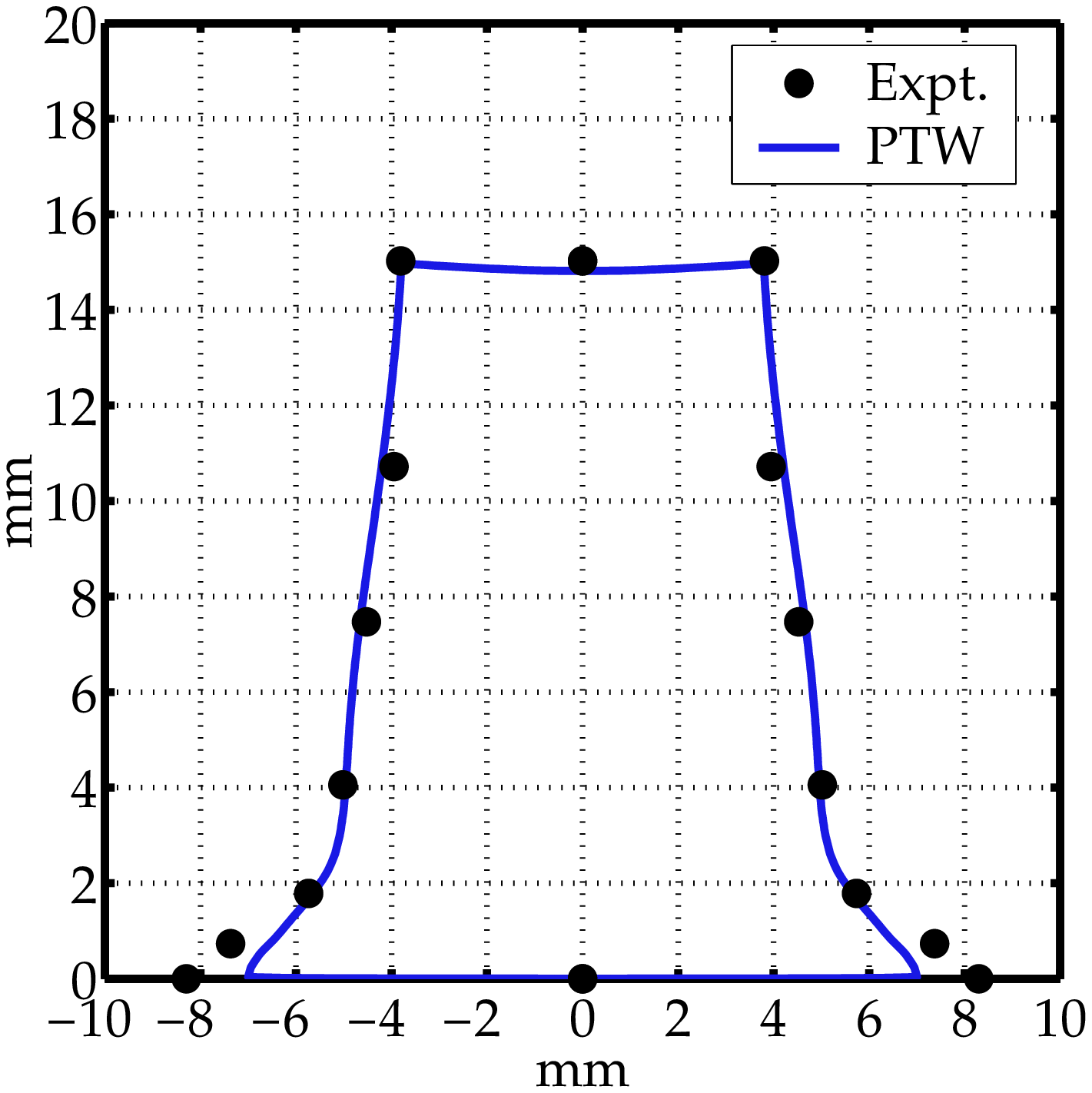}}\\
      (e) Preston-Tonks-Wallace.
    \end{minipage}
    \begin{minipage}[t]{3in}
      \centering
      \scalebox{0.4}{\includegraphics{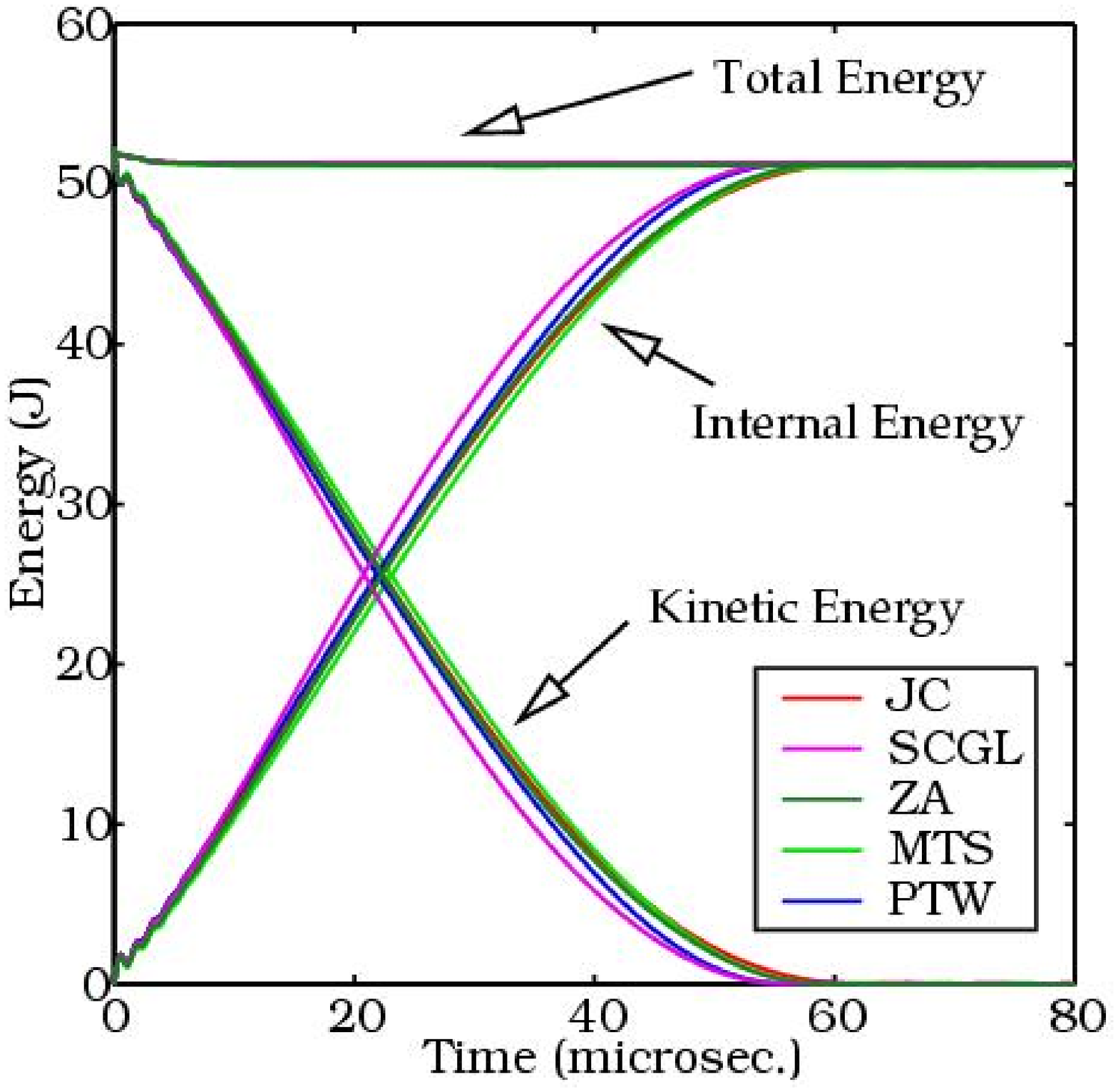}}\\
      (f) Energy versus time.
    \end{minipage}
    \caption{Computed versus experimental profiles for Taylor test Cu-1
             and the computed energy versus time profile.}
    \label{fig:Cu1TaylorProf}
  \end{figure}

  Test Cu-1 is a room temperature test at an initial nominal strain-rate of 
  around 9000/s.  Figures~\ref{fig:Cu1TaylorProf}(a), (b), (c), (d), and (e) 
  show the profiles computed by the JC, SCGL, ZA, MTS, and PTW models, 
  respectively, for test Cu-1.   

  The Johnson-Cook model gives the best match 
  to the experimental data at this temperature (room temperature) if we 
  consider the final length and the final mushroom diameter.  All the other 
  models underestimate the mushroom diameter but predict the final length 
  quite accurately.  The MTS model underestimates the final length.

  The time at which the cylinder loses all its kinetic energy (as 
  predicted by the models) is shown in the energy plot of 
  Figure~\ref{fig:Cu1TaylorProf}(f).  The predicted times vary between
  55 micro secs to 60 micro secs but are essentially the same for all the
  models.  The total energy is conserved relatively well.  The slight
  initial dissipation is the result of the artificial viscosity in the
  numerical algorithm that is used to damp out initial oscillations.

  In Figure~\ref{fig:CuExptTaylorVf} we have seen that the final volume of 
  the cylinder for test Cu-1 is around 5\% larger than the initial volume.  
  We assume that this error is due to errors in digitization.  In that
  case, we have errors of +1\% for measures of length and errors of +2\% for
  measures of area in the experimental profile.  Moments of inertia of 
  areas are expected to have errors of around 7\%.

\begin{figure}[t]
    \centering
    \scalebox{0.40}{\includegraphics{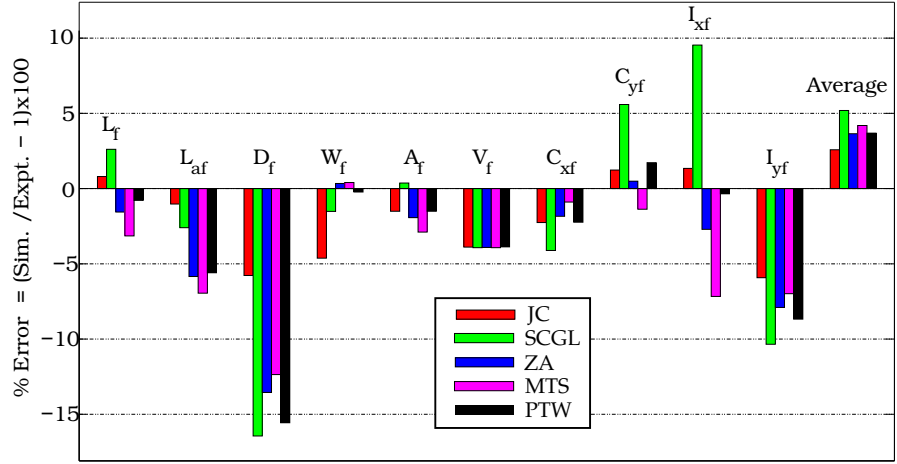}}\\
    \caption{Comparison of error metrics for the five models 
             for Taylor test Cu-1.}
    \label{fig:Cu1Error}
  \end{figure}

  The error metrics for test Cu-1 are shown in Figure~\ref{fig:Cu1Error}.
  The final length ($L_f$) is predicted to within 3\% of the experimental 
  value by all the models.  The Johnson-Cook and Preston-Tonks-Wallace 
  models show the least error.  

  The length of the deformed surface of the cylinder ($L_{af}$) is predicted 
  best by the Johnson-Cook and Steinberg-Cochran-Guinan-Lund models.  The other
  models underestimate the length by more than 5\%.  The final mushroom 
  diameter ($D_f$) is underestimated by 5\% to 15\%.  The Johnson-Cook model 
  does the best for this metric, followed by the Mechanical Threshold Stress
  model.  The width of the bulge ($W_f$) is underestimated by the Johnson-Cook 
  and SCGL models and accurately predicted by the ZA, MTS, and PTW models.  
  The length of the elastic zone ($X_f$) is predicted to be zero by the SCGL, 
  ZA, MTS, and PTW models while the Johnson-Cook model predicts a value of 
  1.5 mm.  Moreover, an accurate estimate of $X_f$ cannot be made 
  from the experimental profile for test Cu-1.  Therefore we do not consider 
  this metric of utility in our comparisons for this test. 

  From Figure~\ref{fig:Cu1Error} we see that the predicted area of 
  the profile ($A_f$) is within 3\% of the experimental value for all 
  the models.  The SCGL model shows the least error in this metric.  If we
  decrease the experimental area by 2\% (in accordance with the assumed
  error in digitization), the Johnson-Cook and PTW models 
  show the least error in this metric. 
  
  The predicted final volume of the cylinder is around 0.8\% larger than 
  the initial volume showing that volume is not preserved accurately by 
  our stress update algorithm.  The error in digitization is around 5\%.
  That gives us a uniform error of 5\% between the experimental and computed
  volume ($V_f$) as can be seen in Figure~\ref{fig:Cu1Error}.

  The locations of the centroids ($C_{xf},C_{yf}$) provide further geometric 
  information about the shapes of the profiles.  These are the first order
  moments of the area.  The computed values are within 2\% of experiment
  except for the MTS model which shows errors of -4\% for $C_{xf}$ and
  +6\% for $C_{yf}$.

  The second moments of the area are shown as $I_{xf}$ and $I_{yf}$ in 
  Figure~\ref{fig:Cu1Error}.  The error in $I_{xf}$ tracks and accentuates
  the error in $L_f$ while the error in $I_{yf}$ tracks the error in 
  $D_f$.  The width of the bulge is included in this metric and it can
  be used the replace metrics such as $L_f$, $D_f$, and $W_f$ for the 
  purpose of comparison.  We notice this tracking behavior when the 
  overall errors are small but not otherwise.

  We have also plotted the arithmetic mean of the absolute value of the
  errors in each of the metrics to get an idea about which model performs best.
  The average error is the least (2.5\%) for the Johnson-Cook model, followed by
  the ZA and PTW models (3.5\%).  The MTS model shows an average error of 4\%
  while the SCGL model shows the largest error (5\%).  If we subtract the 
  digitization error from the experimental values, these errors decrease
  and lie in the range of 2\% to 3\%.  

  In summary, all the models predict profiles that are within the range
  of experimental variation for the test at room temperature.  Additional
  simulations at higher strain-rates (\citet{Banerjee05c}) have confirmed
  that all the models do well for room temperature simulations for strain
  rates ranging from 500 /s to 8000 /s.  We suggest that the simplest model
  should be used for such room temperature simulations and our 
  recommendation is the Zerilli-Armstrong model for copper.

  \subsubsection{Test Cu-2}
\begin{figure}[p]
    \begin{minipage}[t]{3in}
      \centering
      \scalebox{0.4}{\includegraphics{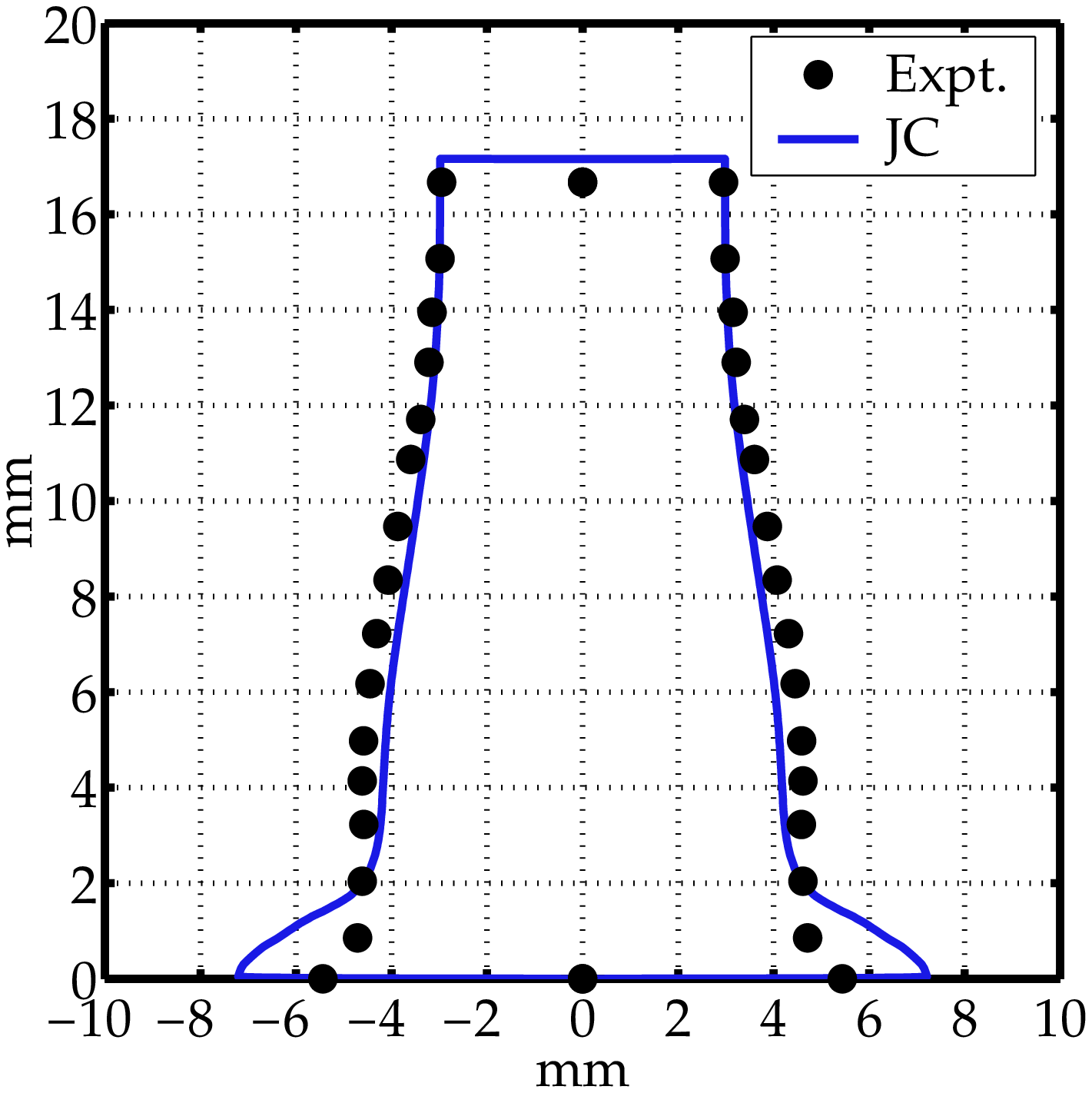}}\\
      (a) Johnson-Cook.
    \end{minipage}
    \begin{minipage}[t]{3in}
      \centering
      \scalebox{0.4}{\includegraphics{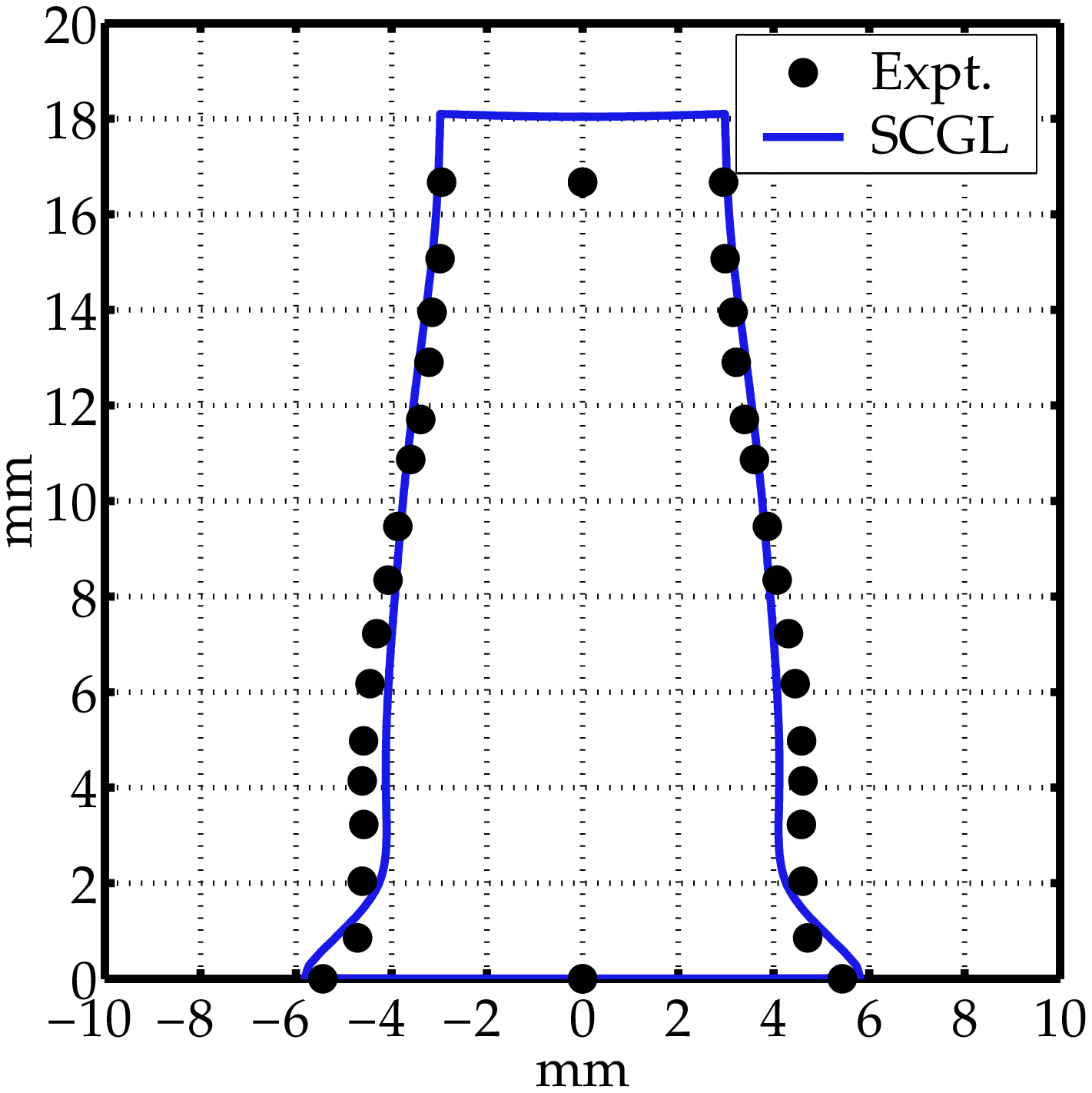}}\\
      (b) Steinberg-Cochran-Guinan-Lund.
    \end{minipage}
    \vspace{15pt}
    \begin{minipage}[t]{3in}
      \centering
      \scalebox{0.4}{\includegraphics{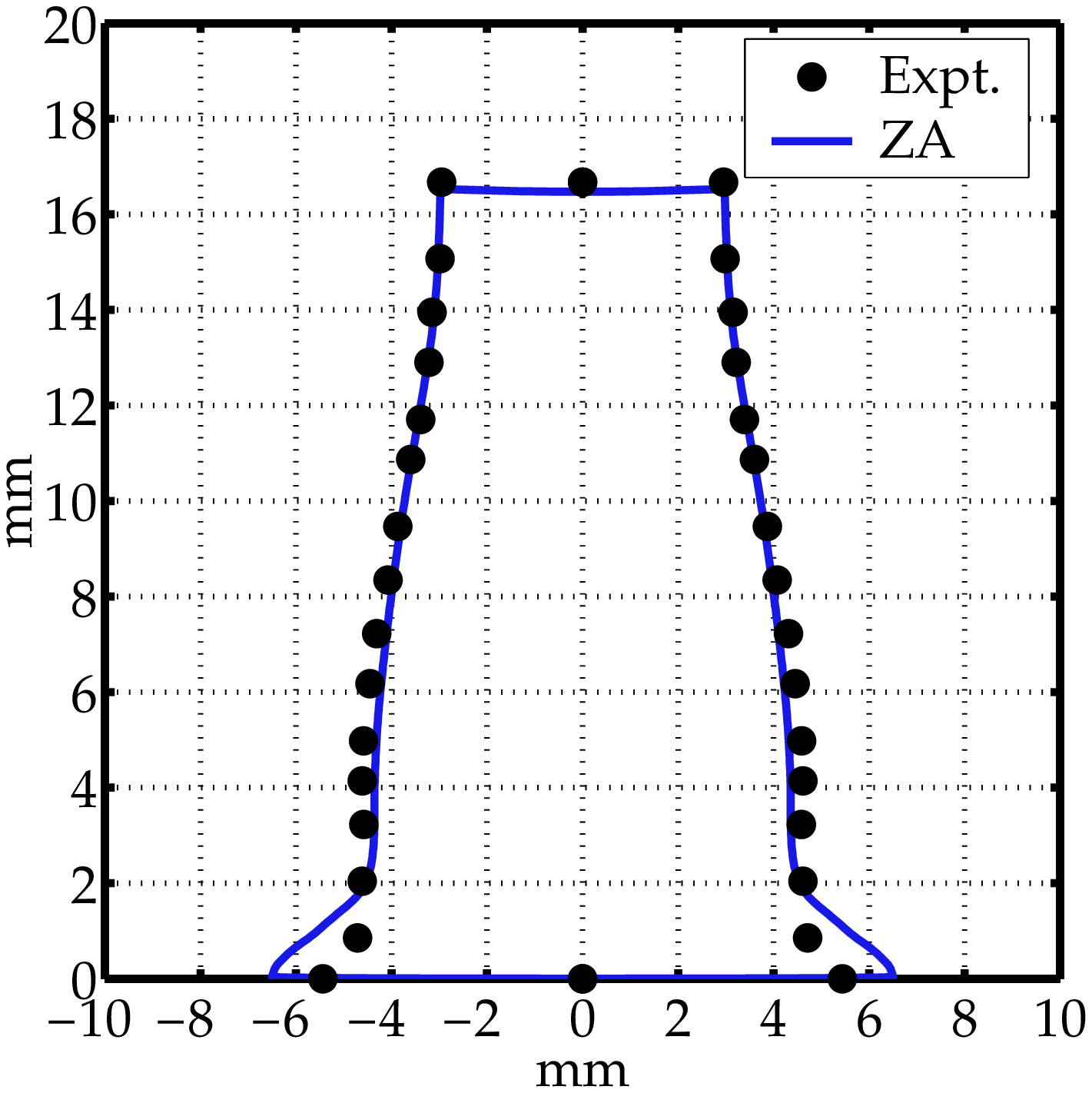}}\\
      (c) Zerilli-Armstrong
    \end{minipage}
    \begin{minipage}[t]{3in}
      \centering
      \scalebox{0.4}{\includegraphics{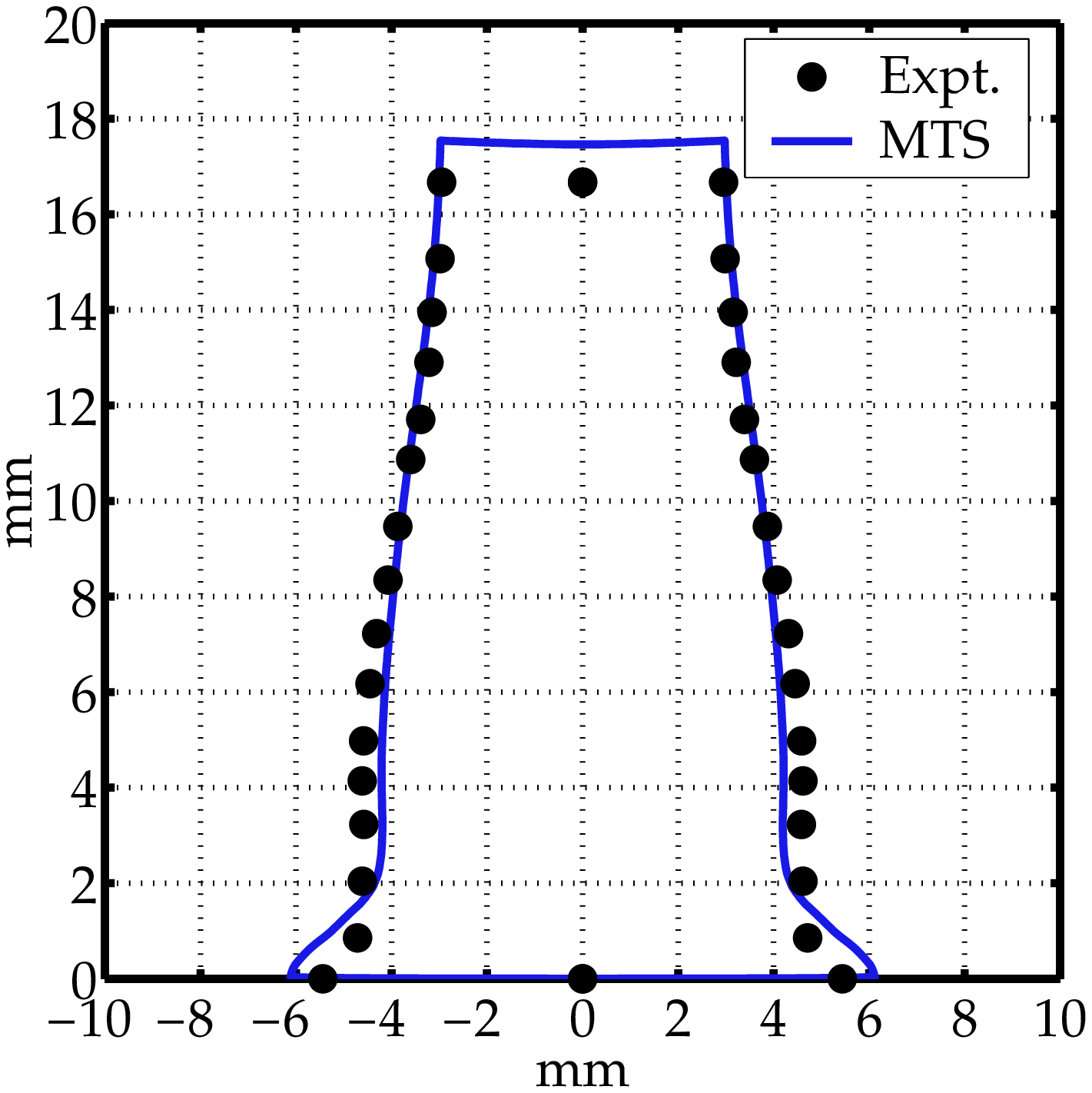}}\\
      (d) Mechanical Threshold Stress.
    \end{minipage}
    \vspace{15pt}
    \begin{minipage}[t]{3in}
      \centering
      \scalebox{0.4}{\includegraphics{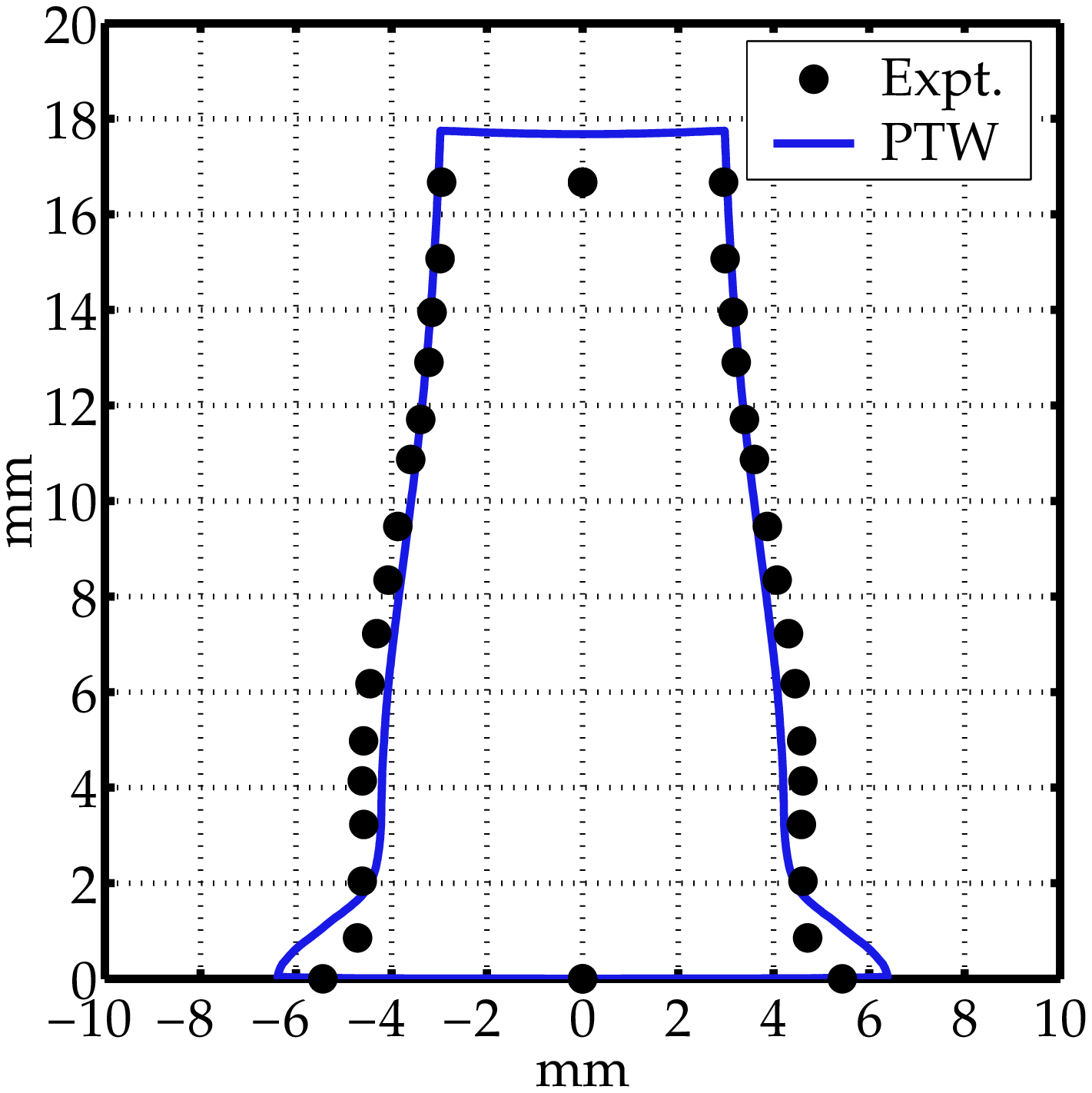}}\\
      (e) Preston-Tonks-Wallace.
    \end{minipage}
    \begin{minipage}[t]{3in}
      \centering
      \scalebox{0.40}{\includegraphics{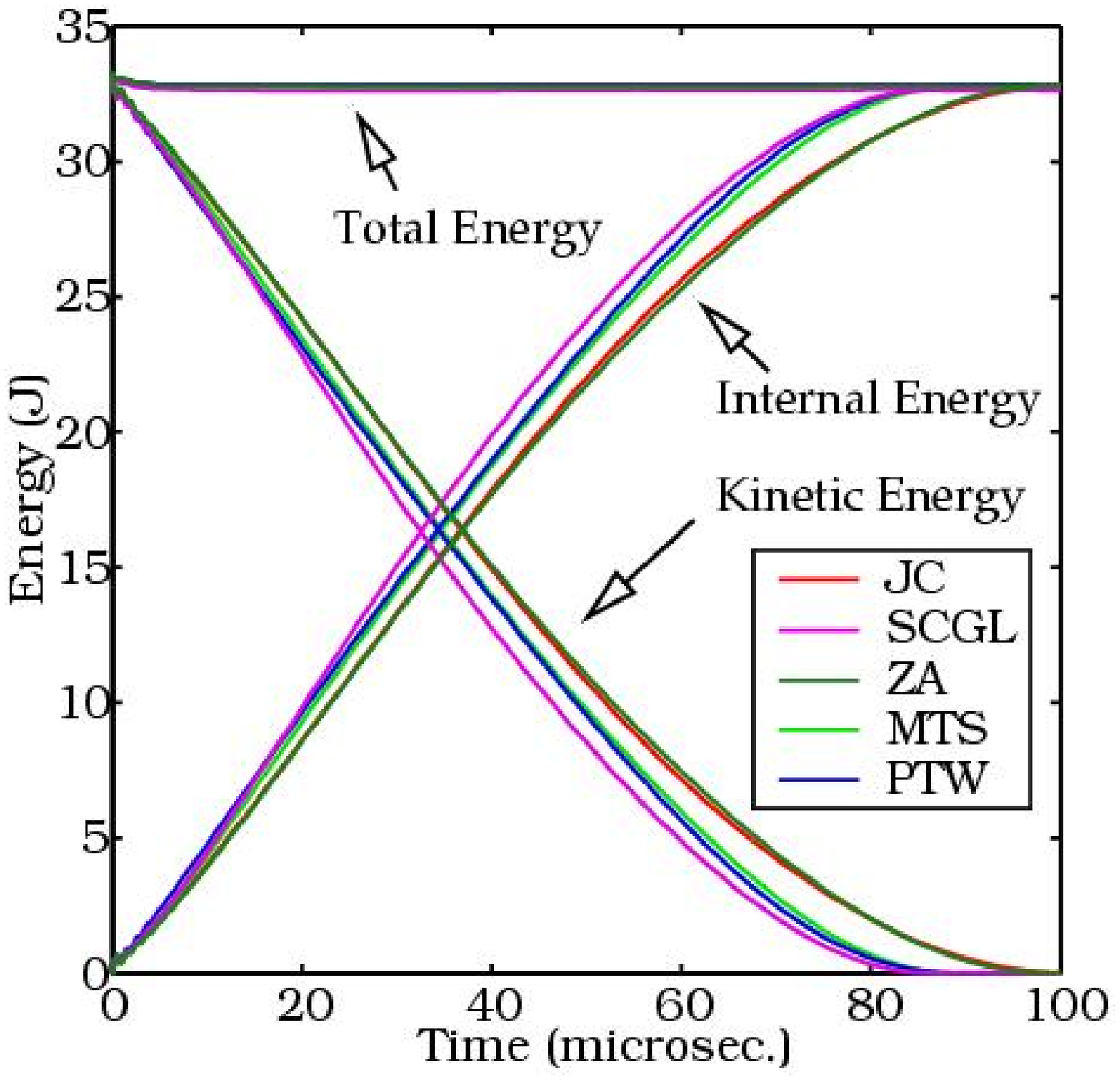}}\\
      (f) Energy versus time.
    \end{minipage}
    \caption{Computed and experimental profiles for Taylor test Cu-2 and
             the computed energy-time profile.}
    \label{fig:Cu3TaylorProf}
  \end{figure}

  Test Cu-2 is at a temperature of 718 K and the initial nominal strain-rate
  is around 6200/s.  Figures~\ref{fig:Cu3TaylorProf}(a), (b), (c), (d), and 
  (e) show the profiles computed by the JC, SCGL, ZA, MTS, and PTW models, 
  respectively, for test Cu-2.  

  In this case, the Johnson-Cook model predicts 
  the final length well but overestimates the mushroom diameter.  The SCGL 
  model overestimates the length but predicts the mushroom diameter well.  The 
  ZA model predicts the overall profile remarkably well except for the mushroom 
  diameter.  The MTS model slightly overestimates both the final length and 
  the mushroom diameter.  The PTW model also performs similarly, except that 
  the error is slightly larger than that for the MTS model. 

  The energy plot for test Cu-2 is shown in Figure~\ref{fig:Cu3TaylorProf}(f).
  In this case, the time of impact predicted by the JC and ZA models is
  around 100 micro secs while that predicted by the SCGL, MTS, and PTW
  models is around 90 micro secs.

\begin{figure}[t]
    \centering
    \scalebox{0.40}{\includegraphics{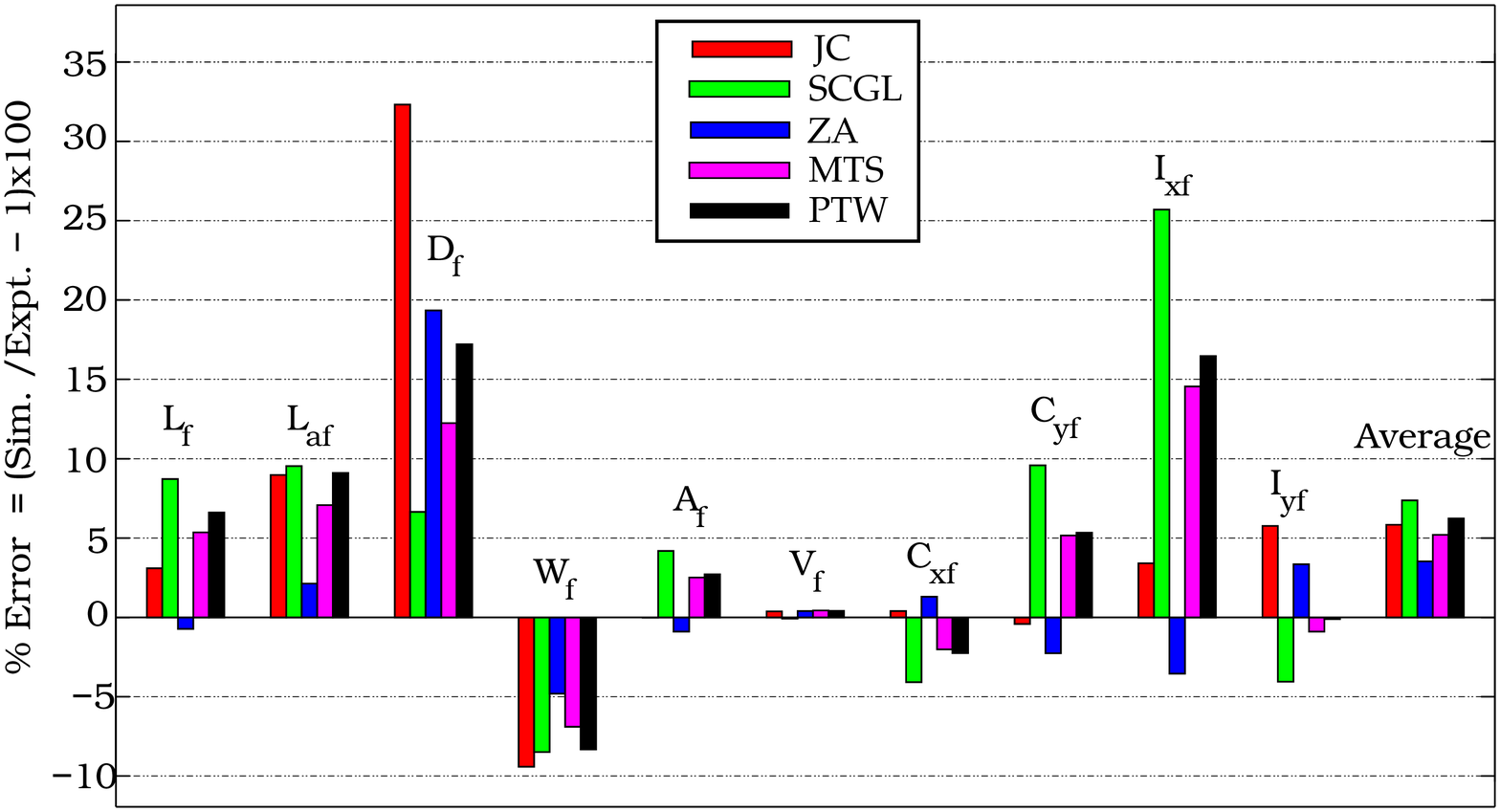}}\\
    \caption{Comparison of error metrics for the five models 
             for Taylor test Cu-2.}
    \label{fig:Cu3Error}
  \end{figure}

  The error metrics for test Cu-2 are shown in Figure~\ref{fig:Cu3Error}.
  In Figure~\ref{fig:CuExptTaylorVf} we have seen that the deformed volume
  computed from the digitized profile is almost exactly equal to the initial
  volume for test Cu-2.  The digitization error can be neglected in this case.

  The least error in the predicted final length ($L_f$) is for the ZA
  model followed by the JC model.  The SCGL model shows the largest
  error in this metric (7\%).  The MTS and PTW models overestimate the
  final length by around 6\%.
  The value of $L_{af}$ is predicted to within 2\% of the experimental 
  value by the ZA model.  The corresponding errors in the other models vary
  from 6\% (MTS) to 9\% (SCGL).
  The mushroom diameter is overestimated by all models.  The JC model
  overestimates this metric by more than 30\%.  The ZA and PTW models 
  overestimate $D_f$ by 17\% to 19\%.  The MTS model overestimates $D_f$
  by 12\%.  The SCGL model does best with an error of 7\%.
  The width of the bulge is underestimated by all the models with errors
  varying between 5\% (ZA) tp 9\% (JC).

  The final area ($A_f$) is predicted almost exactly by the JC model.  The
  ZA model underestimates the area by 1\% while the errors in the other 
  models vary from 2\% to 4\%.  The error in the final volume is less
  than 1\% for all the models.

  The location of the centroid is predicted best by the Johnson-Cook model
  followed by the ZA model.  Both the MTS and PTW models underestimate 
  $C_{xf}$ by 2\% and overestimate $C_{yf}$ by 5\%.  The SCGL model shows
  the largest error for this metric.

  For the second order moments $I_{xf}$, the smallest error is for the 
  Johnson-Cook model followed by the ZA model.  The largest errors are 
  from the SCGL model.  The MTS and PTW models overestimate this metric
  my 15\%.  The PTW model predicts $I_{yf}$ the best, followed by the 
  MTS model showing that the overall shape of the profile is best 
  predicted by these models.  The Johnson-Cook and SCGL models show the
  largest errors in this metric.
  
  On average, the ZA model performs best for test Cu-2 at 718 K with an 
  average error of 4\%.  The
  MTS model shows an average error of 5\% while the JC and PTW models
  show errors of approximately 6\%.  The SCGL model, with an average
  error of approximately 7\%, does the worst.

  \subsubsection{Test Cu-3}
\begin{figure}[p]
    \begin{minipage}[t]{3in}
      \centering
      \scalebox{0.4}{\includegraphics{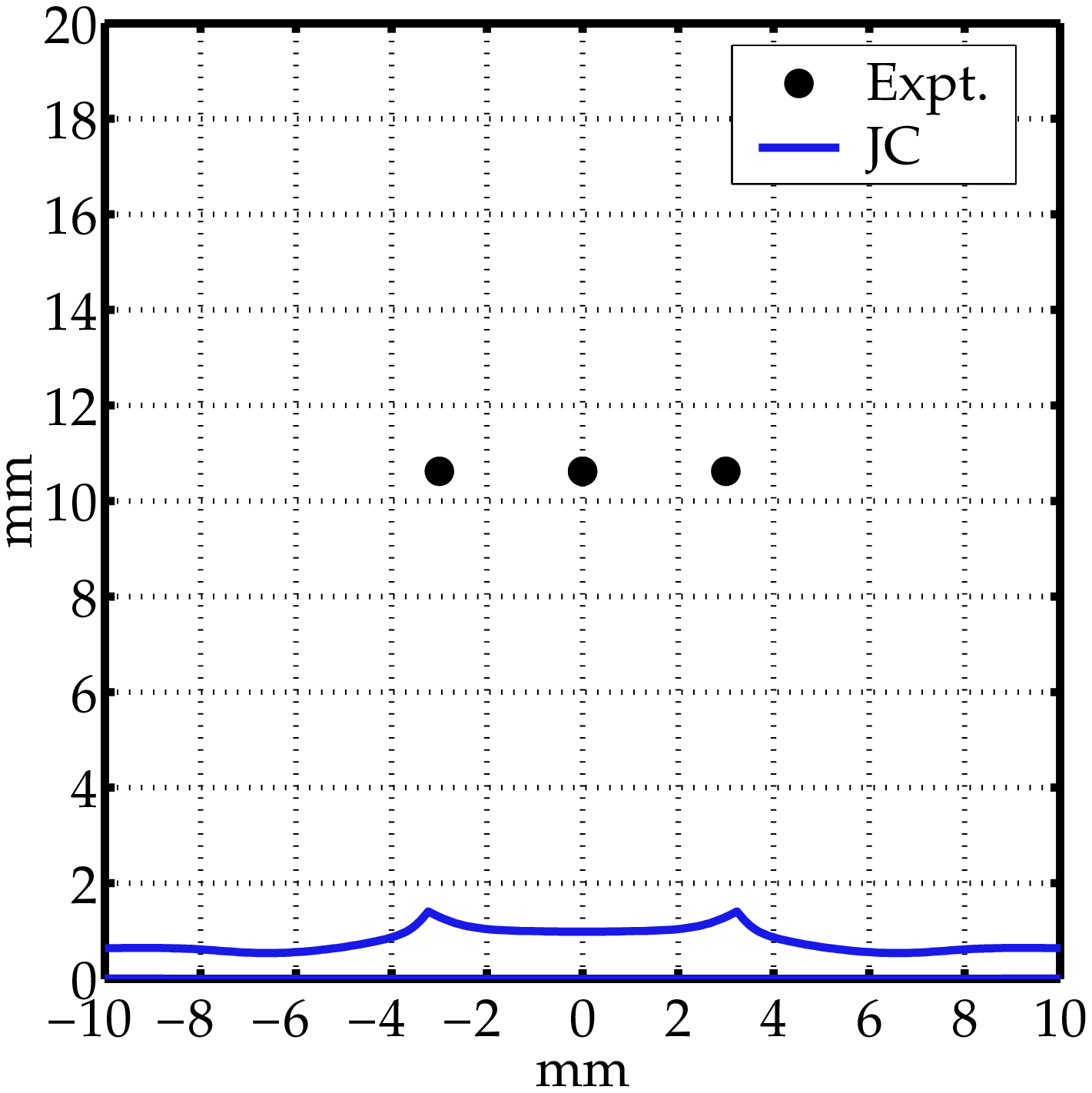}}\\
      (a) Johnson-Cook.
    \end{minipage}
    \begin{minipage}[t]{3in}
      \centering
      \scalebox{0.4}{\includegraphics{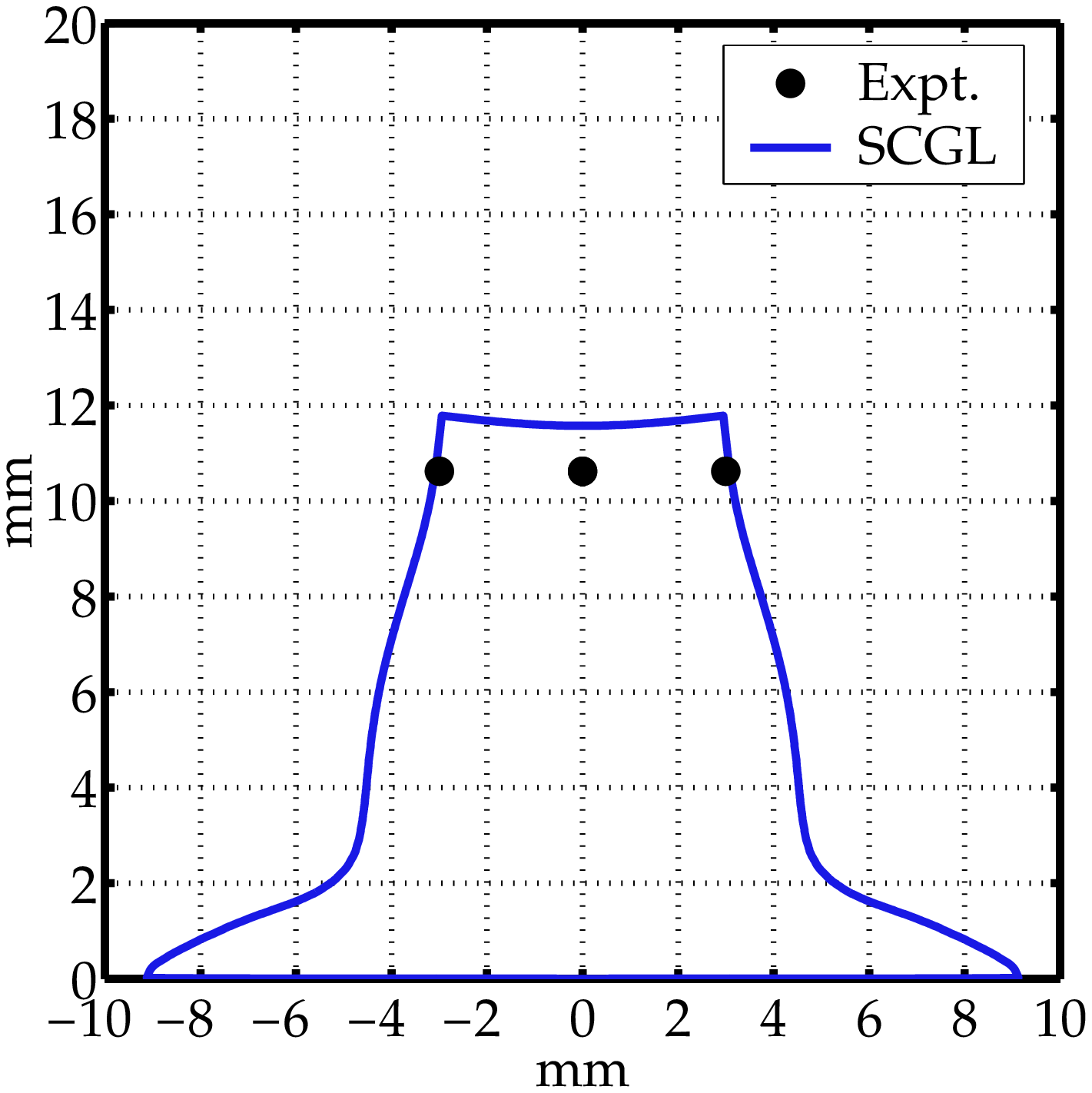}}\\
      (b) Steinberg-Cochran-Guinan-Lund.
    \end{minipage}
    \vspace{15pt}
    \begin{minipage}[t]{3in}
      \centering
      \scalebox{0.4}{\includegraphics{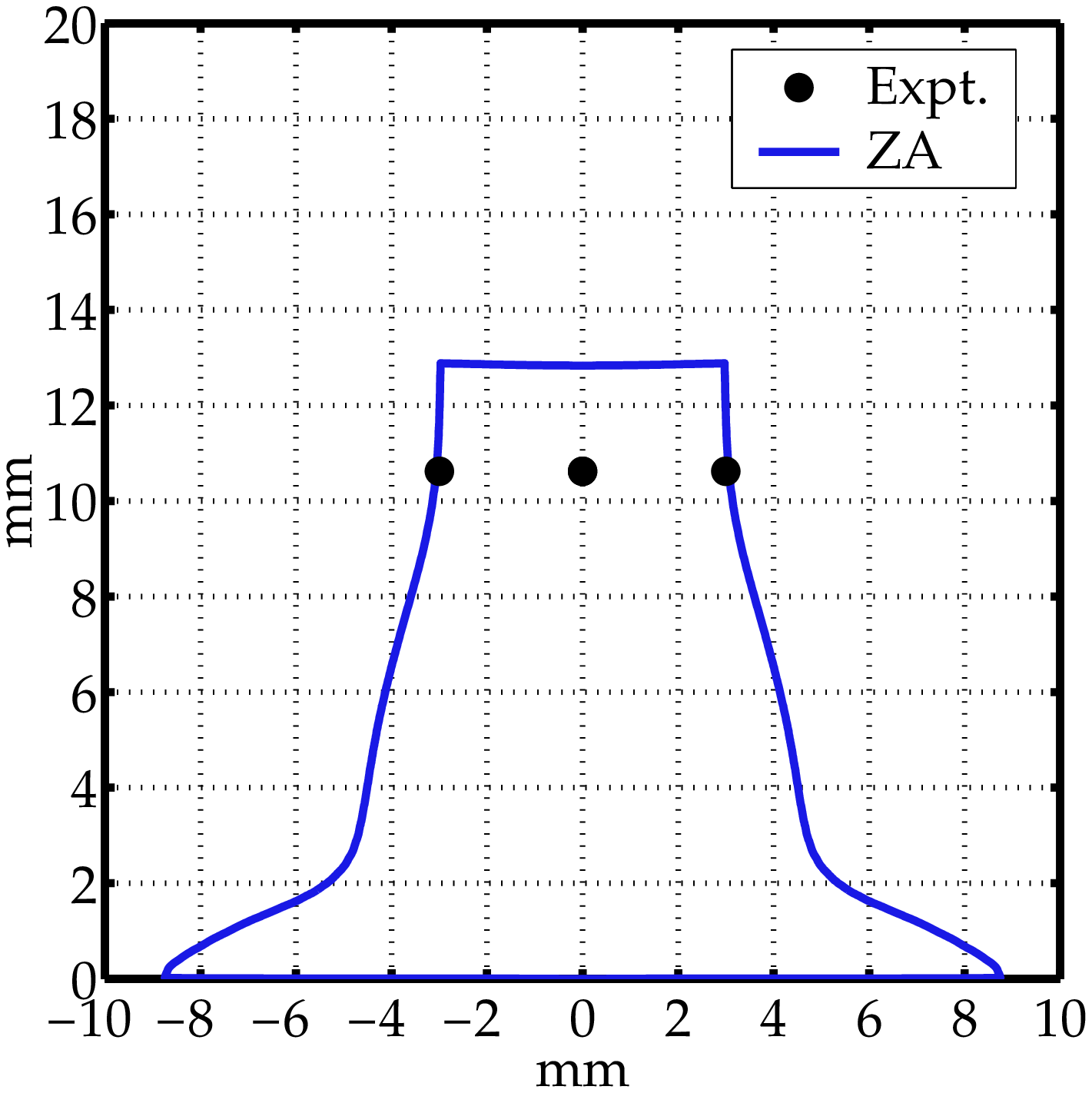}}\\
      (c) Zerilli-Armstrong
    \end{minipage}
    \begin{minipage}[t]{3in}
      \centering
      \scalebox{0.4}{\includegraphics{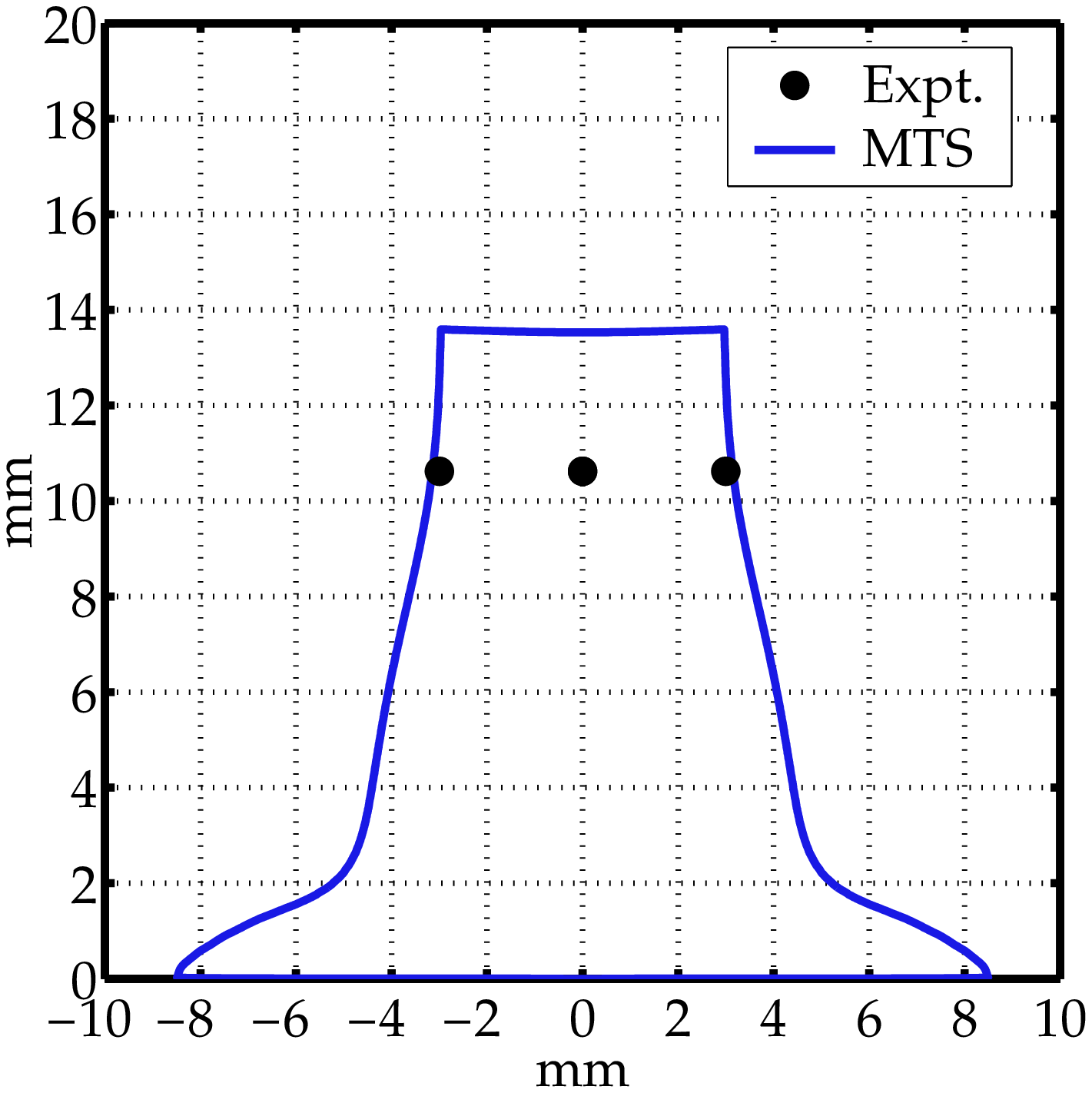}}\\
      (d) Mechanical Threshold Stress.
    \end{minipage}
    \vspace{15pt}
    \begin{minipage}[t]{3in}
      \centering
      \scalebox{0.4}{\includegraphics{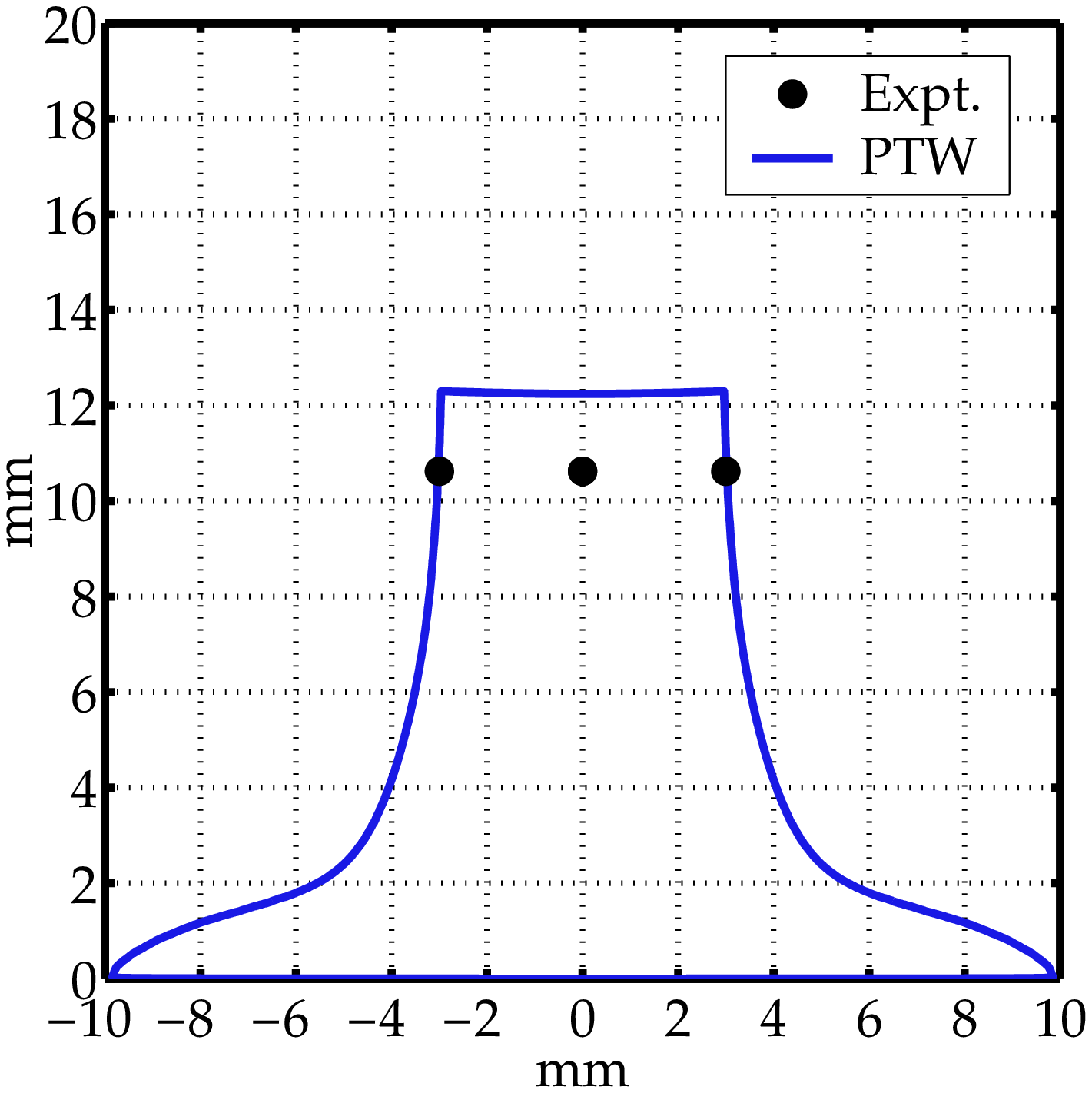}}\\
      (e) Preston-Tonks-Wallace.
    \end{minipage}
    \begin{minipage}[t]{3in}
      \centering
      \scalebox{0.40}{\includegraphics{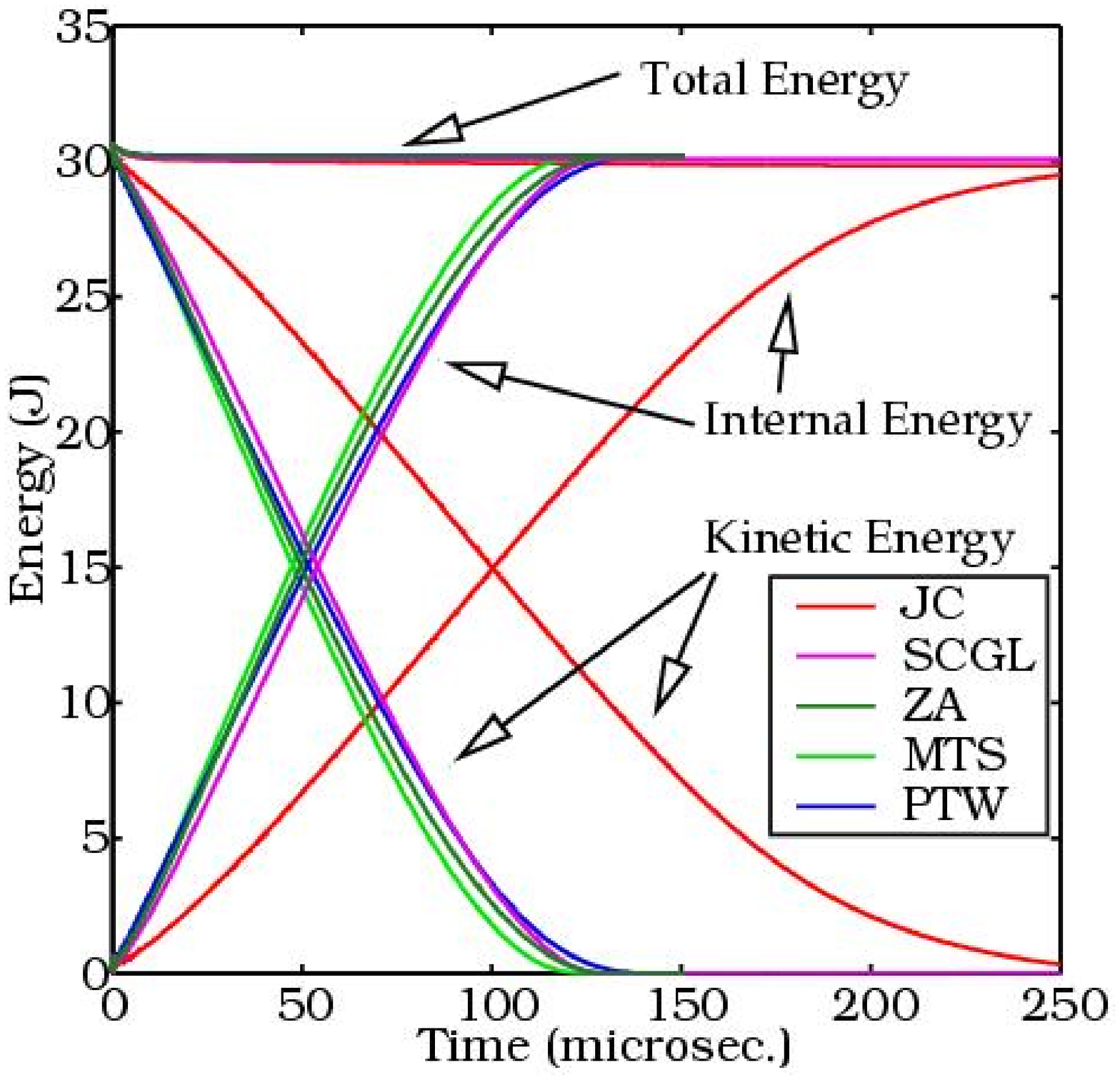}}\\
      (f) Energy vs. time.
    \end{minipage}
    \caption{Computed and experimental profiles for Taylor test Cu-3 and
             the computed energy-time profiles.}
    \label{fig:Cu8TaylorProf}
  \end{figure}

  Test Cu-3 was conducted at 1235 K and at a initial nominal strain-rate of
  approximately 6000 /s.  Figures~\ref{fig:Cu8TaylorProf}(a), (b), (c), (d), 
  and (e) show the profiles computed using the JC, SCGL, ZA, MTS, and 
  PTW models, respectively.  The complete experimental profile of the cylinder 
  was not available for this test.   

  The Johnson-Cook model fails to predict a the deformation of the cylinder at
  this temperature and the material appears to flow along the plane of
  impact.  The SCGL model predicts a reasonably close value of the final 
  length.  However, the low strain-rate part of the SCGL model behaves
  in an unstable manner at some levels of discretization for this test and
  should ideally be discarded in high strain-rate simulations.  The ZA
  model overestimates the final length as does the MTS model.  The PTW
  model predicts a final length that is closer to experiment but does 
  not show the bulge that is characteristic of hardening.  This can be
  seen from the tendency of the model to saturate prematurely as discussed
  in the section on one-dimensional tests.

  The energy plot for test Cu-3 is shown in Figure~\ref{fig:Cu8TaylorProf}(f).
  For this test, the JC model predicts a time of impact greater than 250
  micro secs while the rest of the models predict values between 120 micro secs
  and 130 micro secs.  The reason for the anomalous behavior of the JC
  model is that the rate dependence of the yield stress at high temperature
  is severely underestimated by the JC model.  The nominal strain-rate is 
  around 5000/s for this test at which the yield stress should be considerably
  higher than the 50 MPa that is computed by the JC model.

  We do not have the final profile of the sample for this test and hence
  cannot compare any metrics other than the final length.  The final length
  is predicted most accurately by the SCGL model with an error of 10\%,
  followed by the PTW model (error 15\%) and the ZA model (error 20\%).
  The Johnson-Cook model shown an error of more than 90\%.

  These three sets of tests show that the performance of the models 
  deteriorates with increasing temperature.  However, on average all the
  models predict reasonably accurate profiles for the Taylor impact tests.
  The choice of the model should therefore be dictated by the required 
  computational efficiency and the conditions expected during simulations.

\section{Summary and conclusions}\label{sec:conclude}
  We have compared five flow stress models that are suitable for use in 
  high strain-rate and high temperature simulations using one-dimensional
  tension/compression tests and Taylor impact tests.  We have also evaluated 
  the associated models for shear modulus, melting temperature, 
  and the equation of state.  We observe that during the simulation of
  large plastic deformations at high strain-rates and high temperatures,
  the following should be taken into consideration:
  \begin{enumerate}
    \item The specific heat can be assumed constant when the range of 
          temperatures is small.  However, at temperatures below 250 K 
          or above 750 K, the room temperature value of the specific heat
          may not be appropriate.
    \item The Mie-Gr{\"u}neisen equation of state that we have used is
          valid only up to compressions of 1.3.  A higher order approximation
          should be used if extreme pressures are expected during the 
          simulation.  We note that care should be exercised when this
          equation of state is used for states of large hydrostatic tension.
    \item The physically-based Burakovsky-Preston-Silbar melt temperature 
          model should be used when the material is not well characterized.
          However, the Steinberg-Cochran-Guinan model is quite accurate for
          copper and slightly less computationally expensive.
    \item The Nadal-Le Poac shear modulus model is the most robust model
          for high-temperature high-pressure applications.  However, this
          model (as well as the Steinberg-Cochran-Guinan model) can be 
          unreliable at high hydrostatic tensions.  The MTS shear modulus
          model is the best at temperatures less than 200 K though it is 
          inaccurate at high pressures.
    \item The Johnson-Cook (JC) flow stress model predicts that the strain-rate
          dependence of the yield stress is insignificant at high temperatures.
          This is inaccurate and the model should be recalibrated for 
          high temperature applications.
    \item The contribution of the rate-dependent part of the  
          Steinberg-Cochran-Guinan-Lund (SCGL) model to the total 
          flow stress is small
          enough to be insignificant because it is limited by a constant Peierls
          stress.  The model makes the assumption that 
          $\partial\sigma_y/\partial T = \partial \mu/\partial T$.  This 
          assumption clearly does not hold for copper.  If the model is
          corrected, the high strain-rate yield stresses are predicted quite
          accurately by the model.
    \item The Zerilli-Armstrong (ZA) model is quite accurate at 
          low strain-rates.
          The temperature-dependence of the yield stress predicted by the 
          model is larger than the actual value at high strain-rates with 
          some exceptions.
    \item The Mechanical Threshold Stress (MTS) model shows the 
          correct temperature
          dependence of the yield stress and high temperatures and strain-rates.
          The model is also quite accurate at low strain-rates.  However, 
          the stage-IV hardening at large tensile strains is not predicted
          by the model.
    \item The Preston-Tonks-Wallace (PTW) model is also quite 
          accurate at both low
          and high strain-rates and at high temperatures.  However, the 
          predicted yield stress saturates at a tensile strains of 0.4 and
          the model represents stage-IV hardening worse than the Mechanical
          Threshold Stress model.
    \item The overall error for one-dimensional tests is least for the 
          PTW model.  For the tension tests, the MTS model is the most 
          accurate while the 
          PTW model is the most accurate for compression 
          tests.  The MTS model is the most accurate
          at high strain-rates while the PTW model is the 
          most accurate at low strain-rates.  The PTW model
          is the most accurate for both high- and low-temperature tests at 
          all strain-rates.
    \item For comparing Taylor impact tests, geometric moments of the profile
          are better measures of the shape than arbitrary measures such as
          the width of the bulge at a fixed distance from the end.
    \item At room temperature, all the flow stress models predict the final
          profile of a Taylor impact cylinder quite accurately.  At 718 K,
          the ZA model is the most accurate, followed by the
          MTS model.  At 1235 K, the JC 
          model fails to predict the final profile while the modified
          SCGL and the PTW models
          predict the final length most accurately.  The PTW
          model fails to predict the bulge due to hardening at this  
          temperature due to early saturation of the yield stress.
  \end{enumerate}
  We conclude that no single model can accurately predict the yield stress 
  for the full range of conditions considered in this study.  Care should
  be exercised when choosing the models to be used in a particular simulation.
  If needed, the model should be recalibrated for the range of conditions 
  of interest.

\section*{Acknowledgments}
  This work was supported by the the U.S. Department of Energy through the 
  Center for the Simulation of Accidental Fires and Explosions, under grant 
  W-7405-ENG-48.
\appendix
\section{The Material Point Method}\label{appA}
  The Material Point Method (MPM) ~\citet{Sulsky94,Sulsky95} is a particle 
  method for solid mechanics simulations.  In this method, the state variables 
  of the material are described on Lagrangian particles or ``material points''.
  In addition, a regular, structured Eulerian grid is used as a computational 
  scratch pad to compute spatial gradients and to solve the governing
  conservation equations.  The explicit time-stepping version of the Material 
  Point Method is summarized below.

  It is assumed that an particle state at the beginning of a time step 
  is known.  The mass ($m$), external force ($\Bf^{\Text}$), and 
  velocity ($\Bv$) of the particles are interpolated to the grid using 
  the relations
  \begin{equation}\label{eq:1}
    m_g = \sum_{p} S_{gp}~m_p ~,~~~~
    \Bv_g = (1/m_g)\sum_{p} S_{gp}~m_p~\Bv_p ~,~~~~
    \Bf^{\Text}_g = \sum_{p} S_{gp}~\Bf^{\Text}_p
  \end{equation}
  where the subscript `$g$' indicates a quantity at a grid node and a 
  subscript `$p$' indicates a quantity on a particle.  The symbol $\sum_p$
  indicates a summation over all particles.  The quantity ($S_{gp}$) is 
  the interpolation function of node $g$ evaluated at the position of 
  particle $p$.  

  Next, the velocity gradient at each particle is computed using the 
  grid velocities using the relation
  \begin{equation} \label{eq:2}
    \Grad{\Bv_p} = \sum_g \BG_{gp} \Bv_g
  \end{equation}
  where $\BG_{gp}$ is the gradient of the shape function of node $g$
  evaluated at the position of particle $p$.  The velocity gradient at 
  each particle is used to determine the Cauchy stress ($\Bsig_p$) at the
  particle using a stress update algorithm.

  The internal force at the grid nodes ($\Bf^{\Tint}_g$) is calculated 
  from the divergence of the stress using
  \begin{equation} \label{eq:3}
    \Bf^{\Tint}_g = \sum_p \BG_{gp}~\Bsig_p~V_p
  \end{equation}
  where $V_p$ is the particle volume.
  
  The equation for the conservation of linear momentum is next solved on
  the grid.  This equation can be cast in the form
  \begin{equation} \label{eq:4}
    \Bm_g ~ a_g = \Bf^{\Text}_g - \Bf^{\Tint}_g
  \end{equation}
  where $\Ba_g$ is the acceleration vector at grid node $g$.  

  The velocity vector at node $g$ is updated using an explicit (forward
  Euler) time integration, and the particle velocity and position are then
  updated using grid quantities.  The relevant equations are
  \begin{align} 
     \Bv_g(t+\Delta t) & = \Bv_g(t) + \Ba_g~\Delta t \label{eq:5} \\
     \Bv_p(t+\Delta t) & = \Bv_p(t) + \sum_g S_{gp}~\Ba_g~\Delta t ~;~~~~
     \Bx_p(t+\Delta t) = \Bx_p(t) + \sum_g S_{gp}~\Bv_g~\Delta t \label{eq:6}
  \end{align}

  The above sequence of steps is repeated for each time step.  The above 
  algorithm leads to particularly simple mechanisms for handling contact.
  Details of the interpolants and the contact algorithms can be found 
  in \citet{Bard01} and \citet{Bard04}.  The implicit version of the Material
  Point method can be found in~\citet{Guilkey03}.

\section{Stress Update Algorithm}\label{appB}
  A modified form of a hypoelastic-plastic, semi-implicit elastic-plastic
  stress update algorithm~(\citet{Nemat91,Nemat92,Wang94,Maudlin96,Zocher00}) 
  has been used for the stress update in the simulations presented in this 
  paper.  Details of the socket-based objected oriented form of the 
  complete algorithm can be found in \citet{Banerjee05c}.  We provide a brief 
  description here for the sake of completeness.

  Following ~\citet{Maudlin96}, the rotated spatial rate of deformation 
  tensor ($\Bd$) is decomposed into an elastic part ($\Bd^e$) and a 
  plastic part ($\Bd^p$)
  \begin{equation}
     \Bd = \Bd^e + \Bd^p
  \end{equation}
  If we assume plastic incompressibility ($\Tr{(\Bd^p)} = 0$), we get
  \begin{equation}
     \Beta = \Beta^e + \Beta^p
  \end{equation}
  where $\Beta$, $\Beta^e$, and $\Beta^p$ are the deviatoric parts of $\Bd$,
  $\Bd^e$, and $\Bd^p$, respectively.  For isotropic materials, the hypoelastic
  constitutive equation for deviatoric stress is
  \begin{equation}
    \dot{\Bs} = 2\mu(\Beta - \Beta^p)
  \end{equation}
  where $\Bs$ is the deviatoric part of the stress tensor and $\mu$ is the
  shear modulus.  We assume that the flow stress obeys the Huber-von Mises
  yield condition
  \begin{equation}
    f := \sqrt{\frac{3}{2}}\norm{\Bs} - \sigma_y \le 0  ~~\text{or,}~~
    F := \frac{3}{2} \Bs:\Bs - \sigma_y^2 \le 0 
  \end{equation}
  where $\sigma_y$ is the flow stress.  Assuming an associated flow rule,
  and noting that $\Bd^p = \Beta^p$, we have
  \begin{equation}
    \Beta^p = \Bd^p = \lambda\Partial{f}{\Bsig} 
                    = \Lambda\Partial{F}{\Bsig} = 3\Lambda\Bs
  \end{equation}
  where $\Bsig$ is the stress.  Let $\Bu$ be a tensor proportional to the 
  plastic straining direction, and define $\gamma$ as
  \begin{equation}
    \Bu = \sqrt{3} \frac{\Bs}{\norm{\Bs}}; \quad
    \gamma := \sqrt{3}\Lambda\norm{\Bs}  \quad \Longrightarrow
    \gamma\Bu = 3\Lambda\Bs
  \end{equation}
  Therefore, we have
  \begin{equation} \label{eq:stresseqn}
    \Beta^p = \gamma\Bu; \quad  
    \dot{\Bs} = 2\mu(\Beta - \gamma\Bu)
  \end{equation}
  From the consistency condition, if we assume that the deviatoric stress
  remains constant over a timestep, we get 
  \begin{equation}
    \gamma = \frac{\Bs:\Beta}{\Bs:\Bu}
  \end{equation}
  which provides an initial estimate of the plastic strain-rate.  To obtain
  a semi-implicit update of the stress using equation (\ref{eq:stresseqn}), we
  define
  \begin{equation}\label{eq:taueqn}
    \tau^2 := \frac{3}{2} \Bs:\Bs = \sigma_y^2
  \end{equation}
  Taking a time derivative of equation (\ref{eq:taueqn}) gives us
  \begin{equation}\label{eq:taudot}
    \sqrt{2} \dot{\tau} = \sqrt{3} \frac{\Bs:\dot{\Bs}}{\norm{\Bs}}
  \end{equation}
  Plugging equation (\ref{eq:taudot}) into equation (\ref{eq:stresseqn})$_2$
  we get
  \begin{equation}\label{eq:tau}
    \dot{\tau} = \sqrt{2}\mu(\Bu:\Beta - \gamma\Bu:\Bu)
               = \sqrt{2} \mu (d - 3\gamma)
  \end{equation} 
  where $d = \Bu:\Beta$.  If the initial estimate of the plastic strain-rate
  is that all of the deviatoric strain-rate is plastic, then we get an 
  approximation to $\gamma$, and the corresponding error 
  ($\gamma_{\text{er}}$) given by
  \begin{equation}\label{eq:gammaer}
    \gamma_{\text{approx}} = \frac{d}{3}; \quad
    \gamma_{\text{er}} = \gamma_{\text{approx}} - \gamma = \frac{d}{3} - \gamma
  \end{equation}
  The incremental form of the above equation is
  \begin{equation}\label{eq:delgamma}
    \Delta\gamma = \frac{d^*\Delta t}{3} - \Delta\gamma_{\text{er}}
  \end{equation}
  Integrating equation (\ref{eq:tau}) from time $t_n$ to time $t_{n+1} = 
  t_n + \Delta t$, and using equation (\ref{eq:delgamma}) we get
  \begin{equation}\label{eq:taun}
    \tau_{n+1} = \tau_n + \sqrt{2}\mu(d^*\Delta t - 3\Delta\gamma)
               = \tau_n + 3\sqrt{2}\mu\Delta\gamma_{\text{er}}
  \end{equation}
  where $d^*$ is the average value of $d$ over the timestep.
  Solving for $\Delta\gamma_{\text{er}}$ gives
  \begin{equation}\label{eq:delgammaer}
    \Delta\gamma_{\text{er}} = \cfrac{\tau_{n+1} - \tau_n}{3\sqrt{2}\mu}
      = \cfrac{\sqrt{2}\sigma_y - \sqrt{3}\norm{\Bs_n}}{6\mu}
  \end{equation}
  The direction of the total strain-rate ($\Bu^{\eta}$) and the
  direction of the plastic strain-rate ($\Bu^s$) are given by 
  \begin{equation}
    \Bu^{\eta} = \frac{\Beta}{\norm{\Beta}} ; \quad
    \Bu^{s} = \frac{\Bs}{\norm{\Bs}} 
  \end{equation}
  Let $\theta$ be the fraction of the time increment that sees elastic
  straining.  Then
  \begin{equation}\label{eq:theta}
    \theta = \frac{d^* - 3\gamma_n}{d^*}
  \end{equation}
  where $\gamma_n = d_n/3$ is the value of $\gamma$ at the beginning of the 
  timestep.  We also assume that 
  \begin{equation}\label{eq:dstar}
    d^* = \sqrt{3}\Beta:[(1-\theta)\Bu^{\eta} + \frac{\theta}{2}
              (\Bu^{\eta}+\Bu^{s})]
  \end{equation}
  Plugging equation (\ref{eq:theta}) into equation (\ref{eq:dstar}) we get
  a quadratic equation that can be solved for $d^*$ as follows
  \begin{equation}
    \frac{2}{\sqrt{3}} (d^*)^2 - (\Beta:\Bu^s + \norm{\Beta}) d^*
       + 3\gamma_n (\Beta:\Bu^s - \norm{\Beta}) = 0
  \end{equation}
  The real positive root of the above quadratic equation is taken as the
  estimate for $d$.  The value of $\Delta\gamma$ can now be calculated
  using equations (\ref{eq:delgamma}) and (\ref{eq:delgammaer}).  A
  semi-implicit estimate of the deviatoric stress can be obtained at this
  stage by integrating equation (\ref{eq:stresseqn})$_2$
  \begin{align}
    \tilde{\Bs}_{n+1} & = \Bs_n + 2\mu\left(\eta\Delta t  - \sqrt{3}\Delta\gamma
         \cfrac{\tilde{\Bs}_{n+1}}{\norm{\Bs_{n+1}}}\right) \\
     & = \Bs_n + 2\mu\left(\eta\Delta t  - \frac{3}{\sqrt{2}}\Delta\gamma
         \cfrac{\tilde{\Bs}_{n+1}}{\sigma_y}\right)
  \end{align}
  Solving for $\tilde{\Bs}_{n+1}$, we get
  \begin{equation}
    \tilde{\Bs}_{n+1} = \cfrac{\Bs_{n+1}^{\text{trial}}}
           {1 + 3\sqrt{2}\mu\cfrac{\Delta\gamma}{\sigma_y}}
  \end{equation}
  where $\Bs_{n+1}^{\text{trial}} = \Bs_n + 2\mu\Delta t\Beta$.
  A final radial return adjustment is used to move the stress to the yield
  surface 
  \begin{equation}
    \Bs_{n+1} = \sqrt{\frac{2}{3}}\sigma_y \cfrac{\tilde{\Bs}_{n+1}}
                {\norm{\tilde{\Bs}_{n+1}}}
  \end{equation}
  A pathological situation arises if $\gamma_n = \Bu_n:\Beta_n$ is less
  than or equal to zero or 
  $\Delta\gamma_{\text{er}} \ge \frac{d^*}{3} \Delta t $.  
  This can occur is the rate of plastic deformation
  is small compared to the rate of elastic deformation or if the timestep
  size is too small (see~\citet{Nemat92}).  In such situations, we use a
  locally implicit stress update that uses Newton iterations (as
  discussed in \citet{Simo98}, page 124) to compute $\tilde{\Bs}$.

\section{Computation of Metrics}\label{app:Metric}
  The length of the elastic zone after deformation ($X_f$) is determined
  by checking the deformed diameter with the original diameter of the
  cylinder.  If the difference is greater than 0.003 mm, plastic deformation is
  assumed to have taken place.  The value of $X_f$ is the distance from the
  free end of the cylinder to the first point from the free end where the
  above criterion is met.

  Let the closed polygon representing the final profile of the Taylor cylinder 
  be given by $P = {p_1, p_2, p_3, ...., p_n, p_{n+1} = p_1}$, 
  where $n$ is the number of vertices of the polygon.  We assume that the 
  points are ordered in the counter-clockwise direction.  Each point $p_i$
  has a pair of coordinates ($x_i$, $y_i$).

  Then, the area of the profile ($A_f$) is given by
  \begin{equation}
     A_f = \Half \sum_{i=1}^n (x_i~y_{i+1} - x_{i+1}~y_i) ~.
  \end{equation}

  The centroid of the profile is given by
  \begin{align}
     C_{xf} & = \cfrac{1}{6 A_f} 
       \sum_{i=1}^n (x_i~y_{i+1} - x_{i+1}~y_i) (x_i + x_{i+1}) \\
     C_{yf} & = \cfrac{1}{6 A_f} 
       \sum_{i=1}^n (x_i~y_{i+1} - x_{i+1}~y_i) (y_i + y_{i+1})  ~.
  \end{align}

  The volume of the deformed cylinder is given by the Pappus theorem.  The
  formula for the volume is
  \begin{equation}
     V_f = 2 \pi C_{xf} A_f ~.
  \end{equation}

  The moments of inertia are computed by converting the volume integral into
  a surface integral over the boundary of the profile.  The resulting 
  formulas for the moments of inertia are
  \begin{align}
     I_{xf} & = -\cfrac{1}{12} 
       \sum_{i=1}^n (x_{i+1} - x_i)(y_{i+1} + y_i)(y_{i+1}^2 + y_i^2) \\
     I_{yf} & = \cfrac{1}{12} 
       \sum_{i=1}^n (y_{i+1} - y_i)(x_{i+1} + x_i)(x_{i+1}^2 + x_i^2) ~.
  \end{align}

\bibliographystyle{harvard}
\bibliography{mybiblio}

\end{document}